\documentclass[aps,prd,twocolumn,superscriptaddress]{revtex4-2}
\usepackage[]{mathtools}
\usepackage{bm}
\usepackage{graphicx}
\usepackage{subfigure} 
\usepackage{float}
\usepackage[colorlinks,linkcolor=blue,anchorcolor=blue,citecolor=blue]{hyperref} 
\usepackage{verbatim}
\usepackage{booktabs}
\usepackage{multirow}
\usepackage{color}
\usepackage{mathtools}
\usepackage{longtable}
\usepackage{comment}
\newcommand{\jv}[1]{\textcolor{black}{#1}}
\newcommand{\qh}[1]{\textcolor{black}{#1}}

\begin{document}


\title{Assessing the model waveform accuracy of gravitational waves}
\author{Qian Hu}
\email{q.hu.2@research.gla.ac.uk}
\author{John Veitch}
\email{John.Veitch@glasgow.ac.uk}
\affiliation{Institute for Gravitational Research, School of Physics and Astronomy, University of Glasgow, Glasgow, G12 8QQ, United Kingdom}

\date{\today}

\begin{abstract}
With the improvement in sensitivity of gravitational wave (GW) detectors and the increasing diversity of GW sources, there is a strong need for accurate GW waveform models for data analysis. While the current model accuracy assessments require waveforms generated by numerical relativity (NR) simulations as the ``true waveforms'', in this paper we propose an  assessment approach that does not require NR simulations, which enables us to assess model accuracy everywhere in the parameter space.  By measuring the difference between two waveform models, we derive a necessary condition for a pair of waveform models to both be accurate, for a particular set of parameters. We then apply this method to the parameter estimation samples of the Gravitational-Wave Transient Catalogs GWTC-3 and GWTC-2.1, and find that the waveform accuracy for high signal-to-noise ratio events in some cases fails our assessment criterion. Based on analysis of real events' posterior samples, we discuss the correlation between our quantified accuracy assessments and systematic errors in parameter estimation. We find waveform models that perform worse in our assessment are more likely to give inconsistent estimations. We also investigate waveform accuracy in different parameter regions, and find the accuracy degrades as the spin effects go up, the mass ratio deviates from one,
\jv{or the orbital plane is near-aligned to the line of sight}.
Furthermore, we make predictions of waveform accuracy requirements for future detectors and find the accuracy of current waveform models should be improved by at least 3 orders of magnitude, which is consistent with previous works. 
\end{abstract}

\maketitle
\section{\label{sec1} Introduction}
Over 90 gravitational wave (GW) events have been detected since 2015~\cite{abbott2019:GWTC1GravitationalWaveTransient,abbott2021:GWTC2CompactBinary,abbott2021:GWTC3CompactBinary,theligoscientificcollaboration2021:GWTC2DeepExtended,nitz2019_1OGCFirstOpen,nitz2020_2OGCOpenGravitationalwave,nitz2021_3OGCCatalogGravitational,nitz2021_4OGCCatalogGravitational,Zackay_2021,PhysRevD.101.083030,PhysRevD.100.023011,PhysRevD.100.023007} by Advanced LIGO~\cite{theligoscientificcollaboration2015_AdvancedLIGO} and Advanced Virgo~\cite{acernese2015_AdvancedVirgo2nd}, all of them are from compact binary coalescences (CBCs), where GW waveforms can be modeled by various methods. The data analysis for CBCs such as signal searching~\cite{allen2012:FINDCHIRPAlgorithmDetection} and parameter estimation~\cite{veitch2015:ParameterEstimationCompact,biwer2019:PyCBCInferencePythonbased,lange2018:RapidAccurateParameter,ashton2019:BilbyUserfriendlyBayesian}, are based on the these waveform models of CBCs. Therefore, inaccurate waveforms may cause systematic errors in the scientific interpretation of GW data~\cite{lindblom2008:ModelWaveformAccuracy}. 

{For binary black holes, the evolution can be divided into 3 stages: inspiral, merger, and ringdown, while binary neutron stars and neutron star black hole binaries may exhibit tidal disruption prior to the formation of a final black hole or hypermassive neutron star.} The post-Newtonian (PN) expansion gives a good approximation of the inspiral stage~\cite{blanchet2014_GravitationalRadiationPostnewtonian}, and black hole ringdown can be described by quasi-normal modes~\cite{berti2009_QuasinormalModesBlack}. Other than theoretical approximations that only give the waveform of one of the stages, the most accurate GW waveforms of the whole process of CBCs are generated by numerical relativity (NR)~\cite{jani2016:GeorgiaTechCatalog,mroue2013:Catalog174Binary,boyle2019_SXSCollaborationCatalog}, where the Einstein Field Equations are solved numerically. However, NR waveforms are so expensive to compute that the latest SXS NR waveform catalog contains less than 2000 waveforms~\cite{boyle2019_SXSCollaborationCatalog}. {Besides, NR waveforms are generally short: the inspiral stage is usually calculated for only the last few cycles of the binary (there are exceptions, e.g., Ref.~\cite{Szil_gyi_2015}), plus their sparsity in the parameter space, it is impractical to use them directly in data analysis.} {Coverage of the parameter space is also uneven, as NR waveforms with unequal masses and high spins are more difficult to compute}.

These NR waveforms are therefore used to tune more tractable approaches to waveform modeling, with the aim of minimising the difference between the full NR model and the approximate model.
Several methods exist to compute GW waveforms rapidly,
for example, the \texttt{IMRPhenom}~\cite{ajith2007_PhenomenologicalTemplateFamily,Khan:2015jqa,Pratten:2020fqn,Garcia-Quiros:2020qpx,pratten2021_ComputationallyEfficientModels} family, the \texttt{SEOBNR}~\cite{buonanno1999_EffectiveOnebodyApproach,Damour:2001tu,Bohe:2016gbl,Cotesta:2018fcv,Pan:2013rra,ossokine2020_MultipolarEffectiveonebodyWaveforms} family, \qh{\texttt{TEOBResumS} family~\cite{Nagar:2018zoe,Nagar:2020pcj,Nagar:2019wds}}, and surrogate models~\cite{blackman2015_FastAccuratePrediction,Varma:2019csw,williams2020_PrecessingNumericalRelativity,Field_2014,Purrer:2014fza} like \texttt{NRSur} family. These waveform approximants originate from different ideas of approximation or interpolation, and are calibrated with NR  waveforms or hybridized waveforms of NR simulation and PN approximation. They are widely used in GW data analysis. The state-of-art waveform models from the two families mentioned above, \texttt{IMRPhenomXPHM}~\cite{pratten2021_ComputationallyEfficientModels} and \texttt{SEOBNRv4PHM}~\cite{ossokine2020_MultipolarEffectiveonebodyWaveforms}, are employed in the latest third Gravitational-Wave Transient Catalog (GWTC-3) to extract source properties~\cite{abbott2021:GWTC3CompactBinary}. \texttt{NRSur} and \texttt{TEOBResumS} waveforms are also used in several analyses of LIGO-Virgo data release~\cite{abbott2021:GWTC2CompactBinary, abbott2020:GW190521BinaryBlack,abbott2020:GW190412ObservationBinaryblackhole,abbott2019:GWTC1GravitationalWaveTransient}.

While no approximate waveform will be perfect, we are interested in the question of whether these approximate waveforms are accurate \emph{enough} for the analysis of data from current and future gravitational wave detectors. Ref. \cite{lindblom2008:ModelWaveformAccuracy} gives an accuracy standard of a waveform model used in data analysis under a particular detector noise curve.. It calculates the difference between a model and the ``true waveform'', which in practice is often represented by NR simulations due to their high accuracy, and states the waveform difference should lie within the unit ball centered on the true waveform. Here, the waveform difference is regarded as a vector, and its length can be calculated by the noise-weighted inner product with itself. If the length is less than the unit radius, the detector could not distinguish the model and the true waveform, thus the waveform model is accurate enough for data analysis. The assessment against NR simulations is widely used in the waveform community~\cite{pratten2021_ComputationallyEfficientModels,ossokine2020_MultipolarEffectiveonebodyWaveforms,blackman2015_FastAccuratePrediction,williams2020_PrecessingNumericalRelativity,purrer2020_GravitationalWaveformAccuracy,purrer2020_GravitationalWaveformAccuracy,kumar2016_AccuracyBinaryBlack}. 

However, as mentioned before, the number of NR waveforms is limited, and the assessment against NR is only available on the parameter grids where NR simulations are available. With the improvement of detector sensitivity and accumulating observation time, the diversity of GW sources will increase, and they may be located in parts of the parameter space where waveform approximants have poor or unknown performance. In fact, several intriguing GW events like this have been revealed in GWTC-3. GW191219\_163120 has mass ratio estimated outside of where the waveform models have been calibrated, which results in the uncertainties in its $p_{\mathrm{astro}}$~\cite{abbott2021:GWTC3CompactBinary,theligoscientificcollaboration2021_PopulationMergingCompact}. Parameter estimation of GW200129\_065458 shows notable inconsistencies between the results from two different waveform models \texttt{IMRPhenomXPHM} and \texttt{SEOBNRv4PHM}~\cite{abbott2021:GWTC3CompactBinary,theligoscientificcollaboration2021_PopulationMergingCompact}, which is a source of systematic uncertainty on the presence of orbital precession in this system~\cite{Hannam:2021pit}. {Assessing and mitigating} {waveform systematics for current and future detectors has received considerable attention in recent works~\cite{purrer2020_GravitationalWaveformAccuracy,2020PhRvD.102l4069J,2017PhRvD..96l4041W,ferguson2021_AssessingReadinessNumerical,Kunert:2021hgm,PhysRevD.103.124015}.}

To assess the waveform accuracy in the regions where NR waveforms are not available, we need an alternative approach. In this work, we address this problem by extending the method of ~\cite{lindblom2008:ModelWaveformAccuracy} using the triangle inequality in the noise-weighted inner product space. Instead of calculating the difference between one waveform model and NR simulations, we calculate the difference between two waveform models. We will derive a necessary condition of a pair of the waveform models are both accurate enough, a violation of which means \textit{at least} one of the waveform models is not accurate. Although we can not tell whether one model is inaccurate or both are, the violation of this condition still gives information of waveform model validity to a certain degree, especially when the violation is strong. The model-pair assessment does not require NR simulations, and can be performed anywhere in the parameter space as long as the models are able to generate GW waveforms. 

We will discuss three types of GW waveforms: binary black hole (BBH) waveform, neutron star-black hole (NSBH) waveform, and binary neutron star (BNS) waveform, for compact binaries are the main sources of current GW detection. For BBH waveform, we focus on \texttt{IMRPhenomXPHM} and \texttt{SEOBNRv4PHM} which are used in GWTC-3 and GWTC-2.1 data analysis. We assess their accuracy on the parameter estimation samples of GWTC-3 and GWTC-2.1. We find only part of the samples can pass our assessment, and the overall accuracy performance is on the edge of our criterion. Further analysis and simulations shows the inaccurate samples are basically located in the low mass ratio region (we define mass ratio $q<1$), the high spin region and \qh{the edge-on region ($\theta_\mathrm{JN}\sim \pi/2$)}. Based on this, we conclude that waveform accuracy should be improved by at least 3-4 orders of magnitude for the 3rd generation GW detectors, which is consistent with previous works~\cite{purrer2020_GravitationalWaveformAccuracy}. Besides, thanks to the sufficient amount of BBH events, we are able to perform a population-level analysis on the relation between the difference of the waveform models and the posterior sample inconsistency the different waveforms lead to. We find events with less than 40\% posterior samples that can meet our accuracy standard tend to have inconsistent results from \texttt{IMRPhenomXPHM} and \texttt{SEOBNRv4PHM}. For NSBH and BNS waveform models, we perform similar but simpler analysis, as most of them do not include higher modes and precession effects, which may constrain their validness in data analysis. 

This paper is organized as follows. In Sec.~\ref{sec2}, we introduce our accuracy assessment method, including assessment for detector response in Sec.~\ref{sec2a} and normalization of the waveform difference and its relation to overlap (or mismatch) in Sec.~\ref{sec2c}. In Sec.~\ref{sec3}, we apply our method on the 3 types of waveforms mentioned above. Results of BBH waveforms are showed in Sec.~\ref{sec3a}; NSBH and BNS waveform are showed in Sec.~\ref{sec3b}.
In Sec.~\ref{sec4} we summarize our methods and conclusions.

\section{\label{sec2} Assessing waveform accuracy}
In this section we will introduce the waveform accuracy standard proposed in Ref.~\cite{lindblom2008:ModelWaveformAccuracy}, then extend it to model-pair case. We will discuss different standards \qh{for the detector response and for waveforms at a fixed signal-to-noise ratio (SNR)}, which reflects the intrinsic accuracy in the parameter space.

\subsection{\label{sec2a}Assessment of the detector response}
We will use frequency domain waveforms $h_i(f)$, where $f$ means frequency, $i=0$ denotes the true waveform, and $i=1,2$ denotes the 1st and 2nd waveform models. We define inner product between two frequency series as follows:
\begin{equation}
    \label{innpro0}
    (a \mid b) = 4\int_{0}^{+ \infty} { \frac{a^*(f) b(f)}{S_{n}(f)}}df,
\end{equation}
where \textit{star} means complex conjugate, and $S_{n}(f)$ is the power spectral density (PSD) of the detector \qh{which is defined as}
\begin{equation}
    <n^*(f)n(f')> = \frac{1}{2} S_n(f) \delta (f-f').
\end{equation}
Here $<\dots>$ denotes ensemble average and $n$ is the detector noise. 
Note that Eq.~\ref{innpro0} defines an inner product space, in which frequency series can be treated as vectors. We can define the length (or norm) of a vector:
\begin{equation}
    \| a \| = \sqrt{(a|a)}.
\end{equation}
Some literature defines the inner product as the real part of Eq.~\ref{innpro0}, but there is no difference between two definitions when it comes to the length. As other inner product spaces, the Cauchy-Schwarz inequality and the triangle inequality hold:
\begin{equation}
    \label{csineq}
    \| a \|^2 \| b \|^2 \ge |(a|b)|^2
\end{equation}
\begin{equation}
    \label{triieq}
    \|a\| + \|b\| \ge \|a \pm b\| \ge \mid \|a\| -\|b\| \mid .
\end{equation}

A model waveform can be thought as ``accurate enough'' when the detector can not distinguish it from the real one. Ref.~\cite{lindblom2008:ModelWaveformAccuracy} constructs a waveform family $H$ to quantify the detector's ability to measure the difference between the model and the real waveform, which will be used and extended in this section. \qh{Let $h_0$ be the the true waveform, $h_1$ be the waveform given by the first model, and $\delta h_1 = h_1-h_0$ represents their difference. We construct the following waveform family of the first model}
\begin{equation}
    \label{wavefam}
    H_1(\lambda) = (1-\lambda)h_0 + \lambda h_1 = h_0 + \lambda \delta h_1, ~~~0<\lambda<1,
\end{equation}
where $\lambda$ is a parameter which interpolates between the two models. If the measurement error on $\lambda$ is greater than the length of its domain of definition (i.e. the parametric distance between real and model waveforms), we can claim the detector is not able to distinguish the waveforms, thus the model is accurate enough. The error $\sigma_\lambda$ is given by~\cite{finn1992_DetectionMeasurementGravitational,cutler1994_GravitationalWavesMerging}
\begin{equation}
    \sigma_{\lambda}^{-2}=\left( \frac{\partial H_1}{\partial \lambda} \left| \frac{\partial H_1}{\partial \lambda}\right)\right.=(\delta h_1 \mid \delta h_1).
\end{equation}
Therefore, the accuracy standard for a waveform model is 
\begin{equation}
    \label{deltah1pdeltah1p}
    \| \delta h_1\|^2 = (\delta h_1 \mid \delta h_1)<1.
\end{equation}
Eq.~\ref{deltah1pdeltah1p} implies the waveform difference should lie within a unit ball in the inner product space, any violation of which means the model is not accurate enough.
\jv{Since, $<n|n>=1$~\cite{Damour:2010zb}, another way to understand Eq.~\ref{deltah1pdeltah1p} is that if the distance to the real waveform is longer than the length of detector noise, the detector will be able to tell the error of the model. From this angle, the waveform we are considering here should be the detector response, i.e.,}
\begin{equation}
    \begin{aligned}
        h_0 &= F_+ h^+_0 + F_{\times} h^{\times}_0 \\
        h_1 &= F_+ h^+_1 + F_{\times} h^{\times}_1,
    \end{aligned}
\end{equation}
\qh{where $F_+,F_{\times}$ are antenna response functions determined by source sky direction and detector orientation~\cite{jaranowski1998_DataAnalysisGravitationalwave}. $+,\times$ denote plus and cross polarizations of GWs. We only consider polarizations under Einstein's general relativity in this work.}

\qh{We note that some works ~\cite{lindblom2008:ModelWaveformAccuracy,Santamaria:2010yb,Damour:2010zb} propose a less stringent criterion than Eq.~\ref{deltah1pdeltah1p} by changing the upper limit to $2\epsilon \rho^2$, where $\rho$ is SNR and $\epsilon$ is the maximum tolerated fractional loss in SNR which needs to be appropriately chosen for detection. In this work, we focus more on the waveform systematics in measurement rather than in detection, so we keep using  Eq.~\ref{deltah1pdeltah1p}, i.e., the strict distinguishability criterion.
}

However, to compute $\| \delta h_1\|$, the true waveform $h_0$ is needed, which is usually replaced by the computationally expensive NR simulations that can not span all over the parameter space. As a result, the uncertainties are unknown for the waveforms out of the model's calibration range and this may cause some unknown systematic errors in data analysis. An example is GW191219\_163120~\cite{abbott2021:GWTC3CompactBinary}, of which mass ratio is estimated to be out of the waveform calibration region ($\le 0.041$) so that there are potential uncertainties in its $p_\mathrm{astro}$. 

To avoid being limited by the true waveform $h_0$, we introduce another waveform model $h_2$ to be paired with $h_1$. Although $\delta h_1$ and $\delta h_2$ are unknown, their difference $\delta h_1 - \delta h_2$ can be easily calculated: 
\begin{equation}
    \label{defDeltap}
    \begin{aligned}
        \Delta &= \delta h_1 - \delta h_2  \\
        & = (h_1-h_0) - (h_2-h_0) \\
        & = h_1 - h_2.
    \end{aligned}
\end{equation}
Assuming both of two waveforms are accurate, i.e., they both satisfy Eq.~\ref{deltah1pdeltah1p}, we can obtain an upper limit of $\|\Delta\|$ using the triangle inequality:
\begin{equation}
    \label{Deltap}
    \| \Delta \| \le \|\delta h_1\| + \|\delta h_2\| < 2.
\end{equation}
Eq.~\ref{Deltap} is a necessary condition if $h_1$ and $h_2$ are both accurate. That is to say, if Eq.~\ref{Deltap} is violated, at least one of the waveform models does not satisfy Eq.~\ref{deltah1pdeltah1p}. 

\begin{figure}
    \includegraphics[width=0.45\textwidth]{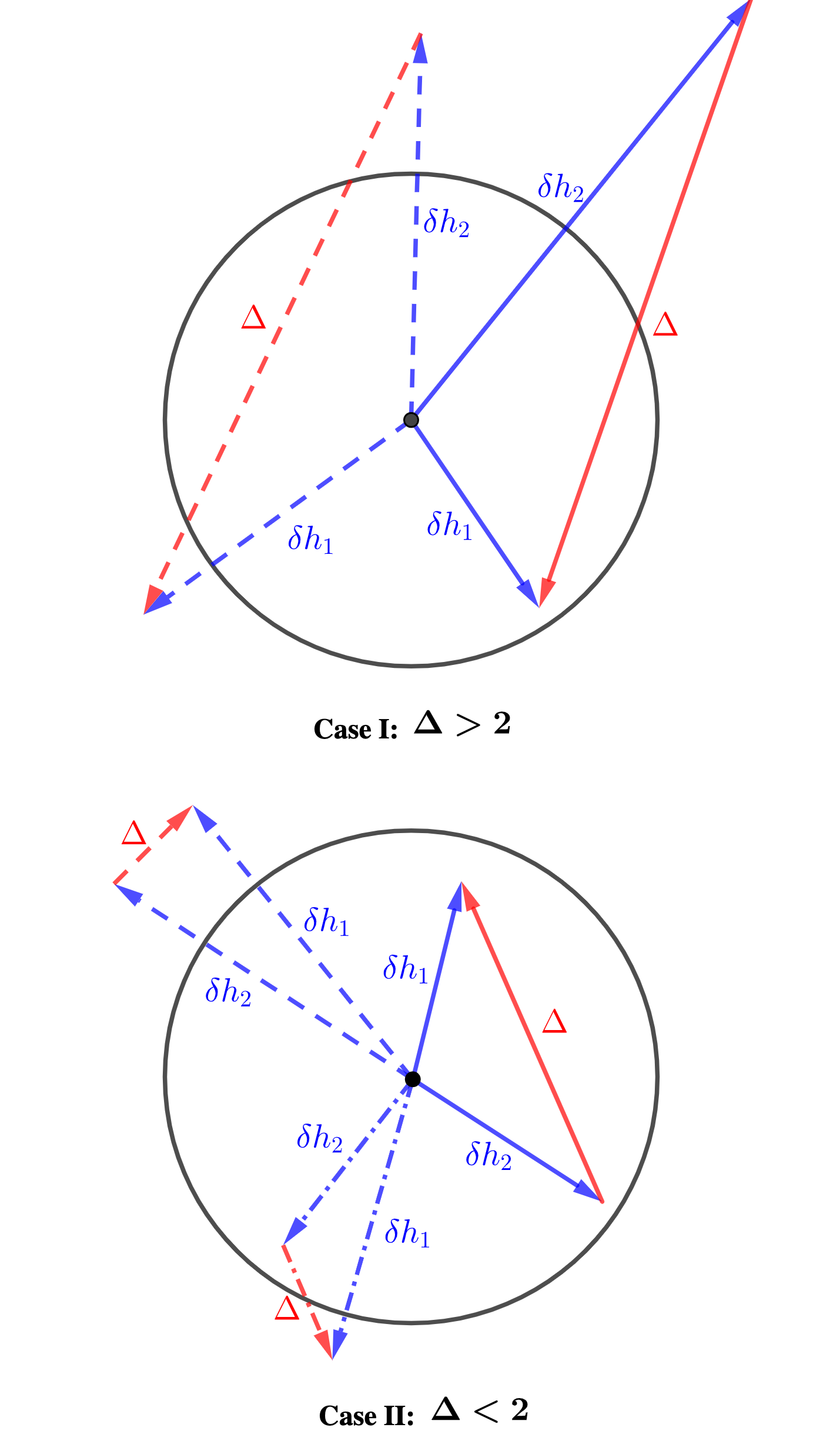}
    \centering
    \caption{\label{vecplot} Vector plots to illustrate all cases of $\Delta$. Blue vectors are the difference between waveform models and the real waveform, and black circles represent the sphere of radius $2$, the upper limit of length of $\delta h_i$ if $h_i$ is accurate ($i=1,2$). Red vectors are $\Delta$, the difference between two waveform models (defined in Eq.~\ref{defDeltap}). Different line styles denote different possibilities. In Case I, $\|\Delta\|$ exceeds the upper limit given by Eq.~\ref{Deltap}, so at least one in $h_1,h_2$ is not accurate enough. In Case II, $\|\Delta\|$ satisfies Eq.~\ref{Deltap}, there may be 0, 1, 2 inaccurate waveforms, corresponding to solid line, dotted-dashed line and dashed line, respectively. We can not determine the accuracy of a waveform pair in Case II.}
\end{figure}

We illustrate possible cases for the $\|\Delta\|$ in Eq.~\ref{Deltap} in the vector plots Fig.~\ref{vecplot}, in which waveforms are treated as vectors in the noise-weighted inner product space. The black circle denotes the sphere of radius $2$. Vectors $\delta h_1$ and $\delta h_2$ denote the difference between the real waveform $h_0$ and the models $h_1$, $h_2$, respectively, and different line styles denote different possibilities. $\delta h_i$ lies in the circle means the i-th model is accurate and satisfies Eq.~\ref{deltah1pdeltah1p}. If the length of $\Delta$ is greater than the upper limit $2$ (the diameter of the black circle), as shown in case I, \textit{at least} one of the waveform model errors can not be put inside the circle, i.e., it does not meet the accuracy standard. However, $\| \Delta \| <2 $ does not mean both of the waveforms are accurate, as shown in Case II. Small $\|\Delta\|$ only implies the two models give similar predictions of the waveform but can not guarantee their accuracy. The key idea of this method is: if two waveforms have significant difference, they can not both be correct.

Eq.~\ref{Deltap} is not a strong criterion; it can not tell which waveform causes the violation (case I), and may miss some waveform errors (case II). Despite this, we suppose it still gives certain information about the correctness of waveform modeling. If $\| \Delta \|>2$, the waveform pair should become less reliable; if $\| \Delta \|\gg 2$, the systematic errors in waveform models should not be neglected as it is highly possible that either of the waveforms is accurate, or one of them has seriously deviated. If $\| \Delta \|< 2$, no evidence of waveform inaccuracy is found by this approach, although we could not exclude the possibility that two waveforms have large but similar errors. The advantage of this method is that it can be performed everywhere in the parameter space, as long as waveform models work in that region.

Eq.\ref{Deltap} can be extended to a detector network by defining inner product between matrices (whose elements are frequency series):
\begin{equation}
    \label{matrixinnerprod}
    \mathbf{C} = (\mathbf{D} |\mathbf{B} ) \Rightarrow C_{j k}=\sum_{p=1}^{n}\left(D_{j p} \mid B_{p k}\right),
\end{equation}
where $\mathbf{D}$ is an $m\times n$ matrix, $\mathbf{B}$ is an $n\times l$ matrix and the result $\mathbf{C}$ is an $m\times l$ matrix. The signal of the network can be denoted as a column vector $\mathbf{h} = (h^{(1)}, h^{(2)}, \dots, h^{(\mathrm{N_d})})^{\mathrm{T}}$, where superscript $(k)$ denotes the k-th detector and $\mathrm{N_d}$ is the number of detectors in the network. We can also subtract two waveform models, and define $\mathbf{h_1} - \mathbf{h_2} = \Delta_{\mathrm{net}}$. The norm of $\Delta_{\mathrm{net}} $ can be calculated
\begin{equation}
    \begin{aligned}
        \|\Delta_{\mathrm{net}} \|^2 &= (\delta\mathbf{h^T} |\delta \mathbf{h}) = \sum_k (\delta h^{(k)} | \delta h^{(k)})\\
        &= \sum_k \left( \Delta^{(k)} \right)^2  < 4 \mathrm{N_d},
    \end{aligned}
\end{equation}
where $F^{(k)}_+, F^{(k)}_{\times}$ are the antenna response functions of the k-th detector. In practice, we can weight the $\Delta$ by the \qh{number of detectors}:
\begin{equation}
    \label{deltaprime}
    \Delta'_{\mathrm{net}} = \frac{\Delta_{\mathrm{net}}}{\sqrt{\mathrm{N_d}}}, 
\end{equation}
so that the $\Delta'_{\mathrm{net}}$ will have an upper limit of 2 {if the waveforms are both accurate enough}. 

\subsection{\label{sec2c}Assessment at fixed SNR}
The two accuracy standards we proposed, Eq.~\ref{Deltap} and Eq.~\ref{deltaprime}, are related to the SNR, as the length of $\Delta$ is proportional to the amplitude of GWs. It is reasonable that the higher the SNR is, the easier it is for detectors to distinguish different waveforms, and the more important systematic errors will be in data analysis. However, SNR depends on not only intrinsic parameters, but also extrinsic parameters \qh{that trivially modulate the amplitude. It is the phase evolution that is critical to reveal physical properties of the source, and is the intrinsic characteristic of a GW waveform~\cite{McWilliams:2010eq} }
We therefore normalize the $\Delta$ with SNR to eliminate the impacts from \qh{amplitudes}. The optimal SNR is defined as $\rho = \sqrt{(h | h)}$~\cite{finn1992_DetectionMeasurementGravitational}, which is also proportional to the amplitude of GWs like $\Delta$. Thus we have $\Delta \propto \rho$. In fact, we have two waveforms to calculate $\Delta$. The normalization factor is chosen as the geometric mean of SNRs from two waveforms, i.e., $\rho_0 = \sqrt{\rho_1\rho_2}$. Take Eq.~\ref{Deltap} as an example, the normalized $\Delta$ is 

\begin{equation}
    \begin{aligned}
        \|\Delta_{\mathrm{SNR}=1}\|^2 &= \frac{(\delta h_1 - \delta h_2 | \delta h_1 - \delta h_2)}{\sqrt{(h_1 | h_1)(h_2 | h_2)}} \\
        &= \frac{( h_1 -  h_2 |  h_1 -  h_2)}{\sqrt{(h_1 | h_1)(h_2 | h_2)}},
    \end{aligned}
\end{equation}
and we simply have 
\begin{equation}
    \label{deltapn}
    \|\Delta_{\mathrm{SNR}=\rho_{0}}\| = \rho_0 \|\Delta_{\mathrm{SNR}=1}\|.
\end{equation}
Eq.~\ref{deltapn} can be used to evaluate waveform accuracy at a fixed SNR. Note the threshold of $\|\Delta_{\mathrm{SNR}=\rho_{0}}\|$ is always 2. 

The normalized $\| \Delta \|$ can be related to the current waveform evaluation variable, overlap $\mathcal{O}$, which is defined as 
\begin{equation}
    \mathcal{O}(h_1, h_2) = \Re \frac{(h_1|h_2)}{\sqrt{(h_1 | h_1)(h_2 | h_2)}},
\end{equation}
where $\Re$ means the real part. $\mathcal{O}$ is between 0 and 1, the higher value represents higher similarities between waveforms $h_1$ and $h_2$. One can define mismatch $\mathcal{M}=1-\mathcal{O}$. Overlap (or the equivalent mismatch) is widely used to assess the correctness of GW waveforms. The state-of-art models of \texttt{IMRPhenom} and \texttt{SEOBNR} families can achieve {mismatches between $10^{-5}$ and $10^{-1}$ }compared with NR simulations~\cite{pratten2021_ComputationallyEfficientModels,ossokine2020_MultipolarEffectiveonebodyWaveforms}, with precession effects and higher modes being taken into consideration. Overlap between two waveform models $\mathcal{O}(h_1, h_2)$ and the length of normalized waveform difference $\|\Delta\|$ have the following relation:
\begin{equation}
    \label{Eq:rela_with_match}
    \|\Delta_{\mathrm{SNR}=1}\|^2 = \frac{\rho_1}{\rho_2} + \frac{\rho_2}{\rho_1} - 2\mathcal{O}(h_1, h_2) \approx 2(1-\mathcal{O}),
\end{equation}
where $\rho_i = \sqrt{(h_i|h_i)},~i=1,2$. $\|\Delta_{\mathrm{SNR}=1}\|$ will decrease with the increase of overlap, and a pair of identical waveforms give $\mathcal{O}=1$ and $\|\Delta\| = 0$. 

We should mention that the inner product in the calculation of waveform difference $\Delta$ (as well as overlap~\cite{kumar2016_AccuracyBinaryBlack}), should be minimized (or, for overlap, maximized) over an arbitrary phase $\phi_0$ and time shift $t_0$, in order to eliminate the kinematical difference between models~\cite{Damour:2010zb}. Considering the sensitive frequency band of current GW detectors, the inner product is integrated from 20\,Hz to 2048\,Hz throughout this paper.

\section{\label{sec3} Applications}
In this section we will apply the accuracy standard Eq.~\ref{deltaprime} and Eq.~\ref{deltapn} to GW waveforms from 3 types of compact binary coalescence: BBH, NSBH, and BNS. We employ the assessment on the GWTC parameter estimation samples and parameter grids we generate; the former aims to investigate whether faulty waveforms were used in GW data analysis and possible systematic error caused by waveform errors, while the latter explores waveforms' performances in different regions of the parameter space. 
Throughout this paper, we ignore the calibration error, which can cause our waveforms {to be} slightly different from those used in GWTC-3 and GWTC-2.1 parameter estimation. The calibration error is typically $<4$ degrees in phase and $<7$\% in amplitude~\cite{Sun:2020wke}, and it acts on both waveform models, so ignoring it will not have large impacts on our results.

\subsection{\label{sec3a}BBH waveforms}
BBH mergers are the most frequent GW events at this stage: Among all 91 GW candidates (36 in GWTC-3~\cite{abbott2021:GWTC3CompactBinary}, 44 in GWTC-2.1~\cite{theligoscientificcollaboration2021:GWTC2DeepExtended} and 11 in GWTC-1~\cite{abbott2019:GWTC1GravitationalWaveTransient}), over 80 of them are confirmed to be BBH events. In the latest data release from LIGO-Virgo-KAGRA (LVK) collaboration, waveform models \texttt{IMRPhenomXPHM} and \texttt{SEOBNRv4PHM} are used for analysis of all the BBH events, including re-analysis of GWTC-1 events published in GWTC-2.1. Due to the low SNR of current NSBH events, the resolution of tidal deformability is poor and no strong sign of matter effects is revealed in data analysis. Besides, higher modes and spin precession eﬀects are more important than matter effects for waveform modeling of NSBHs~\cite{theligoscientificcollaboration2021:ObservationGravitationalWaves}, so  \texttt{IMRPhenomXPHM} and \texttt{SEOBNRv4PHM} are also employed on NSBH events to extract physical information. 

For all the 89 BBH and NSBH events, we use the \qh{cosmologically reweighted} parameter estimation posterior samples from GWTC-3 and GWTC-2.1 data release and calculate $\|\Delta'_{\mathrm{net}}\|$ (Eq.~\ref{deltaprime} ) of the waveform models mentioned above. We use the mixture of \texttt{IMRPhenomXPHM} and \texttt{SEOBNRv4PHM} samples in most events, but in some events \texttt{SEOBNRv4PHM} samples are not  provided~\cite{theligoscientificcollaboration2021:GWTC2DeepExtended}, so we use \texttt{IMRPhenomXPHM} samples to calculate $\|\Delta'_{\mathrm{net}}\|$ between \texttt{IMRPhenomXPHM} and \texttt{SEOBNRv4PHM}. \qh{Samples we use are the same as GWTC-3~\cite{abbott2021:GWTC3CompactBinary} and GWTC-2.1~\cite{theligoscientificcollaboration2021:GWTC2DeepExtended}. For each sample, we generate the waveform (including the detector response) for both models, then apply a time and phase shift on one of them to minimize Eq.~\ref{deltaprime}. The minimized $\|\Delta'_{\mathrm{net}}\|$ is the waveform difference we refer to in the following discussion. } 

When $\|\Delta'_{\mathrm{net}}\|$ is greater than 2 at a sampling point, it implies the difference between \texttt{IMRPhenomXPHM} and \texttt{SEOBNRv4PHM} is so large at this point that they could not both be accurate enough. Furthermore, the difference in waveform will induce a difference in likelihood, and therefore has the potential to affect the results of a parameter estimation algorithm.
This yields a systematic difference in parameter estimates, and so the results from different waveform models may not coincide. Therefore, in addition to $\|\Delta'_{\mathrm{net}}\|$, we also calculate Jensen–Shannon (J-S) \qh{divergence} between \texttt{IMRPhenomXPHM} samples and \texttt{SEOBNRv4PHM} samples (if available). The J-S divergence is a measurement of the similarity between two {probability distributions}
and is used in GWTC-2~\cite{abbott2021:GWTC2CompactBinary}. The greater it is, the {greater the difference between the two distributions}
and there may be potential systematic errors in the data analysis.
Since the J-S divergence for samples from a distribution is easiest to evaluate in one dimension, we choose the greatest J-S divergence among samples for the following parameters: mass ratio $q$, chirp mass $\mathcal{M}$, effective spin $\chi_\mathrm{eff}$ and effective precession spin $\chi_\mathrm{p}$ as a measurement of inconsistency of posterior samples, for they are the major physical parameters to be studied. \qh{We use \texttt{gaussian\_kde} in \texttt{SciPy} to estimate probability density functions. The base of J-S divergence is chosen to be 2, so that the divergence ranges between 0 and 1.}

The full results of the 89 BBH and NSBH events are showed splitly in Tab.~\ref{tab1} and Tab.~\ref{tab2} in App.~\ref{apdxb} for reference. We list the basic information of each event, including some source parameters and network SNR, and statistics we construct, including mean value of $\|\Delta'_{\mathrm{net}}\|$, normalized $\|\Delta'_{\mathrm{net}}\|$ (which equals to $\|\Delta'_{\mathrm{net}}\|/\mathrm{SNR}$), fraction of $\|\Delta'_{\mathrm{net}}\|<2$ samples, and the J-S divergence. We highlight some points in the rest of this subsection.

\subsubsection{Overall accuracy\label{sec3a1}}
\begin{figure*}
    \includegraphics[width=0.95\textwidth]{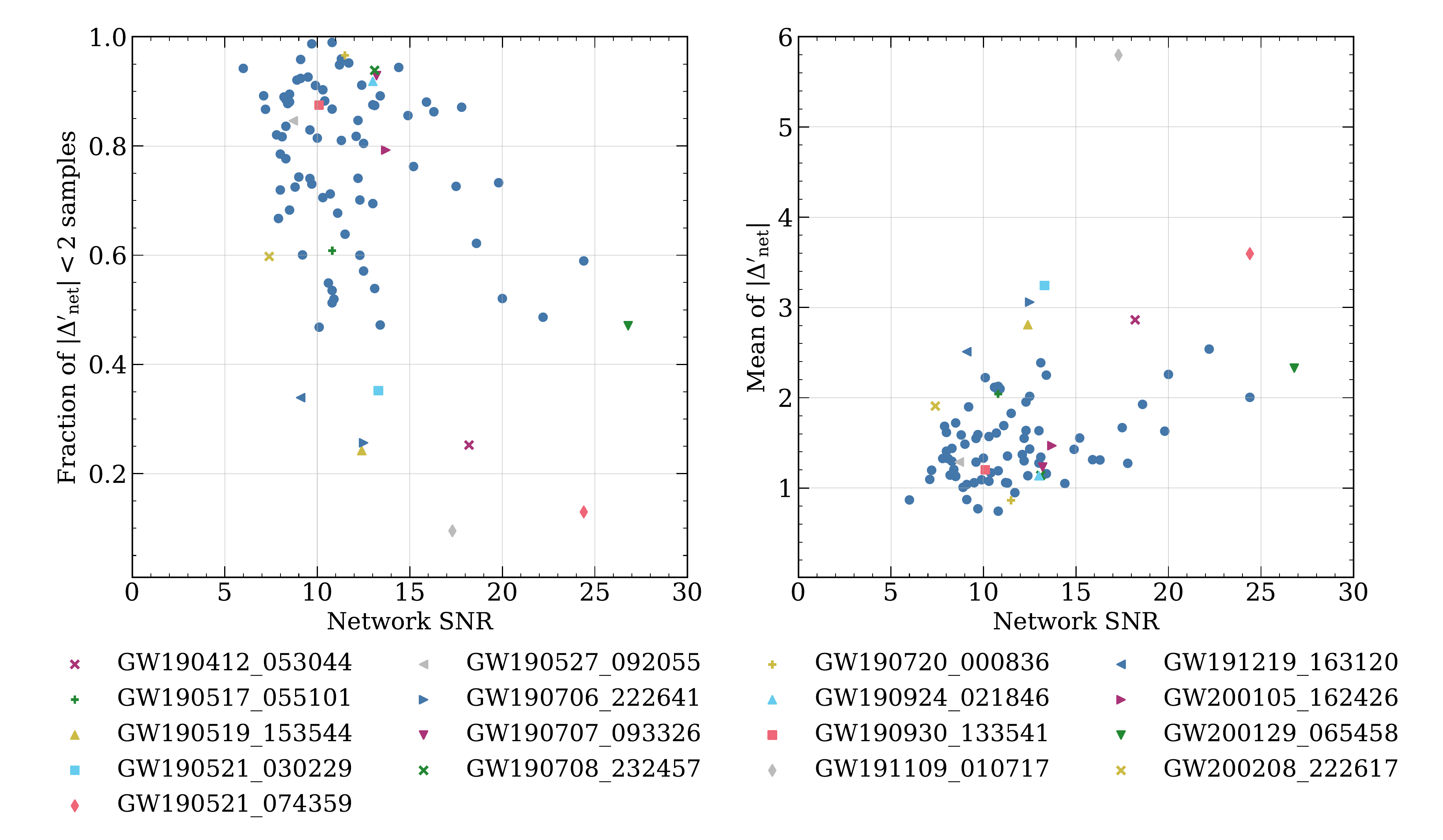}
    \centering
    \caption{\label{pic:delta_vs_SNR}  The left panel shows the relation between fraction of samples that meet our accuracy standard ($\|\Delta'_{\mathrm{net}}\|<2$) and network SNR, and right panel shows mean value of all samples' $\|\Delta'_{\mathrm{net}}\|$ and network SNR. We highlight the events whose $\|\Delta'_{\mathrm{net}}\|<2$ samples fraction is less than 0.4 and whose maximum J-S divergence is greater than $0.1$. These two plots show waveforms of higher SNR events are more likely to violate our waveform accuracy standard, and given the current detector sensitivity, we are already observing some events which violate our assessment criterion. Note the normalized $\|\Delta'_{\mathrm{net}}\|$ can also be read out from the right  panel: it is the slope of the line that connects the origin and each point. We can compare the waveform difference of these events when they have the same SNR by comparing the slope. The numerical values are given in the 7th-10th columns of Tab.~\ref{tab1} and ~\ref{tab2}.}
\end{figure*}

We show the relations between the waveform difference $\|\Delta'_{\mathrm{net}}\|$ of different events and SNR in Fig.~\ref{pic:delta_vs_SNR}, in which each point represents a GW event. We find every event has samples that can not meet the $\|\Delta'_{\mathrm{net}}\|<2$ requirement (left panel), but most events have mean $\|\Delta'_{\mathrm{net}}\|$ around 2 (right panel). This means some waveform pairs used in data analysis can pass (and are near the edge of) our accuracy standard, but violations exist. We could not identify
{whether one or both waveform models is inaccurate}.
Later in Sec.~\ref{sec3a3} we will show it is the samples with large spin or small mass ratio \qh{ or edge-on inclination} that contribute to $\|\Delta'_{\mathrm{net}}\|<2$ fraction. Overall speaking, considering that the violations are generally not strong, we conclude that the current waveform accuracy is around the edge of our assessment standard for the current detector sensitivity which makes detections of SNRs ranging from $8$ to $\sim 30$.

Although the properties of GW sources differ, there is a tendency that large SNR events are more likely to have greater waveform difference (as expected by Eq.~\ref{deltapn}), and have fewer samples that meet the $\|\Delta'_{\mathrm{net}}\|<2$ requirement. This emphasizes the importance of waveform modeling for future GW detections, in which the SNR can reach hundreds to thousands.  We can also make a rough estimation of waveform accuracy requirements for future detectors. The mismatch $\mathcal{M}$ with the ``true'' waveform is widely-used to assess the waveform accuracy, and the relation between $\|\Delta\|$ and $\mathcal{M}$ can be derived with Eqs.~\ref{deltapn} and~\ref{Eq:rela_with_match}:
\begin{equation}
\label{Eq:future_estmate}
\mathcal{M}(h_1,h_2) \approx \frac{1}{2\rho^2_0} \|\Delta_{\mathrm{SNR}=\rho_0} (h_1,h_2) \|^2 
\end{equation}
Eq.~\ref{Eq:future_estmate} gives the mismatch between two waveform models, but limited by the triangle inequality, the mismatch between models $\mathcal{M}(h_1,h_2)$ should be at the same order of magnitude as the mismatch between a model and the real waveform $\mathcal{M}(h_1,h_0)$, under the assumption that both models are well-calibrated by high precision waveforms like NR simulation. 
From our previous discussion we know the $\|\Delta\|$ is around the edge of its upper limit under the current detector sensitivity. If we assume $\|\Delta\|$ is of the same range for future detectors, and SNR is roughly 30–100 times higher, we can determine that the mismatch should decrease 3–4 orders of magnitude. This is consistent with the results reported in Ref.~\cite{purrer2020_GravitationalWaveformAccuracy}.

\subsubsection{Impact on parameter estimation\label{sec3a2}}
From previous discussions, the waveforms generated by posterior samples of GWTC-3 and GTWC-2.1 are mostly within the waveform difference bound, yet there are some exceptions. Seven GW events have more than 60\% posterior samples violating the standard, which means the difference of two waveform models might be too large to ensure their accuracies. The difference of waveforms may result in difference in parameter estimation, indicating systematic errors~\cite{2017PhRvD..96l4041W,purrer2020_GravitationalWaveformAccuracy,cutler2007_LISADetectionsMassive}. 

\begin{figure*}
    \includegraphics[width=0.95\textwidth]{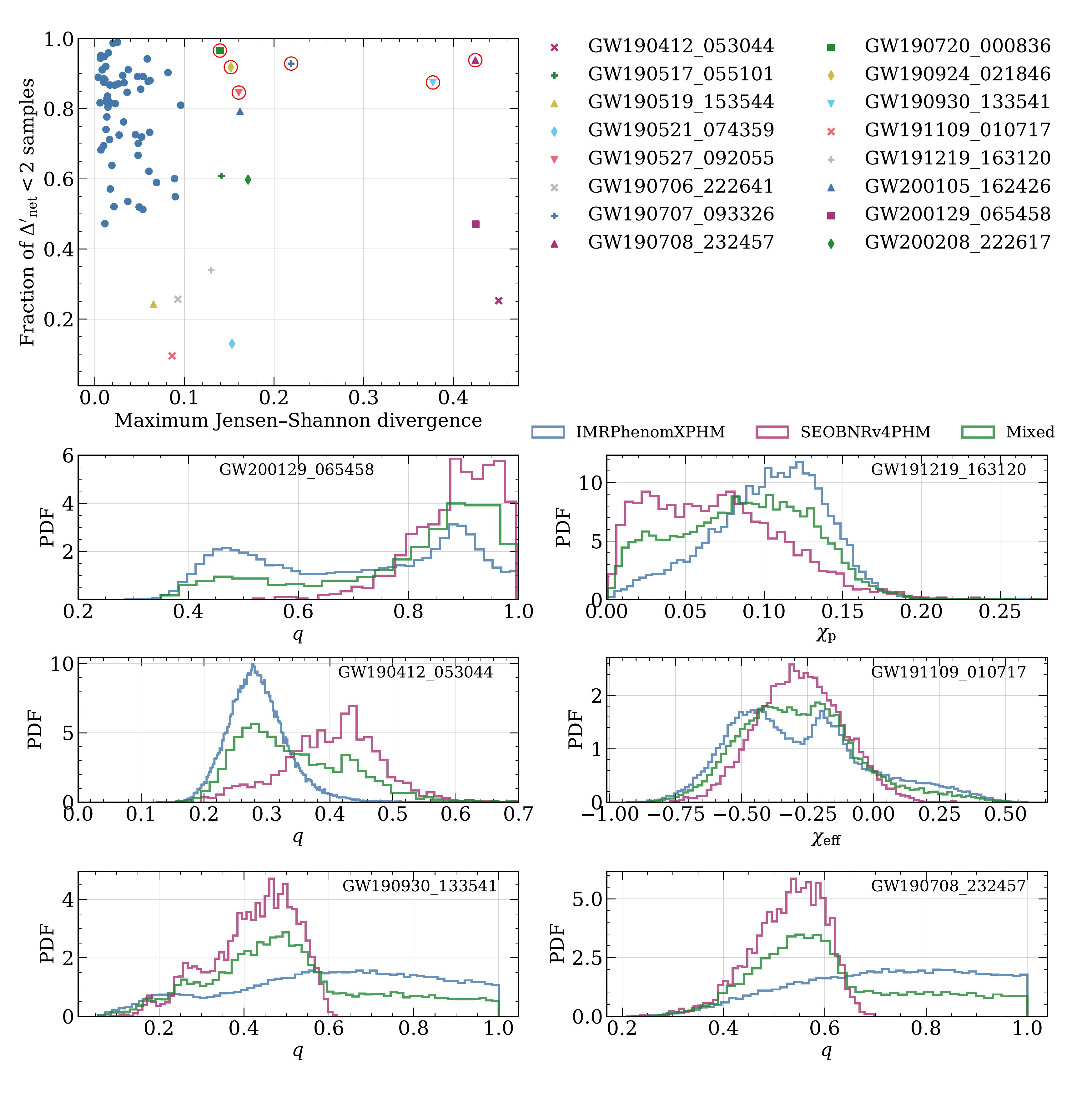}
    \centering
    \caption{\label{pic:js_vs_fraction} First row: We visualize the fraction of samples that meet our accuracy standard and maximum J-S divergence in \{$q,\mathcal{M}, \chi_{\mathrm{eff}}, \chi_{\mathrm{p}}$\} of the two samples (10th and 11th columns of Tab.~\ref{tab1} and ~\ref{tab2}). We highlight the events whose fraction of $\|\Delta'_{\mathrm{net}}\|<2$ samples is less than 40\%, and the events whose maximum J-S divergence is greater than $0.1$. \qh{Some GWTC-2.1 events have nearly flat \texttt{IMRPhenomXPHM} posteriors for mass ratio (as showed in the undermost row), which caused large J-S divergence despite the small waveform difference. We use red circles to label these events.}  \\ Bottom three rows: We show some examples of inconsistent posterior samples; the parameter name and event name are shown in the plots.} 
\end{figure*}

We show the relation between waveform difference and posterior sample consistency (maximum J-S divergence) in Fig.~\ref{pic:js_vs_fraction}, where we can see a weak tendency that events with large waveform difference are more likely to have large J-S divergence, i.e., difference in waveform models may lead to inconsistency in parameter estimation. Particularly, when the fraction of  $\|\Delta'_{\mathrm{net}}\|<2$ samples is below 40\%, the maximum J-S divergence would be greater than the majority of the GW events. This coincides with our expectation. 

However, the inverse statement is not necessarily true. When most posterior samples meet our accuracy standard, it is also possible that two waveform models give inconsistent results. In fact, the waveform error is not the only factor that causes two sets of posterior samples to differ. The behavior of samplers or packages (\texttt{bilby}~\cite{ashton2019:BilbyUserfriendlyBayesian} vs \texttt{RIFT}~\cite{lange2018:RapidAccurateParameter}) and the prior choice (such as high-spin and low-spin prior for neutron stars~\cite{abbott2019:PropertiesBinaryNeutron}) can both influence the consistency between the two posterior samples, although the latter one is not involved in our analysis. {Even if we exclude these factors in a full Bayesian analysis, theoretically, it is the combination of waveform gradients, covariance matrices and waveform difference that contributes to systematic errors in parameter estimation~\cite{cutler2007_LISADetectionsMassive}, not just waveform difference. } {Besides, we use the maximum J-S divergence as the measurement of posterior difference, which might be influenced when the parameter estimation does not work efficiently on some specific parameters. The last row in Fig.~\ref{pic:js_vs_fraction} shows such cases. This makes the correlation between posterior sample consistency and waveform difference more statistically dispersed. }

In the last three rows of Fig.~\ref{pic:js_vs_fraction}, we give some examples of inconsistent posterior samples. GW200129\_065458 has the largest J-S divergence {among GWTC-3 events, and GW190412\_053044 has the largest J-S divergence among GWTC-2.1 events.
The posterior sample inconsistencies of the two events are also reported in GWTC-3~\cite{abbott2021:GWTC3CompactBinary} and GWTC-2.1~\cite{theligoscientificcollaboration2021:GWTC2DeepExtended}. In both events, the result with \texttt{IMRPhenomXPHM} suggests the possibility of a low mass ratio binary, while that with \texttt{SEOBNRv4PHM} does not.} GW191219\_163120 is the lowest mass ratio detection to-date. Its estimated mass ratio is out of the calibration range of waveform models, so potential systematic error may lie in its data analysis~\cite{abbott2021:GWTC3CompactBinary}. In our analysis, GW191219\_163120 does have less posterior samples that pass our assessment than most other events, but it is not the worst one. Besides, its high SNR (26.8) also contributes to waveform difference: its waveform difference becomes small after normalization. This might be caused by the small spins indicated by parameter estimation. Therefore, we suppose the waveform modelling is not that problematic in the low mass ratio and small spin region, but its high SNR reduces model waveform accuracy. We show its estimation of effective precession spin in Fig.~\ref{pic:js_vs_fraction}, in which we see result of \texttt{IMRPhenomXPHM} supports high precession effects in this binary system, while result of \texttt{SEOBNRv4PHM} prefers lower precession effects. GW191109\_010717 produces the largest waveform difference in our analysis. In a later section ~\ref{sec3a3} we will illustrate it might be caused by its special spin effects \qh{and higher modes.} We show its estimation of effective spin in Fig.~\ref{pic:js_vs_fraction}: results from two waveform models show different multimodality. {We also give examples which do not significantly violate our accuracy standard but have inconsistent posterior samples: GW190930\_133541 and GW190708\_232457. Their results from \texttt{IMRPhenomXPHM} seem unable to find the most probable mass ratio. \qh{There are six events having this behaviour in GWTC-2.1, as labeled by red circles in Fig.~\ref{pic:js_vs_fraction}.} Further investigation is needed, but this is beyond the scope of this work.
}

Since most posterior samples in this analysis satisfy or just slightly violate our accuracy standard, and samples from two waveform models, generated by different samplers and packages, are mixed as the final results to counterbalance systematic errors, we suppose the waveform modelling error will not induce significant systematic error in data analysis for current detector sensitivity at the population level, while some special events still need further investigation.

\subsubsection{Waveform difference in different parameter regions\label{sec3a3}}
In Sec~\ref{sec3a}, from the angle of data analysis, we discussed SNR's impact on waveform accuracy. What is more physically interesting is how the waveform accuracy varies with intrinsic properties of the GW source. It is plausible that model accuracy decreases when the system includes some complex processes, such as a highly asymmetric mass ratio, high spin effects, high eccentricity and so forth. \qh{Accuracy may also drop when the contributions from higher modes increase, which usually happens to edge-on binaries~\cite{Biscoveanu:2021nvg,Varma:2016dnf,Colleoni:2020tgc}.}

\begin{figure*}
    \includegraphics[width=1\textwidth]{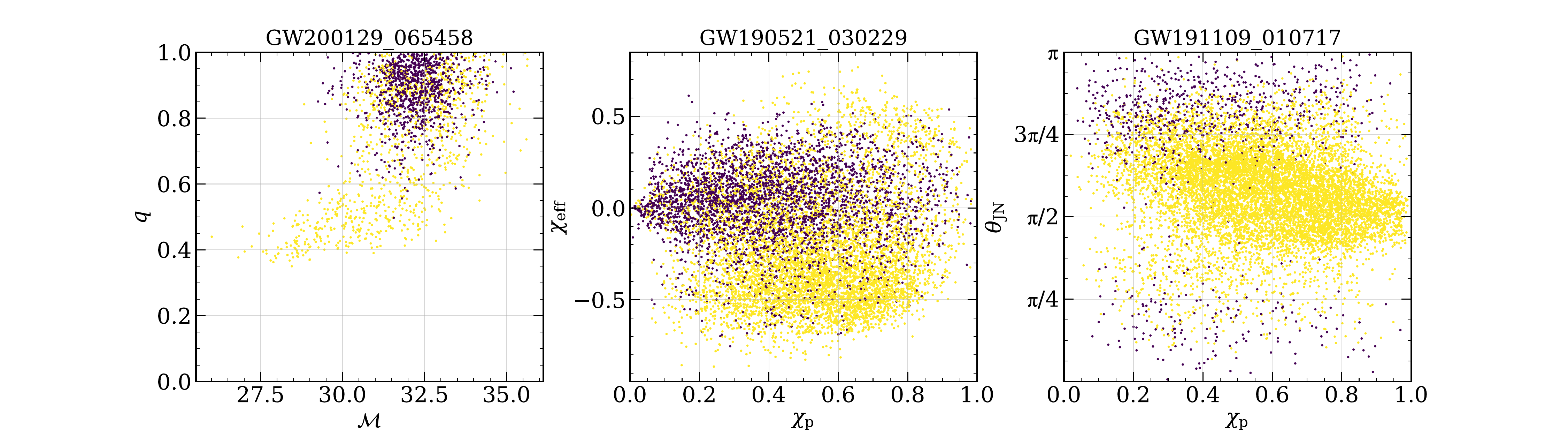}
    \centering
    \caption{\label{pic:para_region_scatter} Posterior scatter of selected events. Yellow points represents samples with $\|\Delta'_{\mathrm{net}}\|>2$, purple points are samples with $\|\Delta'_{\mathrm{net}}\|<2$. \qh{We show three representative events with in two-dimension parameter plane $(\mathcal{M},q)$, $(\chi_{\mathrm{p}},\chi_{\mathrm{eff}})$ and $(\chi_{\mathrm{p}}, \theta_\mathrm{JN})$, respectively. It shows the inaccuracies mainly come from high spin region, low mass ratio region, and edge-on region.}  }
\end{figure*}

We plot posterior samples of selected events and highlight the samples whose waveform difference is greater than 2 in Fig.~\ref{pic:para_region_scatter}. In the nearly equal mass region and small spin region, \texttt{IMRPhenomXPHM} and \texttt{SEOBNRv4PHM} agree with each other and have waveform difference less than 2. However, when mass ratio deviates from $1$, or when spin parameters deviates from $0$, the waveform difference grows and the waveform pair fails to pass the accuracy standard. \qh{For extrinsic parameters, we find that the waveform difference is largest when inclination angle $\theta_{\mathrm{JN}}$ for precessing systems is close to $\pi/2$. We attribute this to two causes: the contribution of higher modes increases when the source is egde-on, and the amplitude modulations caused by precession become increasingly visible, magnifying differences in the way precession is modelled~\cite{Vitale:2014mka,Varma:2016dnf,Fairhurst:2019vut,Colleoni:2020tgc,Biscoveanu:2021nvg,Krishnendu:2021cyi}. This is the reason why events like GW191109\_010717 have a small fraction of posterior samples that pass the accuracy standard: estimations of their parameters mainly lead to low mass ratio, high spin regions or edge-on regions.}

\begin{figure*}
    \includegraphics[width=1\textwidth]{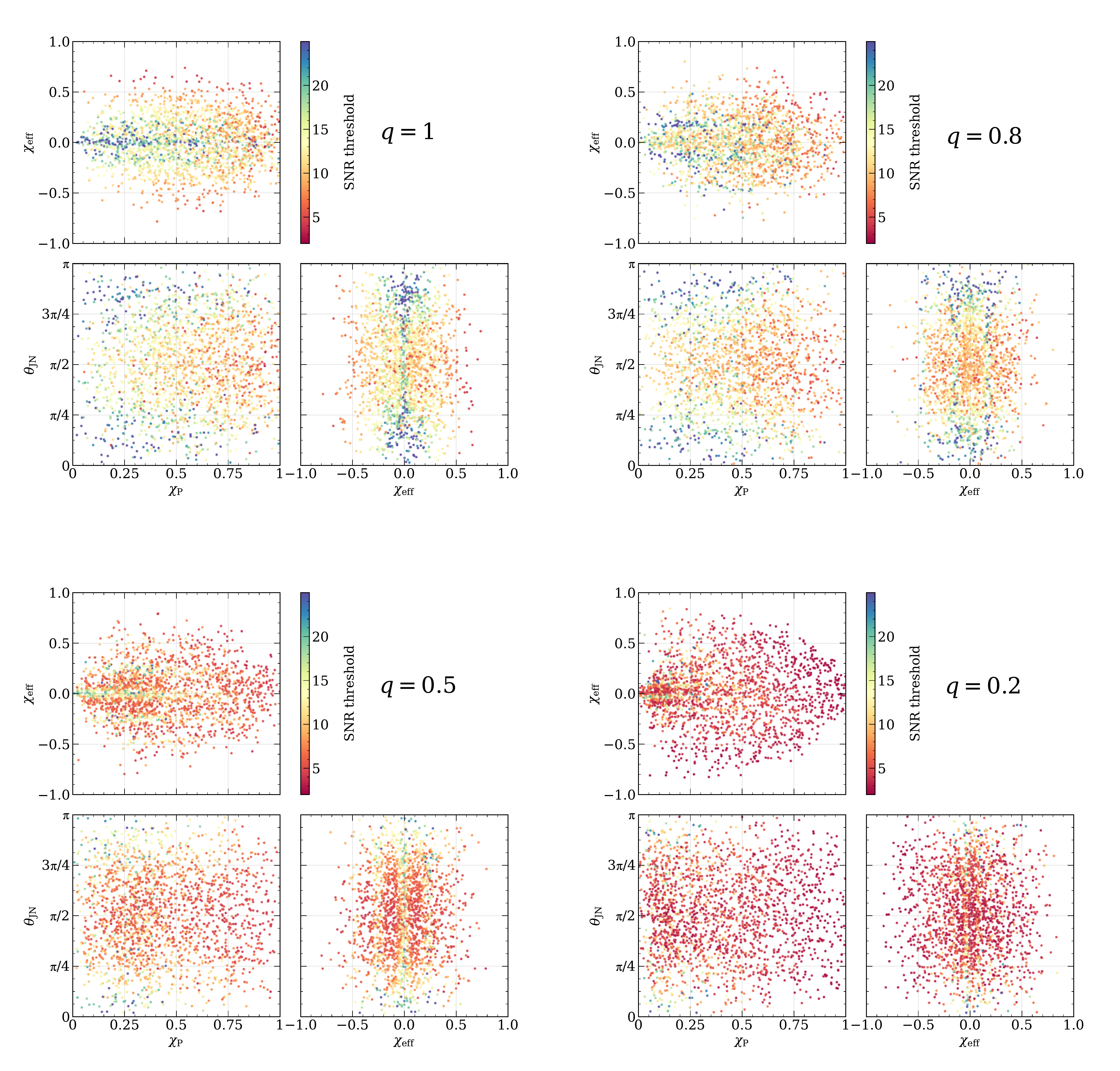}
    \centering
    \caption{\label{pic:sim_scatter} Simulations of random spin \qh{and inclination} BBHs under LIGO design sensitivity. The primary mass is fixed at $30\mathrm{M_{\odot}}$ and mass ratio varies from $1$ to $0.2$, as showed in the top right corner of each figure. We calculate waveform difference $\|\Delta\|$ between \texttt{IMRPhenomXPHM} and \texttt{SEOBNRv4PHM} for each simulation and the SNR when $\|\Delta\|$ reaches 2. The SNR threshold is shown in different colors. \qh{Face-on events }with smaller spins and  equal masses tend to have a higher SNR threshold.}
\end{figure*}

We then perform simulations of BBH events on the design sensitivity of Advanced LIGO~\cite{abbott2020_ProspectsObservingLocalizing}. We set the primary mass at $30\mathrm{M_{\odot}}$, and mass ratio at $1,0.8,0.5$ and $0.2$. The spin of each component is randomly generated: spin magnitude is uniformly distributed between 0 and 1, and spin direction is isotropic. \qh{Inclination angle is isotropic as well. We neglect detector response functions and only include plus polarization here, which will not change our qualitative conclusions.} For each mass ratio we simulate 6000 BBH events and calculate the waveform difference between \texttt{IMRPhenomXPHM} and \texttt{SEOBNRv4PHM}. Since waveform difference $\|\Delta\|$ is proportional to SNR, we introduce a SNR threshold above which $\|\Delta\|$ will be greater than 2. \qh{In Fig.~\ref{pic:sim_scatter}, we plot the distributions of simulation parameters in the style of corner plot for different mass ratios, and the corresponding SNR thresholds in colors. We find the SNR threshold can reach 30 in the low spin and face-on region, but gradually drops below $10$ as the spin parameters increase or $\theta_{\mathrm{JN}}$ tends to $\pi/2$. The change in mass ratio has the same effect, $\|\Delta\|$ can reach $2$ at a smaller spin if the mass ratio is low. However, we find the $q=0.2$ simulations can achieve high SNR threshold for low spin face-on sources, while high-spin or edge-on simulations are more likely to produce low SNR thresholds regardless of the mass ratio. Therefore, for current waveform modeling, spin effects and higher modes may need more improvements than low mass ratio cases. }This coincides with our calculation on the asymmetric mass ratio but small spin event GW191219\_163120. {The disagreement in high-spin CBC waveforms and its impact on parameter estimation is also reported in Ref.~\cite{2017PhRvD..96l4041W}.} 

Our simulation is consistent with the calculation for real events. Given GW events with SNRs ranging from 8 to 30 (for current detector sensitivity), those generated by nearly equal mass systems or low spin systems would have more $\|\Delta'_{\mathrm{net}}\|<2$ samples, while the other events' posterior samples mostly fail our test, like GW191109\_010717. Using the same method in Sec.~\ref{sec3a1} and comparing current SNR threshold with expected SNR of 3rd generation GW detectors, we can also conclude that, in general, the waveform accuracy should be improved for 3 to 4 orders of magnitude. However, for high spin and low mass ratio regions, \qh{as well as higher modes}, the current waveform models may need more improvements. {To calibrate waveform models, these regions might be where NR simulations are most needed for future waveform modelling.}

\subsection{\label{sec3b}NSBH and BNS waveforms}
NSBH and BNS events are much less frequent than BBH events so far - only three events are generally considered as NSBH candidates: GW191219\_163120, GW200105\_162426, and GW200115\_042309, and two are considered as BNS events: GW170817 and GW190425\_081805. Due to the complexity of these systems (e.g., highly asymmetric mass ratio, eccentricity for NSBH binaries, and matter effects for both types), some physical effects are yet to be included in their waveform models. Current NSBH waveform models of \texttt{IMRPhenom} and \texttt{SEOBNR} familes, \texttt{IMRPhenomNSBH} and \texttt{SEOBNRv4\_ROM\_NRTidalv2\_NSBH}~\cite{Pannarale:2015jka}, are calibrated by non-spinning neutron star simulations and only allow aligned spins. For current BNS models, \texttt{IMRPhenomPv2\_NRTidalv2} supports precessing spins while \texttt{SEOBNRv4T\_surrogate} only supports aligned spins. \qh{Recent works have made \texttt{TEOBResumS} able to generate waveforms for precessing BNS systems with higher modes~\cite{Gamba:2021ydi} , as well as waveforms for eccentric BNS systems~\cite{Chiaramello:2020ehz}, but they have not been applied to the GWTC-2.1 and -3.}


For the three NSBH events, we calculate the $\|\Delta'_{\mathrm{net}}\|$ of their posterior samples generated by \texttt{IMRPhenomNSBH} and \texttt{SEOBNRv4\_ROM\_NRTidalv2\_NSBH}. The fraction of $\|\Delta'_{\mathrm{net}}\|<2$ samples are $99.4\%, 99.6\%$ and  $100\%$ for GW191219\_163120, GW200105\_162426, and GW200115\_042309, respectively. Low SNR of these three events may contribute to the small waveform differences, but compared with the BBH events, lacking of precession effects and higher modes should be the decisive factors that make the waveform pair coincide, and it does not necessarily mean these models can describe general NSBH systems with high accuracy. As for BNS events, \texttt{IMRPhenomPv2\_NRTidalv2} is the only model \qh{used in GWTC-2.1 and -3} that includes precession effects, it is not feasible to compare waveform difference of its posterior samples with others. Hence we do not include calculation of BNS waveforms for real events in this work. 

\begin{figure*}
    \includegraphics[width=1\textwidth]{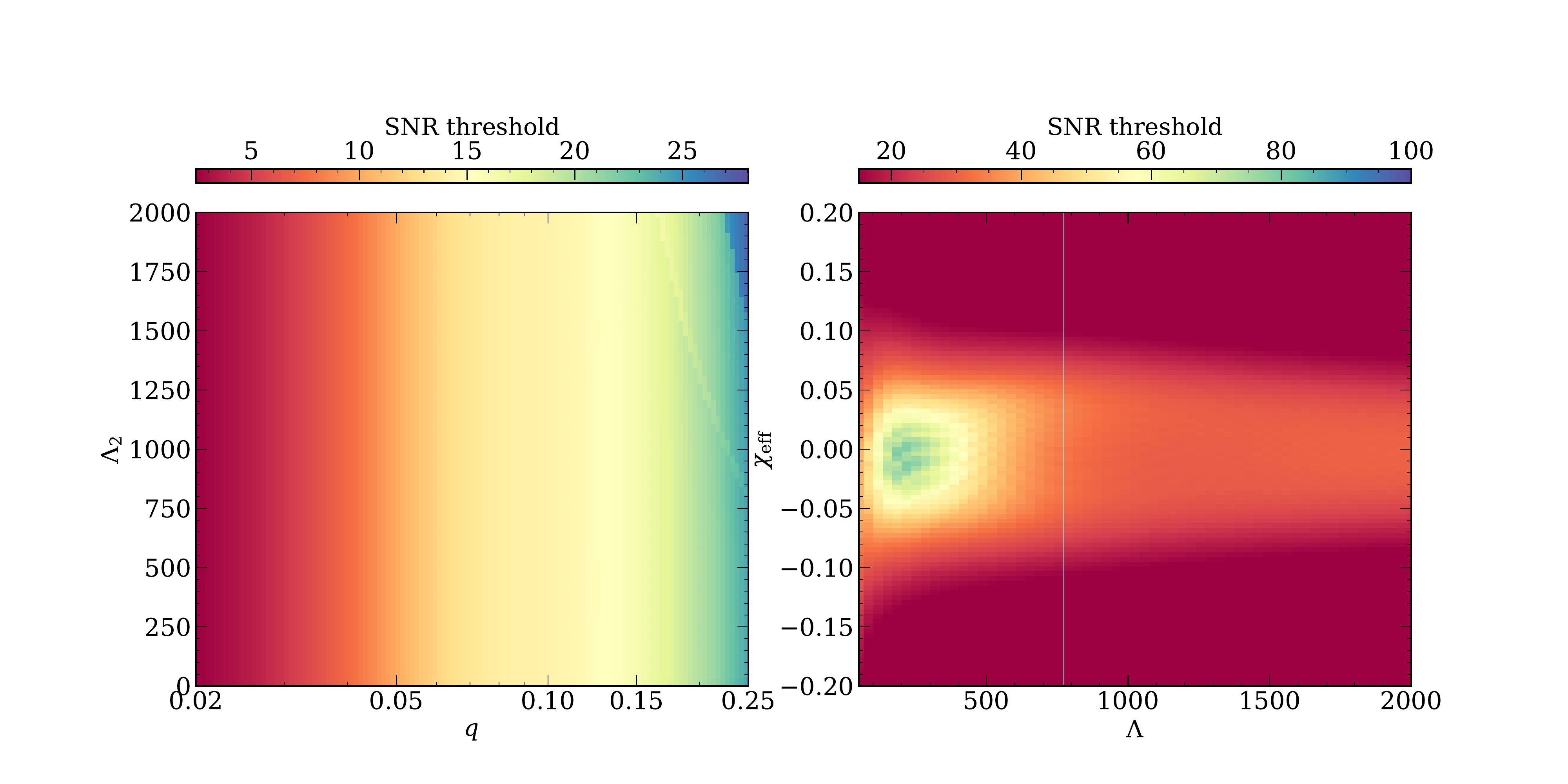}
    \centering
    \caption{\label{pic:simulation_NSBHBNS} Left panel: Simulations of NSBH binaries in the mass ratio $q$ - tidal demformability $\Lambda_2$ plane. The mass of the neutron star is fixed at $1.4~\mathrm{M_{\odot}}$, and we assume both components have zero spin. The colors in the plane represent the SNR threshold for the waveform difference between \texttt{IMRPhenomNSBH} and \texttt{SEOBNRv4\_ROM\_NRTidalv2\_NSBH}, defined in the same way as before.\\
    Right panel: Simulations of BNS binaries in the tidal deformability $\Lambda$ - effective spin $\chi_{\mathrm{eff}}$ plane. We assume both neutron stars are $1.4~\mathrm{M_{\odot}}$ and they have the same spin and tidal deformability parameter. Colors represent SNR threshold for \texttt{IMRPhenomPv2\_NRTidalv2} and \texttt{SEOBNRv4T\_surrogate}.}
\end{figure*}

We perform simulations for NSBH and BNS systems respectively. For NSBH waveform models, we assume zero spin and secondary mass of $1.4~\mathrm{M_{\odot}}$. We change mass ratio between $0.02$ and $0.25$, and tidal deformability parameter between $0$ and $2000$. For BNS, we assume the two neutron stars are exactly the same: same mass $1.4~\mathrm{M_{\odot}}$, same tidal deformability parameter and spin. Then we change spin magnitude between $-0.2$ and $0.2$, and tidal deformability parameter between $0$ and $2000$. \qh{We assume zero inclination for both systems.} The results are shown in Fig.~\ref{pic:simulation_NSBHBNS}.

We find the main disagreement for NSBH waveform models lies in mass ratio, as Fig.~\ref{pic:simulation_NSBHBNS} shows the waveform difference drops with $q$ but is insensitive to the tidal deformability parameter $\Lambda_2$ of the neutron star. The latter is because both approximants use the \texttt{NRTidalv2}~\cite{Dietrich:2019kaq} phase description to model the matter effects. SNR threshold can drop below 5 when $q$ is small, but all the three NSBH candidates have SNRs lower than the thresholds indicated in the corresponding regions in Fig.~\ref{pic:simulation_NSBHBNS}. Note we assume zero spin in this simulation, but non-zero spin samples exist in the three NSBH candidates and would make extra contributions to waveform difference. Therefore, they still have a small fraction of  $\|\Delta'_{\mathrm{net}}\|>2$ samples. Given the SNR threshold in this simulation, NSBH waveform model accuracies (in terms of the mismatch from real waveform) also need an improvement of 3–4 orders of magnitude for future detection, leaving aside the unincluded physical effects.

As for BNS waveforms \texttt{IMRPhenomPv2\_NRTidalv2} and \texttt{SEOBNRv4T\_surrogate}, we change values of $\Lambda$ and spin magnitude. We assume both components have the same aligned spin and mass, so the individual spin magnitude is equal to the effective spin. We find two waveform models agree with each other quite well in the $\Lambda<500$, $\chi_{\mathrm{eff}}<0.05$ region, with SNR thresholds up to 100. This is the region that coincides with our current understanding of neutron stars. However, when spin increases, the SNR threshold can drop below 20. This also implies accuracy of future waveform models should be improved by several orders of magnitude.

We do not discuss further about NSBH and BNS waveform models, for we suppose the number of real events is not enough for us to perform analysis on the population level, and further work on precession, higher modes, and even eccentricity should be done for more NSBH and BNS waveform models.

\section{\label{sec4} Conclusions}
In this work, we developed a diagnostic test for the presence of waveform mismodelling. This extends the work of Ref.~\cite{lindblom2008:ModelWaveformAccuracy} to realistic analyses. While Ref.~\cite{lindblom2008:ModelWaveformAccuracy} suggests a waveform model should have error (as a vector) shorter than 1 to be accurate enough, we introduce two waveform models and find their difference should be shorter than 2 if they are both accurate enough. This method frees accuracy evaluation from the unknown true waveform, and it enables the evaluation to be performed in larger, continuous regions in the parameter space: the regions where waveform models can work, rather than where NR simulations are done. We should note that our method can only tell the existence of inaccurate waveform models. It can not tell which one (or both) is (are) inaccurate if the pair fail, or guarantee any accuracy when the pair do not fail. The key idea is: If two models have significant difference, they can not be both accurate enough, but when the difference is small, we can not rule out the possibility that two models are making similar mistakes. 

For BBH waveform models, we choose the statr-of-art models from \texttt{IMRPhenom} and \texttt{SEOBNR} family, \texttt{IMRPhenomXPHM} and \texttt{SEOBNRv4PHM} for illustration. We have applied our test to existing parameter estimates from the GWTC-3 and GWTC-2.1 (which used the waveform models mentioned above), and found differences in the results of data analysis from different waveform models. The samples that fail our test are mostly located in the low mass ratio, high spin \qh{or egde-on} regions in the parameter space, which means waveform models become less reliable in these regions. Our simulations agree with this: the waveform difference between \texttt{IMRPhenomXPHM} and \texttt{SEOBNRv4PHM} can reach the threshold 2 when SNR is less than 10 in high spin regions and edge-on regions; waveform difference increases in low mass ratio region the as well. We also note that \qh{spin effects and inclinations (higher modes) are more problematic for waveform modelling than mass ratio}. This points out where NR simulations are needed most for future waveform calibration.

We have investigated the correlation between waveform difference and inconsistency of parameter estimation samples given by different waveform models. The latter is measured by the J-S divergence. For the GWTC-3 and GWTC-2.1 posterior samples, we find when the fraction of $\Delta<2$ samples is less than 40\%, it is more likely to obtain a J-S divergence larger than most other events, which is a sign of underlying systematic errors caused by waveform error. We also note that the inverse is not necessarily true, as the waveform model is not the only factor that can influence the generation of posterior samples, { but nonetheless it is always helpful to have one of the factors checked}. Since multi-waveform analysis is becoming a standard way of reducing systematic errors in parameter estimation of GW sources, we suggest that waveform difference analysis can be used as a real-time quantitative check in the parameter estimation workflow. 

For NSBH waveforms, we select \texttt{IMRPhenomNSBH} and \texttt{SEOBNRv4\_ROM\_NRTidalv2\_NSBH}, the two models used in GWTC-3 and GWTC-2.1 parameter estimation. The posterior samples of the 3 NSBH candidates have small NSBH waveform difference compared to BBH waveforms. We credit this to the fact that these waveform models do not include non-aligned spins or higher modes as BBH waveforms. As expected, we find waveform difference increase when mass ratio decreases in our simulation. The SNR threshold drops below 10 when mass ratio is less than 0.05, indicating that more calibrations are needed for this region, leaving aside the lack of some other physical effects.

For BNS waveforms, we have not applied our test on real events samples, for only \texttt{IMRPhenomPv2\_NRTidalv2} is used in GWTC-2.1 and -3 data analysis, and we could not find another comparable model to be paired with it. We simulate aligned spin BNSs for \texttt{IMRPhenomPv2\_NRTidalv2} and \texttt{SEOBNRv4T\_surrogate} instead.  We find the systematic differences between the approximants we examined are small in the region where $\Lambda<500$, and $|\chi_\mathrm{eff}|<0.05$, which should be the case for our current understanding of neutron stars. However, there are some differences when $\Lambda<50$ where the waveforms appear to diverge again. In the high spin regions, the SNR threshold drops below 20, which can not meet future high SNR detections. 

The waveform difference is related to the widely used overlap (or mismatch) through Eqs.~\ref{Eq:rela_with_match} and \ref{Eq:future_estmate}. If we assume two models are well calibrated and have comparable errors, we can give a rough estimate of future waveform accuracy requirement. This complements previous works on waveform accuracy~\cite{purrer2020_GravitationalWaveformAccuracy}. Looking at the SNR thresholds for the three types of waveforms, we know the current waveform accuracy is not enough for future high SNR detections where SNR can reaches up to 1000. The mismatch from the real waveform needed to be reduced by at least 3 orders of magnitude. This is consistent with previous work.

\qh{Finally, this method can be extended to more complex GW waveform models for future GW detection, such as waveforms including eccentricity. We can perform this analysis as long as there are at least two waveform models with the similar accuracy and which include the same physical parameters. Our method can work beyond the NR calibration range, thus it can be efficient way to study the waveform models' extrapolation performance. This may also be a guide of where NR simulations are most needed in the parameter space.}  

\begin{acknowledgments}
The authors would like to thank Daniel Williams, Michael P\"{u}rrer, Christopher Berry, Ik Siong Heng, Rossella Gamba and Jacob Lange for helpful discussions and suggestions. Daniel also helped us use the GWTC-3 and GWTC-2.1 PE data release. {The authors are grateful for computational resources provided by the LIGO Lab at Caltech which is supported by National Science Foundation Grants PHY-0757058 and PHY-0823459.} QH is supported by CSC. JV is supported by STFC grant ST/V005634/1.

\end{acknowledgments}

\appendix

\section{Full results of BBH waveform in GWTC-3 and GWTC-2.1 \label{apdxb}}

\begin{table*}[]
\begin{ruledtabular}
\begin{tabular}{ccccccccccc}
Event name           & $m_1$                 & $m_2$              & $\chi_\mathrm{eff}$            & $\chi_\mathrm{p}$       &$\theta_\mathrm{JN}$       & SNR  & Mean & $\mathrm{Mean_{Norm}}$ & Fraction & Max J-S Div. \\
\hline
    GW150914\_095045 & $37.99_{-2.89}^{+5.11}$     & $32.96_{-5.03}^{+3.18}$    & $-0.04_{-0.14}^{+0.12}$ & $0.51_{-0.38}^{+0.35}$ & $2.70_{-0.71}^{+0.32}$ & 24.4 & 2.0  & 0.08            & 0.59     & 0.069   \\
GW151012\_095443 & $29.62_{-6.87}^{+17.08}$    & $16.30_{-6.02}^{+5.73}$    & $0.12_{-0.21}^{+0.28}$  & $0.36_{-0.27}^{+0.43}$ & $1.70_{-1.40}^{+1.13}$ & 10   & 1.33 & 0.13            & 0.81     & 0.023   \\
GW151226\_033853 & $15.50_{-4.00}^{+12.19}$    & $8.18_{-3.02}^{+2.62}$     & $0.20_{-0.08}^{+0.23}$  & $0.52_{-0.35}^{+0.36}$ & $0.88_{-0.66}^{+2.00}$ & 13.1 & 2.39 & 0.18            & 0.54     & N/A     \\
GW170104\_101158 & $34.78_{-4.85}^{+7.73}$     & $25.25_{-5.60}^{+4.43}$    & $-0.04_{-0.19}^{+0.15}$ & $0.40_{-0.31}^{+0.40}$ & $1.10_{-0.86}^{+1.79}$ & 13   & 1.28 & 0.1             & 0.88     & 0.01    \\
GW170608\_020116 & $11.41_{-1.49}^{+4.36}$     & $8.38_{-2.06}^{+1.21}$     & $0.05_{-0.05}^{+0.13}$  & $0.32_{-0.24}^{+0.41}$ & $2.37_{-2.06}^{+0.58}$ & 14.9 & 1.43 & 0.1             & 0.86     & 0.051   \\
GW170729\_185629 & $77.88_{-14.48}^{+17.89}$   & $44.13_{-17.46}^{+18.11}$  & $0.29_{-0.33}^{+0.25}$  & $0.39_{-0.29}^{+0.40}$ & $1.35_{-1.03}^{+1.44}$ & 10.8 & 2.08 & 0.19            & 0.54     & 0.037   \\
GW170809\_082821 & $41.19_{-6.27}^{+9.63}$     & $29.25_{-6.46}^{+5.63}$    & $0.07_{-0.17}^{+0.17}$  & $0.39_{-0.30}^{+0.44}$ & $2.61_{-0.59}^{+0.39}$ & 12.4 & 1.14 & 0.09            & 0.91     & 0.037   \\
GW170814\_103043 & $34.70_{-3.52}^{+6.01}$     & $28.03_{-4.59}^{+3.22}$    & $0.08_{-0.12}^{+0.13}$  & $0.48_{-0.37}^{+0.38}$ & $0.69_{-0.48}^{+1.92}$ & 15.9 & 1.31 & 0.08            & 0.88     & 0.062   \\
GW170818\_022509 & $42.14_{-5.04}^{+7.99}$     & $33.42_{-6.14}^{+5.07}$    & $-0.06_{-0.22}^{+0.19}$ & $0.56_{-0.41}^{+0.34}$ & $2.46_{-0.50}^{+0.47}$ & 11.3 & 1.06 & 0.09            & 0.96     & 0.015   \\
GW170823\_131358 & $51.67_{-7.85}^{+11.92}$    & $39.42_{-11.03}^{+8.11}$   & $0.05_{-0.22}^{+0.21}$  & $0.47_{-0.35}^{+0.41}$ & $1.73_{-1.48}^{+1.16}$ & 11.5 & 1.83 & 0.16            & 0.64     & 0.019   \\
GW190408\_181802 & $31.70_{-3.88}^{+6.96}$     & $23.83_{-5.01}^{+3.56}$    & $-0.03_{-0.17}^{+0.13}$ & $0.37_{-0.29}^{+0.41}$ & $1.01_{-0.79}^{+1.85}$ & 14.4 & 1.05 & 0.07            & 0.94     & 0.006   \\
GW190412\_053044 & $31.76_{-6.60}^{+6.81}$     & $10.34_{-1.55}^{+2.19}$    & $0.21_{-0.13}^{+0.12}$  & $0.19_{-0.12}^{+0.22}$ & $0.92_{-0.40}^{+1.69}$ & 18.2 & 2.86 & 0.16            & 0.25     & 0.45    \\
GW190413\_052954 & $52.79_{-10.02}^{+15.00}$   & $37.90_{-11.57}^{+9.95}$   & $-0.04_{-0.32}^{+0.27}$ & $0.44_{-0.33}^{+0.42}$ & $0.79_{-0.58}^{+2.01}$ & 8.5  & 1.13 & 0.13            & 0.88     & N/A     \\
GW190413\_134308 & $83.12_{-15.64}^{+19.30}$   & $50.15_{-23.80}^{+19.47}$  & $-0.01_{-0.38}^{+0.28}$ & $0.55_{-0.41}^{+0.36}$ & $1.85_{-1.52}^{+1.01}$ & 10.3 & 1.57 & 0.15            & 0.71     & N/A     \\
GW190421\_213856 & $60.79_{-8.71}^{+13.06}$    & $46.94_{-14.55}^{+9.04}$   & $-0.10_{-0.27}^{+0.21}$ & $0.45_{-0.34}^{+0.41}$ & $2.03_{-1.70}^{+0.85}$ & 9.7  & 1.59 & 0.16            & 0.73     & N/A     \\
GW190503\_185404 & $53.32_{-10.28}^{+12.17}$   & $36.45_{-12.62}^{+10.17}$  & $-0.05_{-0.30}^{+0.23}$ & $0.43_{-0.33}^{+0.40}$ & $2.50_{-0.62}^{+0.46}$ & 12.2 & 1.55 & 0.13            & 0.74     & 0.013   \\
GW190512\_180714 & $29.34_{-6.72}^{+6.90}$     & $15.83_{-2.97}^{+4.64}$    & $0.02_{-0.14}^{+0.13}$  & $0.26_{-0.20}^{+0.41}$ & $1.79_{-1.50}^{+1.07}$ & 12.2 & 1.3  & 0.11            & 0.85     & 0.036   \\
GW190513\_205428 & $49.99_{-12.78}^{+14.75}$   & $25.60_{-6.98}^{+10.88}$   & $0.16_{-0.22}^{+0.29}$  & $0.35_{-0.26}^{+0.43}$ & $0.79_{-0.58}^{+2.01}$ & 12.3 & 1.64 & 0.13            & 0.7      & 0.048   \\
GW190514\_065416 & $66.96_{-12.57}^{+21.39}$   & $47.33_{-18.23}^{+12.11}$  & $-0.08_{-0.35}^{+0.29}$ & $0.45_{-0.33}^{+0.42}$ & $1.47_{-1.20}^{+1.38}$ & 8.3  & 1.44 & 0.17            & 0.78     & 0.013   \\
GW190517\_055101 & $52.92_{-9.99}^{+14.76}$    & $32.75_{-12.25}^{+9.55}$   & $0.49_{-0.28}^{+0.21}$  & $0.55_{-0.32}^{+0.31}$ & $2.12_{-1.18}^{+0.70}$ & 10.8 & 2.04 & 0.19            & 0.61     & 0.141   \\
GW190519\_153544 & $94.80_{-12.40}^{+15.84}$   & $59.88_{-18.85}^{+16.88}$  & $0.33_{-0.24}^{+0.20}$  & $0.45_{-0.28}^{+0.36}$ & $1.61_{-1.02}^{+0.95}$ & 12.4 & 2.81 & 0.23            & 0.24     & 0.066   \\
GW190521\_030229 & $152.36_{-17.62}^{+31.65}$  & $89.65_{-52.17}^{+48.96}$  & $-0.14_{-0.45}^{+0.50}$ & $0.49_{-0.35}^{+0.33}$ & $1.38_{-1.07}^{+1.40}$ & 13.3 & 3.24 & 0.24            & 0.35     & N/A     \\
GW190521\_074359 & $52.19_{-5.39}^{+7.65}$     & $40.36_{-7.16}^{+5.87}$    & $0.10_{-0.13}^{+0.13}$  & $0.39_{-0.30}^{+0.37}$ & $1.48_{-1.10}^{+1.27}$ & 24.4 & 3.6  & 0.15            & 0.13     & 0.153   \\
GW190527\_092055 & $51.05_{-9.27}^{+29.88}$    & $32.36_{-13.36}^{+11.95}$  & $0.10_{-0.22}^{+0.22}$  & $0.36_{-0.28}^{+0.47}$ & $1.15_{-0.87}^{+1.65}$ & 8.7  & 1.29 & 0.15            & 0.85     & 0.161   \\
GW190602\_175927 & $106.81_{-17.99}^{+24.90}$  & $67.63_{-31.56}^{+23.03}$  & $0.12_{-0.28}^{+0.25}$  & $0.45_{-0.34}^{+0.43}$ & $2.13_{-1.87}^{+0.79}$ & 12.3 & 1.95 & 0.16            & 0.6      & N/A     \\
GW190620\_030421 & $87.13_{-17.28}^{+23.93}$   & $53.35_{-24.30}^{+17.64}$  & $0.34_{-0.29}^{+0.22}$  & $0.48_{-0.32}^{+0.38}$ & $2.07_{-1.74}^{+0.82}$ & 10.9 & 2.1  & 0.19            & 0.52     & 0.049   \\
GW190630\_185205 & $41.40_{-6.50}^{+8.12}$     & $28.24_{-5.63}^{+5.79}$    & $0.10_{-0.13}^{+0.14}$  & $0.33_{-0.24}^{+0.36}$ & $1.41_{-1.16}^{+1.43}$ & 15.2 & 1.55 & 0.1             & 0.76     & 0.032   \\
GW190701\_203306 & $74.56_{-10.85}^{+15.98}$   & $56.08_{-17.93}^{+12.08}$  & $-0.08_{-0.31}^{+0.23}$ & $0.44_{-0.33}^{+0.41}$ & $0.58_{-0.42}^{+0.55}$ & 11.7 & 0.95 & 0.08            & 0.95     & 0.007   \\
GW190706\_222641 & $117.53_{-18.55}^{+22.57}$  & $63.74_{-27.42}^{+27.20}$  & $0.28_{-0.31}^{+0.25}$  & $0.47_{-0.33}^{+0.38}$ & $1.86_{-1.50}^{+0.96}$ & 12.5 & 3.06 & 0.24            & 0.26     & 0.093   \\
GW190707\_093326 & $14.07_{-2.33}^{+3.01}$     & $9.28_{-1.48}^{+1.69}$     & $-0.04_{-0.09}^{+0.10}$ & $0.28_{-0.22}^{+0.39}$ & $2.12_{-1.89}^{+0.81}$ & 13.2 & 1.23 & 0.09            & 0.93     & 0.219   \\
GW190708\_232457 & $23.41_{-5.07}^{+4.97}$     & $13.70_{-2.17}^{+3.57}$    & $0.05_{-0.10}^{+0.10}$  & $0.26_{-0.20}^{+0.44}$ & $1.39_{-1.18}^{+1.54}$ & 13.1 & 1.14 & 0.09            & 0.94     & 0.424   \\
GW190719\_215514 & $58.83_{-16.34}^{+75.12}$   & $32.78_{-16.38}^{+14.38}$  & $0.25_{-0.32}^{+0.33}$  & $0.45_{-0.32}^{+0.39}$ & $1.61_{-1.33}^{+1.26}$ & 8    & 1.41 & 0.18            & 0.79     & N/A     \\
GW190720\_000836 & $16.57_{-3.94}^{+6.36}$     & $8.76_{-2.10}^{+2.44}$     & $0.19_{-0.11}^{+0.14}$  & $0.29_{-0.20}^{+0.39}$ & $2.59_{-1.99}^{+0.41}$ & 11.5 & 0.86 & 0.08            & 0.97     & 0.14    \\
GW190727\_060333 & $58.87_{-8.01}^{+13.07}$    & $46.18_{-13.24}^{+8.22}$   & $0.09_{-0.27}^{+0.25}$  & $0.50_{-0.37}^{+0.38}$ & $1.54_{-1.27}^{+1.35}$ & 12.1 & 1.37 & 0.11            & 0.82     & 0.017   \\
GW190728\_064510 & $14.56_{-2.60}^{+8.19}$     & $9.39_{-2.91}^{+1.92}$     & $0.13_{-0.07}^{+0.19}$  & $0.29_{-0.20}^{+0.39}$ & $1.11_{-0.90}^{+1.77}$ & 13.4 & 1.16 & 0.09            & 0.89     & 0.048   \\
GW190731\_140936 & $64.91_{-10.80}^{+15.63}$   & $46.27_{-17.54}^{+12.67}$  & $0.07_{-0.25}^{+0.28}$  & $0.41_{-0.32}^{+0.43}$ & $1.23_{-0.98}^{+1.63}$ & 8.5  & 1.13 & 0.13            & 0.89     & 0.031   \\
GW190803\_022701 & $57.71_{-8.92}^{+13.23}$    & $42.96_{-13.81}^{+9.20}$   & $-0.01_{-0.28}^{+0.23}$ & $0.44_{-0.34}^{+0.42}$ & $0.91_{-0.70}^{+1.93}$ & 9.1  & 0.87 & 0.1             & 0.96     & N/A     \\
GW190814\_211039 & $24.48_{-1.43}^{+1.55}$     & $2.72_{-0.11}^{+0.11}$     & $0.00_{-0.07}^{+0.07}$  & $0.04_{-0.03}^{+0.04}$ & $0.90_{-0.24}^{+1.40}$ & 22.2 & 2.54 & 0.11            & 0.49     & N/A     \\
GW190828\_063405 & $43.27_{-4.65}^{+7.36}$     & $35.40_{-6.98}^{+4.69}$    & $0.15_{-0.16}^{+0.15}$  & $0.43_{-0.32}^{+0.41}$ & $2.38_{-2.02}^{+0.57}$ & 16.3 & 1.31 & 0.08            & 0.86     & N/A     \\
GW190828\_065509 & $30.42_{-8.29}^{+8.31}$     & $13.44_{-2.71}^{+4.82}$    & $0.05_{-0.17}^{+0.16}$  & $0.26_{-0.20}^{+0.42}$ & $1.86_{-1.52}^{+0.96}$ & 11.1 & 1.69 & 0.15            & 0.68     & N/A     \\
GW190910\_112807 & $56.54_{-6.47}^{+8.69}$     & $44.50_{-9.24}^{+7.47}$    & $-0.00_{-0.20}^{+0.17}$ & $0.38_{-0.30}^{+0.43}$ & $1.62_{-1.25}^{+1.16}$ & 13.4 & 2.25 & 0.17            & 0.47     & 0.011   \\
GW190915\_235702 & $43.05_{-6.00}^{+11.11}$    & $32.50_{-7.95}^{+5.62}$    & $-0.03_{-0.24}^{+0.19}$ & $0.56_{-0.40}^{+0.34}$ & $1.84_{-1.48}^{+0.99}$ & 13   & 1.63 & 0.13            & 0.69     & 0.01    \\
GW190924\_021846 & $9.78_{-2.00}^{+4.83}$      & $5.67_{-1.61}^{+1.35}$     & $0.03_{-0.08}^{+0.20}$  & $0.25_{-0.19}^{+0.41}$ & $0.84_{-0.64}^{+1.95}$ & 13   & 1.13 & 0.09            & 0.92     & 0.152   \\
GW190929\_012149 & $101.70_{-18.32}^{+25.99}$  & $41.64_{-18.10}^{+25.15}$  & $-0.03_{-0.28}^{+0.23}$ & $0.31_{-0.25}^{+0.51}$ & $1.45_{-0.97}^{+1.18}$ & 10.1 & 2.22 & 0.22            & 0.47     & N/A     \\
GW190930\_133541 & $16.36_{-4.57}^{+9.27}$     & $7.98_{-2.34}^{+2.78}$     & $0.19_{-0.16}^{+0.22}$  & $0.30_{-0.21}^{+0.42}$ & $0.72_{-0.55}^{+2.06}$ & 10.1 & 1.2  & 0.12            & 0.87     & 0.377   \\
\end{tabular}
\caption{\label{tab1}The first half of our BBH analysis results, including 10 GWTC-1 events and 35 GWTC2.1 events that are included in GWTC-2. First seven columns are basic information of the events: event names in YYMMDD\_HHMMSS form, component masses $m_{1,2}$ in detector frame (which have a difference of factor $1+z$ from Ref.~\cite{abbott2021:GWTC3CompactBinary,theligoscientificcollaboration2021:GWTC2DeepExtended}, $z$ is the cosmological redshift), effective spin $\chi_{\mathrm{eff}}$, eﬀective precession spin $\chi_{\mathrm{p}}$, inclination angle $\theta_{\mathrm{JN}}$ and network SNR. \qh{The parameters are showed by 50\% percentile and 90\% confidence error bar, but note $\theta_{\mathrm{JN}}$ usually has a bimodal distribution, the 50\% percentile might be misleading.}
Last four columns are statistics we construct: mean value and normalized mean value of $\|\Delta'_{\mathrm{net}}\|$ among all the samples (latter one is simply mean value divided by network SNR), fraction of $\|\Delta'_{\mathrm{net}}\|<2$ samples, and the maximum J-S divergence between samples of $q$, $\mathcal{M}$, $\chi_\mathrm{eff}$ and $\chi_\mathrm{p}$ . It is labeled as ``N/A'' if result from one of the waveforms is not available in GWTC-2.1 data release.}
\end{ruledtabular}
\end{table*}

\begin{table*}[]
\begin{ruledtabular}
\begin{tabular}{ccccccccccc}
Event name           & $m_1$                 & $m_2$              & $\chi_\mathrm{eff}$            & $\chi_\mathrm{p}$      &$\theta_\mathrm{JN}$         & SNR  & Mean & $\mathrm{Mean_{Norm}}$ & Fraction & Max J-S Div. \\
\hline
GW190403\_051519 & $185.98_{-58.13}^{+37.48}$  & $44.64_{-24.08}^{+61.34}$  & $0.68_{-0.43}^{+0.16}$  & $0.32_{-0.22}^{+0.38}$ & $1.84_{-1.66}^{+1.12}$ & 8    & 1.62 & 0.2             & 0.72     & 0.052   \\
    GW190426\_190642 & $178.48_{-31.54}^{+87.83}$  & $132.65_{-63.94}^{+32.32}$ & $0.23_{-0.41}^{+0.42}$  & $0.51_{-0.36}^{+0.37}$ & $2.08_{-1.72}^{+0.80}$ & 9.6  & 1.55 & 0.16            & 0.74     & N/A     \\
    GW190725\_174728 & $14.23_{-3.62}^{+12.21}$    & $7.59_{-3.00}^{+2.40}$     & $-0.04_{-0.16}^{+0.36}$ & $0.37_{-0.28}^{+0.46}$ & $1.00_{-0.74}^{+1.79}$ & 9.1  & 1.04 & 0.11            & 0.92     & N/A     \\
    GW190805\_211137 & $87.19_{-15.28}^{+24.88}$   & $59.60_{-23.31}^{+16.40}$  & $0.37_{-0.39}^{+0.29}$  & $0.50_{-0.32}^{+0.34}$ & $1.00_{-0.74}^{+1.75}$ & 8.3  & 1.3  & 0.16            & 0.84     & 0.014   \\
    GW190916\_200658 & $78.32_{-21.54}^{+32.30}$   & $42.37_{-21.18}^{+24.07}$  & $0.20_{-0.31}^{+0.33}$  & $0.37_{-0.28}^{+0.43}$ & $1.61_{-1.35}^{+1.27}$ & 8.2  & 1.14 & 0.14            & 0.89     & 0.004   \\
    GW190917\_114630 & $11.15_{-4.48}^{+3.72}$     & $2.35_{-0.48}^{+1.21}$     & $-0.08_{-0.43}^{+0.21}$ & $0.17_{-0.13}^{+0.42}$ & $1.35_{-1.16}^{+1.60}$ & 9.5  & 1.06 & 0.11            & 0.93     & N/A     \\
    GW190925\_232845 & $24.69_{-3.17}^{+7.70}$     & $18.46_{-4.16}^{+2.68}$    & $0.09_{-0.15}^{+0.16}$  & $0.39_{-0.30}^{+0.43}$ & $0.77_{-0.58}^{+2.07}$ & 9.9  & 1.09 & 0.11            & 0.91     & 0.008   \\
    GW190926\_050336 & $63.55_{-14.11}^{+31.78}$   & $31.96_{-14.17}^{+21.57}$  & $-0.02_{-0.33}^{+0.25}$ & $0.37_{-0.29}^{+0.48}$ & $1.67_{-1.19}^{+1.03}$ & 9    & 1.49 & 0.17            & 0.74     & N/A     \\
    GW191103\_012549 & $14.03_{-2.34}^{+7.42}$     & $9.42_{-2.85}^{+1.79}$     & $0.21_{-0.10}^{+0.16}$  & $0.40_{-0.26}^{+0.41}$ & $1.38_{-1.14}^{+1.52}$ & 8.9  & 1.01 & 0.11            & 0.92     & 0.012   \\
    GW191105\_143521 & $13.00_{-1.78}^{+4.54}$     & $9.36_{-2.19}^{+1.45}$     & $-0.02_{-0.09}^{+0.13}$ & $0.30_{-0.24}^{+0.45}$ & $1.07_{-0.85}^{+1.82}$ & 9.7  & 0.77 & 0.08            & 0.99     & 0.02    \\
    GW191109\_010717 & $81.16_{-8.89}^{+12.89}$    & $59.72_{-17.43}^{+15.58}$  & $-0.29_{-0.31}^{+0.42}$ & $0.63_{-0.37}^{+0.29}$ & $1.91_{-1.18}^{+0.87}$ & 17.3 & 5.8  & 0.34            & 0.1      & 0.086   \\
    GW191113\_071753 & $36.10_{-16.18}^{+14.71}$   & $7.31_{-1.57}^{+6.49}$     & $0.00_{-0.29}^{+0.37}$  & $0.20_{-0.16}^{+0.54}$ & $1.70_{-1.32}^{+1.08}$ & 7.9  & 1.68 & 0.21            & 0.67     & 0.048   \\
    GW191126\_115259 & $15.71_{-2.51}^{+7.24}$     & $10.75_{-2.98}^{+1.94}$    & $0.21_{-0.11}^{+0.15}$  & $0.39_{-0.26}^{+0.40}$ & $1.71_{-1.48}^{+1.20}$ & 8.3  & 1.14 & 0.14            & 0.89     & 0.01    \\
    GW191127\_050227 & $86.41_{-37.31}^{+60.12}$   & $38.45_{-25.26}^{+31.09}$  & $0.18_{-0.36}^{+0.34}$  & $0.52_{-0.41}^{+0.41}$ & $1.46_{-1.16}^{+1.40}$ & 9.2  & 1.9  & 0.21            & 0.6      & 0.089   \\
    GW191129\_134029 & $12.29_{-2.26}^{+4.87}$     & $7.80_{-1.94}^{+1.67}$     & $0.06_{-0.08}^{+0.16}$  & $0.26_{-0.19}^{+0.36}$ & $1.73_{-1.46}^{+1.16}$ & 13.1 & 1.34 & 0.1             & 0.87     & 0.033   \\
    GW191204\_110529 & $36.20_{-5.66}^{+15.49}$    & $26.21_{-7.54}^{+5.17}$    & $0.05_{-0.27}^{+0.26}$  & $0.52_{-0.39}^{+0.38}$ & $1.57_{-1.24}^{+1.24}$ & 8.8  & 1.59 & 0.18            & 0.72     & 0.027   \\
    GW191204\_171526 & $13.44_{-1.98}^{+3.77}$     & $9.29_{-1.84}^{+1.54}$     & $0.16_{-0.05}^{+0.08}$  & $0.39_{-0.26}^{+0.35}$ & $2.26_{-2.00}^{+0.66}$ & 17.5 & 1.67 & 0.1             & 0.73     & 0.045   \\
    GW191215\_223052 & $33.48_{-4.68}^{+9.29}$     & $24.46_{-5.17}^{+4.07}$    & $-0.04_{-0.21}^{+0.17}$ & $0.50_{-0.38}^{+0.37}$ & $1.20_{-0.85}^{+1.50}$ & 11.2 & 1.06 & 0.09            & 0.95     & 0.01    \\
    GW191216\_213338 & $12.95_{-2.38}^{+4.92}$     & $8.23_{-1.99}^{+1.73}$     & $0.11_{-0.06}^{+0.13}$  & $0.23_{-0.16}^{+0.35}$ & $2.50_{-0.81}^{+0.44}$ & 18.6 & 1.93 & 0.1             & 0.62     & 0.06    \\
    GW191219\_163120 & $34.73_{-2.68}^{+2.27}$     & $1.30_{-0.05}^{+0.08}$     & $-0.00_{-0.09}^{+0.07}$ & $0.09_{-0.07}^{+0.07}$ & $1.76_{-1.49}^{+1.13}$ & 9.1  & 2.51 & 0.28            & 0.34     & 0.13    \\
    GW191222\_033537 & $67.15_{-9.54}^{+14.73}$    & $52.41_{-15.40}^{+10.67}$  & $-0.04_{-0.25}^{+0.20}$ & $0.41_{-0.32}^{+0.41}$ & $1.62_{-1.33}^{+1.24}$ & 12.5 & 2.02 & 0.16            & 0.57     & 0.017   \\
    GW191230\_180458 & $82.70_{-13.10}^{+19.46}$   & $62.85_{-21.42}^{+13.88}$  & $-0.05_{-0.31}^{+0.26}$ & $0.52_{-0.39}^{+0.38}$ & $2.03_{-1.67}^{+0.85}$ & 10.4 & 1.17 & 0.11            & 0.88     & 0.011   \\
    GW200105\_162426 & $9.57_{-1.82}^{+1.85}$      & $2.02_{-0.25}^{+0.35}$     & $0.00_{-0.18}^{+0.13}$  & $0.09_{-0.07}^{+0.17}$ & $1.54_{-1.22}^{+1.28}$ & 13.7 & 1.47 & 0.11            & 0.79     & 0.162   \\
    GW200112\_155838 & $44.01_{-5.16}^{+8.22}$     & $35.18_{-7.42}^{+5.11}$    & $0.06_{-0.15}^{+0.15}$  & $0.39_{-0.30}^{+0.39}$ & $0.88_{-0.68}^{+2.04}$ & 19.8 & 1.63 & 0.08            & 0.73     & 0.061   \\
    GW200115\_042309 & $6.30_{-2.69}^{+2.15}$      & $1.53_{-0.30}^{+0.91}$     & $-0.15_{-0.42}^{+0.24}$ & $0.20_{-0.16}^{+0.34}$ & $0.62_{-0.43}^{+1.94}$ & 11.3 & 1.35 & 0.12            & 0.81     & 0.096   \\
    GW200128\_022011 & $65.05_{-9.41}^{+15.99}$    & $51.16_{-13.35}^{+10.32}$  & $0.12_{-0.25}^{+0.24}$  & $0.57_{-0.40}^{+0.34}$ & $1.38_{-1.06}^{+1.46}$ & 10.6 & 2.12 & 0.2             & 0.55     & 0.09    \\
    GW200129\_065458 & $40.25_{-3.33}^{+12.23}$    & $34.06_{-10.82}^{+3.32}$   & $0.11_{-0.16}^{+0.11}$  & $0.52_{-0.37}^{+0.42}$ & $0.66_{-0.41}^{+0.59}$ & 26.8 & 2.33 & 0.09            & 0.47     & 0.425   \\
    GW200202\_154313 & $11.02_{-1.51}^{+3.83}$     & $7.99_{-1.85}^{+1.21}$     & $0.04_{-0.06}^{+0.13}$  & $0.28_{-0.22}^{+0.40}$ & $2.57_{-0.59}^{+0.42}$ & 10.8 & 0.74 & 0.07            & 0.99     & 0.025   \\
    GW200208\_130117 & $52.95_{-8.40}^{+12.23}$    & $38.51_{-11.18}^{+8.64}$   & $-0.07_{-0.27}^{+0.22}$ & $0.38_{-0.29}^{+0.41}$ & $2.53_{-0.58}^{+0.44}$ & 10.8 & 1.19 & 0.11            & 0.87     & 0.017   \\
    GW200208\_222617 & $83.38_{-48.68}^{+171.77}$  & $21.91_{-11.81}^{+13.04}$  & $0.45_{-0.44}^{+0.43}$  & $0.41_{-0.30}^{+0.37}$ & $1.54_{-1.22}^{+1.30}$ & 7.4  & 1.91 & 0.26            & 0.6      & 0.171   \\
    GW200209\_085452 & $55.70_{-9.88}^{+14.96}$    & $42.95_{-13.60}^{+11.00}$  & $-0.12_{-0.30}^{+0.24}$ & $0.51_{-0.37}^{+0.39}$ & $1.74_{-1.44}^{+1.12}$ & 9.6  & 1.29 & 0.13            & 0.83     & 0.014   \\
    GW200210\_092254 & $28.68_{-5.24}^{+8.45}$     & $3.38_{-0.52}^{+0.52}$     & $0.02_{-0.21}^{+0.22}$  & $0.15_{-0.12}^{+0.22}$ & $2.31_{-1.97}^{+0.60}$ & 8.4  & 1.2  & 0.14            & 0.88     & 0.06    \\
    GW200216\_220804 & $84.36_{-21.07}^{+28.39}$   & $50.74_{-30.85}^{+22.50}$  & $0.10_{-0.36}^{+0.34}$  & $0.45_{-0.35}^{+0.42}$ & $0.89_{-0.69}^{+1.87}$ & 8.1  & 1.32 & 0.16            & 0.82     & 0.006   \\
    GW200219\_094415 & $58.57_{-8.95}^{+13.46}$    & $44.41_{-14.14}^{+9.26}$   & $-0.08_{-0.29}^{+0.23}$ & $0.48_{-0.35}^{+0.40}$ & $1.19_{-0.92}^{+1.59}$ & 10.7 & 1.61 & 0.15            & 0.71     & 0.017   \\
    GW200220\_061928 & $165.56_{-27.93}^{+62.24}$  & $120.51_{-55.27}^{+29.47}$ & $0.06_{-0.38}^{+0.40}$  & $0.50_{-0.37}^{+0.37}$ & $1.24_{-0.96}^{+1.55}$ & 7.2  & 1.2  & 0.17            & 0.87     & 0.022   \\
    GW200220\_124850 & $64.13_{-10.34}^{+16.51}$   & $46.89_{-16.61}^{+11.54}$  & $-0.07_{-0.33}^{+0.27}$ & $0.49_{-0.37}^{+0.39}$ & $1.66_{-1.34}^{+1.17}$ & 8.5  & 1.72 & 0.2             & 0.68     & 0.007   \\
    GW200224\_222234 & $52.30_{-5.37}^{+9.08}$     & $42.82_{-9.89}^{+5.83}$    & $0.10_{-0.15}^{+0.15}$  & $0.49_{-0.36}^{+0.37}$ & $0.62_{-0.45}^{+0.55}$ & 20   & 2.26 & 0.11            & 0.52     & 0.022   \\
    GW200225\_060421 & $23.58_{-3.25}^{+5.65}$     & $17.20_{-4.55}^{+3.05}$    & $-0.12_{-0.28}^{+0.17}$ & $0.53_{-0.38}^{+0.34}$ & $1.31_{-1.00}^{+1.47}$ & 12.5 & 1.43 & 0.11            & 0.8      & 0.015   \\
    GW200302\_015811 & $48.45_{-9.58}^{+10.33}$    & $25.50_{-7.63}^{+12.02}$   & $0.01_{-0.26}^{+0.25}$  & $0.37_{-0.28}^{+0.45}$ & $1.34_{-1.01}^{+1.43}$ & 10.8 & 2.13 & 0.2             & 0.51     & 0.054   \\
    GW200306\_093714 & $39.52_{-10.80}^{+21.47}$   & $20.98_{-9.91}^{+8.86}$    & $0.32_{-0.46}^{+0.28}$  & $0.43_{-0.31}^{+0.39}$ & $1.12_{-0.87}^{+1.74}$ & 7.8  & 1.33 & 0.17            & 0.82     & 0.013   \\
    GW200308\_173609 & $143.98_{-85.60}^{+299.28}$ & $52.82_{-33.18}^{+83.32}$  & $0.16_{-0.49}^{+0.58}$  & $0.41_{-0.30}^{+0.42}$ & $1.55_{-1.27}^{+1.19}$ & 7.1  & 1.1  & 0.15            & 0.89     & 0.054   \\
    GW200311\_115853 & $41.83_{-4.59}^{+8.21}$     & $34.00_{-7.26}^{+4.68}$    & $-0.02_{-0.20}^{+0.16}$ & $0.45_{-0.35}^{+0.40}$ & $0.55_{-0.40}^{+0.52}$ & 17.8 & 1.27 & 0.07            & 0.87     & 0.026   \\
    GW200316\_215756 & $15.98_{-3.33}^{+12.23}$    & $9.52_{-3.49}^{+2.32}$     & $0.13_{-0.10}^{+0.27}$  & $0.29_{-0.20}^{+0.38}$ & $2.32_{-1.85}^{+0.58}$ & 10.3 & 1.08 & 0.1             & 0.9      & 0.082   \\
    GW200322\_091133 & $105.87_{-84.00}^{+392.13}$ & $26.23_{-18.59}^{+58.87}$  & $0.08_{-0.47}^{+0.51}$  & $0.50_{-0.41}^{+0.36}$ & $1.66_{-1.12}^{+1.04}$ & 6    & 0.87 & 0.14            & 0.94     & 0.059     
\end{tabular}
\caption{\label{tab2}The second half of our BBH analysis results, including 8 new events in GWTC-2.1 (compared to GWTC-2) and 36 events in GWTC-3. Columns have the same meaning as Tab.~\ref{tab1}.}
\end{ruledtabular}
\end{table*}

\bibliography{refs.bib}

\begin{thebibliography}{77}%
\makeatletter
\providecommand \@ifxundefined [1]{%
 \@ifx{#1\undefined}
}%
\providecommand \@ifnum [1]{%
 \ifnum #1\expandafter \@firstoftwo
 \else \expandafter \@secondoftwo
 \fi
}%
\providecommand \@ifx [1]{%
 \ifx #1\expandafter \@firstoftwo
 \else \expandafter \@secondoftwo
 \fi
}%
\providecommand \natexlab [1]{#1}%
\providecommand \enquote  [1]{``#1''}%
\providecommand \bibnamefont  [1]{#1}%
\providecommand \bibfnamefont [1]{#1}%
\providecommand \citenamefont [1]{#1}%
\providecommand \href@noop [0]{\@secondoftwo}%
\providecommand \href [0]{\begingroup \@sanitize@url \@href}%
\providecommand \@href[1]{\@@startlink{#1}\@@href}%
\providecommand \@@href[1]{\endgroup#1\@@endlink}%
\providecommand \@sanitize@url [0]{\catcode `\\12\catcode `\$12\catcode
  `\&12\catcode `\#12\catcode `\^12\catcode `\_12\catcode `\%12\relax}%
\providecommand \@@startlink[1]{}%
\providecommand \@@endlink[0]{}%
\providecommand \url  [0]{\begingroup\@sanitize@url \@url }%
\providecommand \@url [1]{\endgroup\@href {#1}{\urlprefix }}%
\providecommand \urlprefix  [0]{URL }%
\providecommand \Eprint [0]{\href }%
\providecommand \doibase [0]{https://doi.org/}%
\providecommand \selectlanguage [0]{\@gobble}%
\providecommand \bibinfo  [0]{\@secondoftwo}%
\providecommand \bibfield  [0]{\@secondoftwo}%
\providecommand \translation [1]{[#1]}%
\providecommand \BibitemOpen [0]{}%
\providecommand \bibitemStop [0]{}%
\providecommand \bibitemNoStop [0]{.\EOS\space}%
\providecommand \EOS [0]{\spacefactor3000\relax}%
\providecommand \BibitemShut  [1]{\csname bibitem#1\endcsname}%
\let\auto@bib@innerbib\@empty
\bibitem [{\citenamefont {{LIGO Scientific Collaboration and Virgo
  Collaboration}}(2019)}]{abbott2019:GWTC1GravitationalWaveTransient}%
  \BibitemOpen
  \bibfield  {author} {\bibinfo {author} {\bibnamefont {{LIGO Scientific
  Collaboration and Virgo Collaboration}}},\ }\bibfield  {title} {\bibinfo
  {title} {{{GWTC-1}}: {{A Gravitational-Wave Transient Catalog}} of {{Compact
  Binary Mergers Observed}} by {{LIGO}} and {{Virgo}} during the {{First}} and
  {{Second Observing Runs}}},\ }\href
  {https://doi.org/10.1103/PhysRevX.9.031040} {\bibfield  {journal} {\bibinfo
  {journal} {Physical Review X}\ }\textbf {\bibinfo {volume} {9}},\ \bibinfo
  {pages} {031040} (\bibinfo {year} {2019})}\BibitemShut {NoStop}%
\bibitem [{\citenamefont {Abbott}\ \emph
  {et~al.}(2021{\natexlab{a}})\citenamefont {Abbott} \emph
  {et~al.}}]{abbott2021:GWTC2CompactBinary}%
  \BibitemOpen
  \bibfield  {author} {\bibinfo {author} {\bibfnamefont {R.}~\bibnamefont
  {Abbott}} \emph {et~al.},\ }\bibfield  {title} {\bibinfo {title} {{{GWTC-2}}:
  {{Compact Binary Coalescences Observed}} by {{LIGO}} and {{Virgo}} during the
  {{First Half}} of the {{Third Observing Run}}},\ }\href
  {https://doi.org/10.1103/PhysRevX.11.021053} {\bibfield  {journal} {\bibinfo
  {journal} {Physical Review X}\ }\textbf {\bibinfo {volume} {11}},\ \bibinfo
  {pages} {1–54} (\bibinfo {year} {2021}{\natexlab{a}})}\BibitemShut
  {NoStop}%
\bibitem [{\citenamefont {Abbott}\ \emph
  {et~al.}(2021{\natexlab{b}})\citenamefont {Abbott} \emph
  {et~al.}}]{abbott2021:GWTC3CompactBinary}%
  \BibitemOpen
  \bibfield  {author} {\bibinfo {author} {\bibfnamefont {R.}~\bibnamefont
  {Abbott}} \emph {et~al.},\ }\href@noop {} {\bibinfo {title} {{{GWTC-3}}:
  {{Compact Binary Coalescences Observed}} by {{LIGO}} and {{Virgo During}} the
  {{Second Part}} of the {{Third Observing Run}}}} (\bibinfo {year}
  {2021}{\natexlab{b}}),\ \Eprint {https://arxiv.org/abs/2111.03606}
  {arXiv:2111.03606} \BibitemShut {NoStop}%
\bibitem [{\citenamefont {{The LIGO Scientific Collaboration}}\ and\
  \citenamefont {{The Virgo
  Collaboration}}(2021)}]{theligoscientificcollaboration2021:GWTC2DeepExtended}%
  \BibitemOpen
  \bibfield  {author} {\bibinfo {author} {\bibnamefont {{The LIGO Scientific
  Collaboration}}}\ and\ \bibinfo {author} {\bibnamefont {{The Virgo
  Collaboration}}},\ }\href@noop {} {\bibinfo {title} {{{GWTC-2}}.1: {{Deep
  Extended Catalog}} of {{Compact Binary Coalescences Observed}} by {{LIGO}}
  and {{Virgo During}} the {{First Half}} of the {{Third Observing Run}}}}
  (\bibinfo {year} {2021}),\ \Eprint {https://arxiv.org/abs/2108.01045}
  {arXiv:2108.01045 [gr-qc]} \BibitemShut {NoStop}%
\bibitem [{\citenamefont {Nitz}\ \emph {et~al.}(2019)\citenamefont {Nitz},
  \citenamefont {Capano}, \citenamefont {Nielsen}, \citenamefont {Reyes},
  \citenamefont {White}, \citenamefont {Brown},\ and\ \citenamefont
  {Krishnan}}]{nitz2019_1OGCFirstOpen}%
  \BibitemOpen
  \bibfield  {author} {\bibinfo {author} {\bibfnamefont {A.~H.}\ \bibnamefont
  {Nitz}}, \bibinfo {author} {\bibfnamefont {C.}~\bibnamefont {Capano}},
  \bibinfo {author} {\bibfnamefont {A.~B.}\ \bibnamefont {Nielsen}}, \bibinfo
  {author} {\bibfnamefont {S.}~\bibnamefont {Reyes}}, \bibinfo {author}
  {\bibfnamefont {R.}~\bibnamefont {White}}, \bibinfo {author} {\bibfnamefont
  {D.~A.}\ \bibnamefont {Brown}},\ and\ \bibinfo {author} {\bibfnamefont
  {B.}~\bibnamefont {Krishnan}},\ }\bibfield  {title} {\bibinfo {title}
  {1-{{OGC}}: {{The First Open Gravitational-wave Catalog}} of {{Binary
  Mergers}} from {{Analysis}} of {{Public Advanced LIGO Data}}},\ }\href
  {https://doi.org/10.3847/1538-4357/ab0108} {\bibfield  {journal} {\bibinfo
  {journal} {The Astrophysical Journal}\ }\textbf {\bibinfo {volume} {872}},\
  \bibinfo {pages} {195} (\bibinfo {year} {2019})}\BibitemShut {NoStop}%
\bibitem [{\citenamefont {Nitz}\ \emph {et~al.}(2020)\citenamefont {Nitz},
  \citenamefont {Dent}, \citenamefont {Davies}, \citenamefont {Kumar},
  \citenamefont {Capano}, \citenamefont {Harry}, \citenamefont {Mozzon},
  \citenamefont {Nuttall}, \citenamefont {Lundgren},\ and\ \citenamefont
  {T{\'a}pai}}]{nitz2020_2OGCOpenGravitationalwave}%
  \BibitemOpen
  \bibfield  {author} {\bibinfo {author} {\bibfnamefont {A.~H.}\ \bibnamefont
  {Nitz}}, \bibinfo {author} {\bibfnamefont {T.}~\bibnamefont {Dent}}, \bibinfo
  {author} {\bibfnamefont {G.~S.}\ \bibnamefont {Davies}}, \bibinfo {author}
  {\bibfnamefont {S.}~\bibnamefont {Kumar}}, \bibinfo {author} {\bibfnamefont
  {C.~D.}\ \bibnamefont {Capano}}, \bibinfo {author} {\bibfnamefont
  {I.}~\bibnamefont {Harry}}, \bibinfo {author} {\bibfnamefont
  {S.}~\bibnamefont {Mozzon}}, \bibinfo {author} {\bibfnamefont
  {L.}~\bibnamefont {Nuttall}}, \bibinfo {author} {\bibfnamefont
  {A.}~\bibnamefont {Lundgren}},\ and\ \bibinfo {author} {\bibfnamefont
  {M.}~\bibnamefont {T{\'a}pai}},\ }\bibfield  {title} {\bibinfo {title}
  {2-{{OGC}}: {{Open Gravitational-wave Catalog}} of {{Binary Mergers}} from
  {{Analysis}} of {{Public Advanced LIGO}} and {{Virgo Data}}},\ }\href
  {https://doi.org/10.3847/1538-4357/ab733f} {\bibfield  {journal} {\bibinfo
  {journal} {The Astrophysical Journal}\ }\textbf {\bibinfo {volume} {891}},\
  \bibinfo {pages} {123} (\bibinfo {year} {2020})}\BibitemShut {NoStop}%
\bibitem [{\citenamefont {Nitz}\ \emph
  {et~al.}(2021{\natexlab{a}})\citenamefont {Nitz}, \citenamefont {Capano},
  \citenamefont {Kumar}, \citenamefont {Wang}, \citenamefont {Kastha},
  \citenamefont {Sch{\"a}fer}, \citenamefont {Dhurkunde},\ and\ \citenamefont
  {Cabero}}]{nitz2021_3OGCCatalogGravitational}%
  \BibitemOpen
  \bibfield  {author} {\bibinfo {author} {\bibfnamefont {A.~H.}\ \bibnamefont
  {Nitz}}, \bibinfo {author} {\bibfnamefont {C.~D.}\ \bibnamefont {Capano}},
  \bibinfo {author} {\bibfnamefont {S.}~\bibnamefont {Kumar}}, \bibinfo
  {author} {\bibfnamefont {Y.-F.}\ \bibnamefont {Wang}}, \bibinfo {author}
  {\bibfnamefont {S.}~\bibnamefont {Kastha}}, \bibinfo {author} {\bibfnamefont
  {M.}~\bibnamefont {Sch{\"a}fer}}, \bibinfo {author} {\bibfnamefont
  {R.}~\bibnamefont {Dhurkunde}},\ and\ \bibinfo {author} {\bibfnamefont
  {M.}~\bibnamefont {Cabero}},\ }\bibfield  {title} {\bibinfo {title}
  {3-{{OGC}}: {{Catalog}} of gravitational waves from compact-binary mergers},\
  }\href {https://doi.org/10.3847/1538-4357/ac1c03} {\bibfield  {journal}
  {\bibinfo  {journal} {The Astrophysical Journal}\ }\textbf {\bibinfo {volume}
  {922}},\ \bibinfo {pages} {76} (\bibinfo {year} {2021}{\natexlab{a}})},\
  \Eprint {https://arxiv.org/abs/2105.09151} {arXiv:2105.09151} \BibitemShut
  {NoStop}%
\bibitem [{\citenamefont {Nitz}\ \emph
  {et~al.}(2021{\natexlab{b}})\citenamefont {Nitz}, \citenamefont {Kumar},
  \citenamefont {Wang}, \citenamefont {Kastha}, \citenamefont {Wu},
  \citenamefont {Sch{\"a}fer}, \citenamefont {Dhurkunde},\ and\ \citenamefont
  {Capano}}]{nitz2021_4OGCCatalogGravitational}%
  \BibitemOpen
  \bibfield  {author} {\bibinfo {author} {\bibfnamefont {A.~H.}\ \bibnamefont
  {Nitz}}, \bibinfo {author} {\bibfnamefont {S.}~\bibnamefont {Kumar}},
  \bibinfo {author} {\bibfnamefont {Y.-F.}\ \bibnamefont {Wang}}, \bibinfo
  {author} {\bibfnamefont {S.}~\bibnamefont {Kastha}}, \bibinfo {author}
  {\bibfnamefont {S.}~\bibnamefont {Wu}}, \bibinfo {author} {\bibfnamefont
  {M.}~\bibnamefont {Sch{\"a}fer}}, \bibinfo {author} {\bibfnamefont
  {R.}~\bibnamefont {Dhurkunde}},\ and\ \bibinfo {author} {\bibfnamefont
  {C.~D.}\ \bibnamefont {Capano}},\ }\bibfield  {title} {\bibinfo {title}
  {4-{{OGC}}: {{Catalog}} of gravitational waves from compact-binary mergers},\
  }\href {http://arxiv.org/abs/2112.06878} {\bibfield  {journal} {\bibinfo
  {journal} {arXiv}\ } (\bibinfo {year} {2021}{\natexlab{b}})},\ \Eprint
  {https://arxiv.org/abs/2112.06878} {arXiv:2112.06878} \BibitemShut {NoStop}%
\bibitem [{\citenamefont {Zackay}\ \emph {et~al.}(2021)\citenamefont {Zackay},
  \citenamefont {Dai}, \citenamefont {Venumadhav}, \citenamefont {Roulet},\
  and\ \citenamefont {Zaldarriaga}}]{Zackay_2021}%
  \BibitemOpen
  \bibfield  {author} {\bibinfo {author} {\bibfnamefont {B.}~\bibnamefont
  {Zackay}}, \bibinfo {author} {\bibfnamefont {L.}~\bibnamefont {Dai}},
  \bibinfo {author} {\bibfnamefont {T.}~\bibnamefont {Venumadhav}}, \bibinfo
  {author} {\bibfnamefont {J.}~\bibnamefont {Roulet}},\ and\ \bibinfo {author}
  {\bibfnamefont {M.}~\bibnamefont {Zaldarriaga}},\ }\bibfield  {title}
  {\bibinfo {title} {Detecting gravitational waves with disparate detector
  responses: Two new binary black hole mergers},\ }\bibfield  {journal}
  {\bibinfo  {journal} {Physical Review D}\ }\textbf {\bibinfo {volume}
  {104}},\ \href {https://doi.org/10.1103/physrevd.104.063030}
  {10.1103/physrevd.104.063030} (\bibinfo {year} {2021})\BibitemShut {NoStop}%
\bibitem [{\citenamefont {Venumadhav}\ \emph {et~al.}(2020)\citenamefont
  {Venumadhav}, \citenamefont {Zackay}, \citenamefont {Roulet}, \citenamefont
  {Dai},\ and\ \citenamefont {Zaldarriaga}}]{PhysRevD.101.083030}%
  \BibitemOpen
  \bibfield  {author} {\bibinfo {author} {\bibfnamefont {T.}~\bibnamefont
  {Venumadhav}}, \bibinfo {author} {\bibfnamefont {B.}~\bibnamefont {Zackay}},
  \bibinfo {author} {\bibfnamefont {J.}~\bibnamefont {Roulet}}, \bibinfo
  {author} {\bibfnamefont {L.}~\bibnamefont {Dai}},\ and\ \bibinfo {author}
  {\bibfnamefont {M.}~\bibnamefont {Zaldarriaga}},\ }\bibfield  {title}
  {\bibinfo {title} {{New binary black hole mergers in the second observing run
  of Advanced LIGO and Advanced Virgo}},\ }\href
  {https://doi.org/10.1103/PhysRevD.101.083030} {\bibfield  {journal} {\bibinfo
   {journal} {Phys. Rev. D}\ }\textbf {\bibinfo {volume} {101}},\ \bibinfo
  {pages} {083030} (\bibinfo {year} {2020})}\BibitemShut {NoStop}%
\bibitem [{\citenamefont {Venumadhav}\ \emph {et~al.}(2019)\citenamefont
  {Venumadhav}, \citenamefont {Zackay}, \citenamefont {Roulet}, \citenamefont
  {Dai},\ and\ \citenamefont {Zaldarriaga}}]{PhysRevD.100.023011}%
  \BibitemOpen
  \bibfield  {author} {\bibinfo {author} {\bibfnamefont {T.}~\bibnamefont
  {Venumadhav}}, \bibinfo {author} {\bibfnamefont {B.}~\bibnamefont {Zackay}},
  \bibinfo {author} {\bibfnamefont {J.}~\bibnamefont {Roulet}}, \bibinfo
  {author} {\bibfnamefont {L.}~\bibnamefont {Dai}},\ and\ \bibinfo {author}
  {\bibfnamefont {M.}~\bibnamefont {Zaldarriaga}},\ }\bibfield  {title}
  {\bibinfo {title} {New search pipeline for compact binary mergers: Results
  for binary black holes in the first observing run of advanced ligo},\ }\href
  {https://doi.org/10.1103/PhysRevD.100.023011} {\bibfield  {journal} {\bibinfo
   {journal} {Phys. Rev. D}\ }\textbf {\bibinfo {volume} {100}},\ \bibinfo
  {pages} {023011} (\bibinfo {year} {2019})}\BibitemShut {NoStop}%
\bibitem [{\citenamefont {Zackay}\ \emph {et~al.}(2019)\citenamefont {Zackay},
  \citenamefont {Venumadhav}, \citenamefont {Dai}, \citenamefont {Roulet},\
  and\ \citenamefont {Zaldarriaga}}]{PhysRevD.100.023007}%
  \BibitemOpen
  \bibfield  {author} {\bibinfo {author} {\bibfnamefont {B.}~\bibnamefont
  {Zackay}}, \bibinfo {author} {\bibfnamefont {T.}~\bibnamefont {Venumadhav}},
  \bibinfo {author} {\bibfnamefont {L.}~\bibnamefont {Dai}}, \bibinfo {author}
  {\bibfnamefont {J.}~\bibnamefont {Roulet}},\ and\ \bibinfo {author}
  {\bibfnamefont {M.}~\bibnamefont {Zaldarriaga}},\ }\bibfield  {title}
  {\bibinfo {title} {Highly spinning and aligned binary black hole merger in
  the advanced ligo first observing run},\ }\href
  {https://doi.org/10.1103/PhysRevD.100.023007} {\bibfield  {journal} {\bibinfo
   {journal} {Phys. Rev. D}\ }\textbf {\bibinfo {volume} {100}},\ \bibinfo
  {pages} {023007} (\bibinfo {year} {2019})}\BibitemShut {NoStop}%
\bibitem [{\citenamefont {Aasi}\ \emph {et~al.}(2015)\citenamefont {Aasi} \emph
  {et~al.}}]{theligoscientificcollaboration2015_AdvancedLIGO}%
  \BibitemOpen
  \bibfield  {author} {\bibinfo {author} {\bibfnamefont {J.}~\bibnamefont
  {Aasi}} \emph {et~al.},\ }\bibfield  {title} {\bibinfo {title} {Advanced
  {{LIGO}}},\ }\href {https://doi.org/10.1088/0264-9381/32/7/074001} {\bibfield
   {journal} {\bibinfo  {journal} {Classical and Quantum Gravity}\ }\textbf
  {\bibinfo {volume} {32}},\ \bibinfo {pages} {074001} (\bibinfo {year}
  {2015})},\ \Eprint {https://arxiv.org/abs/1411.4547} {arXiv:1411.4547}
  \BibitemShut {NoStop}%
\bibitem [{\citenamefont {Acernese}\ \emph {et~al.}(2015)\citenamefont
  {Acernese} \emph {et~al.}}]{acernese2015_AdvancedVirgo2nd}%
  \BibitemOpen
  \bibfield  {author} {\bibinfo {author} {\bibfnamefont {F.}~\bibnamefont
  {Acernese}} \emph {et~al.},\ }\bibfield  {title} {\bibinfo {title} {Advanced
  {{Virgo}}: A 2nd generation interferometric gravitational wave detector},\
  }\href {https://doi.org/10.1088/0264-9381/32/2/024001} {\bibfield  {journal}
  {\bibinfo  {journal} {Classical and Quantum Gravity}\ }\textbf {\bibinfo
  {volume} {32}},\ \bibinfo {pages} {024001} (\bibinfo {year} {2015})},\
  \Eprint {https://arxiv.org/abs/1408.3978} {arXiv:1408.3978} \BibitemShut
  {NoStop}%
\bibitem [{\citenamefont {Allen}\ \emph {et~al.}(2012)\citenamefont {Allen},
  \citenamefont {Anderson}, \citenamefont {Brady}, \citenamefont {Brown},\ and\
  \citenamefont {Creighton}}]{allen2012:FINDCHIRPAlgorithmDetection}%
  \BibitemOpen
  \bibfield  {author} {\bibinfo {author} {\bibfnamefont {B.}~\bibnamefont
  {Allen}}, \bibinfo {author} {\bibfnamefont {W.~G.}\ \bibnamefont {Anderson}},
  \bibinfo {author} {\bibfnamefont {P.~R.}\ \bibnamefont {Brady}}, \bibinfo
  {author} {\bibfnamefont {D.~A.}\ \bibnamefont {Brown}},\ and\ \bibinfo
  {author} {\bibfnamefont {J.~D.~E.}\ \bibnamefont {Creighton}},\ }\bibfield
  {title} {\bibinfo {title} {{{FINDCHIRP}}: {{An}} algorithm for detection of
  gravitational waves from inspiraling compact binaries},\ }\href
  {https://doi.org/10.1103/PhysRevD.85.122006} {\bibfield  {journal} {\bibinfo
  {journal} {Physical Review D}\ }\textbf {\bibinfo {volume} {85}},\ \bibinfo
  {pages} {122006} (\bibinfo {year} {2012})}\BibitemShut {NoStop}%
\bibitem [{\citenamefont {Veitch}\ \emph {et~al.}(2015)\citenamefont {Veitch}
  \emph {et~al.}}]{veitch2015:ParameterEstimationCompact}%
  \BibitemOpen
  \bibfield  {author} {\bibinfo {author} {\bibfnamefont {J.}~\bibnamefont
  {Veitch}} \emph {et~al.},\ }\bibfield  {title} {\bibinfo {title} {Parameter
  estimation for compact binaries with ground-based gravitational-wave
  observations using the {{LALInference}} software library},\ }\href
  {https://doi.org/10.1103/PhysRevD.91.042003} {\bibfield  {journal} {\bibinfo
  {journal} {Physical Review D}\ }\textbf {\bibinfo {volume} {91}},\ \bibinfo
  {pages} {1–25} (\bibinfo {year} {2015})}\BibitemShut {NoStop}%
\bibitem [{\citenamefont {Biwer}\ \emph {et~al.}(2019)\citenamefont {Biwer},
  \citenamefont {Capano}, \citenamefont {De}, \citenamefont {Cabero},
  \citenamefont {Brown}, \citenamefont {Nitz},\ and\ \citenamefont
  {Raymond}}]{biwer2019:PyCBCInferencePythonbased}%
  \BibitemOpen
  \bibfield  {author} {\bibinfo {author} {\bibfnamefont {C.~M.}\ \bibnamefont
  {Biwer}}, \bibinfo {author} {\bibfnamefont {C.~D.}\ \bibnamefont {Capano}},
  \bibinfo {author} {\bibfnamefont {S.}~\bibnamefont {De}}, \bibinfo {author}
  {\bibfnamefont {M.}~\bibnamefont {Cabero}}, \bibinfo {author} {\bibfnamefont
  {D.~A.}\ \bibnamefont {Brown}}, \bibinfo {author} {\bibfnamefont {A.~H.}\
  \bibnamefont {Nitz}},\ and\ \bibinfo {author} {\bibfnamefont
  {V.}~\bibnamefont {Raymond}},\ }\bibfield  {title} {\bibinfo {title}
  {{{PyCBC}} inference: A python-based parameter estimation toolkit for compact
  binary coalescence signals},\ }\bibfield  {journal} {\bibinfo  {journal}
  {Publications of the Astronomical Society of the Pacific}\ }\textbf {\bibinfo
  {volume} {131}},\ \href {https://doi.org/10.1088/1538-3873/aaef0b}
  {10.1088/1538-3873/aaef0b} (\bibinfo {year} {2019})\BibitemShut {NoStop}%
\bibitem [{\citenamefont {Lange}\ \emph {et~al.}(2018)\citenamefont {Lange},
  \citenamefont {O'Shaughnessy},\ and\ \citenamefont
  {Rizzo}}]{lange2018:RapidAccurateParameter}%
  \BibitemOpen
  \bibfield  {author} {\bibinfo {author} {\bibfnamefont {J.}~\bibnamefont
  {Lange}}, \bibinfo {author} {\bibfnamefont {R.}~\bibnamefont
  {O'Shaughnessy}},\ and\ \bibinfo {author} {\bibfnamefont {M.}~\bibnamefont
  {Rizzo}},\ }\href {http://arxiv.org/abs/1805.10457} {\bibinfo {title} {Rapid
  and accurate parameter inference for coalescing, precessing compact
  binaries}} (\bibinfo {year} {2018}),\ \Eprint
  {https://arxiv.org/abs/1805.10457} {arXiv:1805.10457} \BibitemShut {NoStop}%
\bibitem [{\citenamefont {Ashton}\ \emph {et~al.}(2019)\citenamefont {Ashton}
  \emph {et~al.}}]{ashton2019:BilbyUserfriendlyBayesian}%
  \BibitemOpen
  \bibfield  {author} {\bibinfo {author} {\bibfnamefont {G.}~\bibnamefont
  {Ashton}} \emph {et~al.},\ }\bibfield  {title} {\bibinfo {title} {{BILBY: A
  user-friendly Bayesian inference library for gravitational-wave astronomy}},\
  }\href {https://doi.org/10.3847/1538-4365/ab06fc} {\bibfield  {journal}
  {\bibinfo  {journal} {Astrophys. J. Suppl.}\ }\textbf {\bibinfo {volume}
  {241}},\ \bibinfo {pages} {27} (\bibinfo {year} {2019})},\ \Eprint
  {https://arxiv.org/abs/1811.02042} {arXiv:1811.02042 [astro-ph.IM]}
  \BibitemShut {NoStop}%
\bibitem [{\citenamefont {Lindblom}\ \emph {et~al.}(2008)\citenamefont
  {Lindblom}, \citenamefont {Owen},\ and\ \citenamefont
  {Brown}}]{lindblom2008:ModelWaveformAccuracy}%
  \BibitemOpen
  \bibfield  {author} {\bibinfo {author} {\bibfnamefont {L.}~\bibnamefont
  {Lindblom}}, \bibinfo {author} {\bibfnamefont {B.~J.}\ \bibnamefont {Owen}},\
  and\ \bibinfo {author} {\bibfnamefont {D.~A.}\ \bibnamefont {Brown}},\
  }\bibfield  {title} {\bibinfo {title} {Model waveform accuracy standards for
  gravitational wave data analysis},\ }\href
  {https://doi.org/10.1103/PhysRevD.78.124020} {\bibfield  {journal} {\bibinfo
  {journal} {Physical Review D}\ }\textbf {\bibinfo {volume} {78}},\ \bibinfo
  {pages} {1–12} (\bibinfo {year} {2008})}\BibitemShut {NoStop}%
\bibitem [{\citenamefont
  {Blanchet}(2014)}]{blanchet2014_GravitationalRadiationPostnewtonian}%
  \BibitemOpen
  \bibfield  {author} {\bibinfo {author} {\bibfnamefont {L.}~\bibnamefont
  {Blanchet}},\ }\bibfield  {title} {\bibinfo {title} {Gravitational radiation
  from post-newtonian sources and inspiralling compact binaries},\ }\href
  {https://doi.org/10.12942/lrr-2014-2} {\bibfield  {journal} {\bibinfo
  {journal} {Living Reviews in Relativity}\ }\textbf {\bibinfo {volume} {17}},\
  \bibinfo {pages} {1–185} (\bibinfo {year} {2014})}\BibitemShut {NoStop}%
\bibitem [{\citenamefont {Berti}\ \emph {et~al.}(2009)\citenamefont {Berti},
  \citenamefont {Cardoso},\ and\ \citenamefont
  {Starinets}}]{berti2009_QuasinormalModesBlack}%
  \BibitemOpen
  \bibfield  {author} {\bibinfo {author} {\bibfnamefont {E.}~\bibnamefont
  {Berti}}, \bibinfo {author} {\bibfnamefont {V.}~\bibnamefont {Cardoso}},\
  and\ \bibinfo {author} {\bibfnamefont {A.~O.}\ \bibnamefont {Starinets}},\
  }\bibfield  {title} {\bibinfo {title} {Quasinormal modes of black holes and
  black branes},\ }\href {https://doi.org/10.1088/0264-9381/26/16/163001}
  {\bibfield  {journal} {\bibinfo  {journal} {Classical and Quantum Gravity}\
  }\textbf {\bibinfo {volume} {26}},\ \bibinfo {pages} {163001} (\bibinfo
  {year} {2009})}\BibitemShut {NoStop}%
\bibitem [{\citenamefont {Jani}\ \emph {et~al.}(2016)\citenamefont {Jani},
  \citenamefont {Healy}, \citenamefont {Clark}, \citenamefont {London},
  \citenamefont {Laguna},\ and\ \citenamefont
  {Shoemaker}}]{jani2016:GeorgiaTechCatalog}%
  \BibitemOpen
  \bibfield  {author} {\bibinfo {author} {\bibfnamefont {K.}~\bibnamefont
  {Jani}}, \bibinfo {author} {\bibfnamefont {J.}~\bibnamefont {Healy}},
  \bibinfo {author} {\bibfnamefont {J.~A.}\ \bibnamefont {Clark}}, \bibinfo
  {author} {\bibfnamefont {L.}~\bibnamefont {London}}, \bibinfo {author}
  {\bibfnamefont {P.}~\bibnamefont {Laguna}},\ and\ \bibinfo {author}
  {\bibfnamefont {D.}~\bibnamefont {Shoemaker}},\ }\bibfield  {title} {\bibinfo
  {title} {Georgia tech catalog of gravitational waveforms},\ }\href
  {https://doi.org/10.1088/0264-9381/33/20/204001} {\bibfield  {journal}
  {\bibinfo  {journal} {Classical and Quantum Gravity}\ }\textbf {\bibinfo
  {volume} {33}},\ \bibinfo {pages} {1–8} (\bibinfo {year}
  {2016})}\BibitemShut {NoStop}%
\bibitem [{\citenamefont {Mrou{\'e}}\ \emph {et~al.}(2013)\citenamefont
  {Mrou{\'e}} \emph {et~al.}}]{mroue2013:Catalog174Binary}%
  \BibitemOpen
  \bibfield  {author} {\bibinfo {author} {\bibfnamefont {A.~H.}\ \bibnamefont
  {Mrou{\'e}}} \emph {et~al.},\ }\bibfield  {title} {\bibinfo {title} {Catalog
  of 174 {{Binary Black Hole Simulations}} for {{Gravitational Wave
  Astronomy}}},\ }\href {https://doi.org/10.1103/PhysRevLett.111.241104}
  {\bibfield  {journal} {\bibinfo  {journal} {Physical Review Letters}\
  }\textbf {\bibinfo {volume} {111}},\ \bibinfo {pages} {241104} (\bibinfo
  {year} {2013})}\BibitemShut {NoStop}%
\bibitem [{\citenamefont {Boyle}\ \emph {et~al.}(2019)\citenamefont {Boyle}
  \emph {et~al.}}]{boyle2019_SXSCollaborationCatalog}%
  \BibitemOpen
  \bibfield  {author} {\bibinfo {author} {\bibfnamefont {M.}~\bibnamefont
  {Boyle}} \emph {et~al.},\ }\bibfield  {title} {\bibinfo {title} {The {{SXS}}
  collaboration catalog of binary black hole simulations},\ }\href
  {https://doi.org/10.1088/1361-6382/ab34e2} {\bibfield  {journal} {\bibinfo
  {journal} {Classical and Quantum Gravity}\ }\textbf {\bibinfo {volume}
  {36}},\ \bibinfo {pages} {195006} (\bibinfo {year} {2019})}\BibitemShut
  {NoStop}%
\bibitem [{\citenamefont {Szil{\'{a}}gyi}\ \emph {et~al.}(2015)\citenamefont
  {Szil{\'{a}}gyi}, \citenamefont {Blackman}, \citenamefont {Buonanno},
  \citenamefont {Taracchini}, \citenamefont {Pfeiffer}, \citenamefont {Scheel},
  \citenamefont {Chu}, \citenamefont {Kidder},\ and\ \citenamefont
  {Pan}}]{Szil_gyi_2015}%
  \BibitemOpen
  \bibfield  {author} {\bibinfo {author} {\bibfnamefont {B.}~\bibnamefont
  {Szil{\'{a}}gyi}}, \bibinfo {author} {\bibfnamefont {J.}~\bibnamefont
  {Blackman}}, \bibinfo {author} {\bibfnamefont {A.}~\bibnamefont {Buonanno}},
  \bibinfo {author} {\bibfnamefont {A.}~\bibnamefont {Taracchini}}, \bibinfo
  {author} {\bibfnamefont {H.~P.}\ \bibnamefont {Pfeiffer}}, \bibinfo {author}
  {\bibfnamefont {M.~A.}\ \bibnamefont {Scheel}}, \bibinfo {author}
  {\bibfnamefont {T.}~\bibnamefont {Chu}}, \bibinfo {author} {\bibfnamefont
  {L.~E.}\ \bibnamefont {Kidder}},\ and\ \bibinfo {author} {\bibfnamefont
  {Y.}~\bibnamefont {Pan}},\ }\bibfield  {title} {\bibinfo {title} {Approaching
  the post-newtonian regime with numerical relativity: A compact-object binary
  simulation spanning 350 gravitational-wave cycles},\ }\bibfield  {journal}
  {\bibinfo  {journal} {Physical Review Letters}\ }\textbf {\bibinfo {volume}
  {115}},\ \href {https://doi.org/10.1103/physrevlett.115.031102}
  {10.1103/physrevlett.115.031102} (\bibinfo {year} {2015})\BibitemShut
  {NoStop}%
\bibitem [{\citenamefont {Ajith}\ \emph {et~al.}(2007)\citenamefont {Ajith}
  \emph {et~al.}}]{ajith2007_PhenomenologicalTemplateFamily}%
  \BibitemOpen
  \bibfield  {author} {\bibinfo {author} {\bibfnamefont {P.}~\bibnamefont
  {Ajith}} \emph {et~al.},\ }\bibfield  {title} {\bibinfo {title} {A
  phenomenological template family for black-hole coalescence waveforms},\
  }\href {https://doi.org/10.1088/0264-9381/24/19/S31} {\bibfield  {journal}
  {\bibinfo  {journal} {Classical and Quantum Gravity}\ }\textbf {\bibinfo
  {volume} {24}},\ \bibinfo {pages} {S689} (\bibinfo {year}
  {2007})}\BibitemShut {NoStop}%
\bibitem [{\citenamefont {Khan}\ \emph {et~al.}(2016)\citenamefont {Khan},
  \citenamefont {Husa}, \citenamefont {Hannam}, \citenamefont {Ohme},
  \citenamefont {P\"urrer}, \citenamefont {Jim\'enez~Forteza},\ and\
  \citenamefont {Boh\'e}}]{Khan:2015jqa}%
  \BibitemOpen
  \bibfield  {author} {\bibinfo {author} {\bibfnamefont {S.}~\bibnamefont
  {Khan}}, \bibinfo {author} {\bibfnamefont {S.}~\bibnamefont {Husa}}, \bibinfo
  {author} {\bibfnamefont {M.}~\bibnamefont {Hannam}}, \bibinfo {author}
  {\bibfnamefont {F.}~\bibnamefont {Ohme}}, \bibinfo {author} {\bibfnamefont
  {M.}~\bibnamefont {P\"urrer}}, \bibinfo {author} {\bibfnamefont
  {X.}~\bibnamefont {Jim\'enez~Forteza}},\ and\ \bibinfo {author}
  {\bibfnamefont {A.}~\bibnamefont {Boh\'e}},\ }\bibfield  {title} {\bibinfo
  {title} {{Frequency-domain gravitational waves from nonprecessing black-hole
  binaries. II. A phenomenological model for the advanced detector era}},\
  }\href {https://doi.org/10.1103/PhysRevD.93.044007} {\bibfield  {journal}
  {\bibinfo  {journal} {Phys. Rev. D}\ }\textbf {\bibinfo {volume} {93}},\
  \bibinfo {pages} {044007} (\bibinfo {year} {2016})},\ \Eprint
  {https://arxiv.org/abs/1508.07253} {arXiv:1508.07253 [gr-qc]} \BibitemShut
  {NoStop}%
\bibitem [{\citenamefont {Pratten}\ \emph {et~al.}(2020)\citenamefont
  {Pratten}, \citenamefont {Husa}, \citenamefont {Garcia-Quiros}, \citenamefont
  {Colleoni}, \citenamefont {Ramos-Buades}, \citenamefont {Estelles},\ and\
  \citenamefont {Jaume}}]{Pratten:2020fqn}%
  \BibitemOpen
  \bibfield  {author} {\bibinfo {author} {\bibfnamefont {G.}~\bibnamefont
  {Pratten}}, \bibinfo {author} {\bibfnamefont {S.}~\bibnamefont {Husa}},
  \bibinfo {author} {\bibfnamefont {C.}~\bibnamefont {Garcia-Quiros}}, \bibinfo
  {author} {\bibfnamefont {M.}~\bibnamefont {Colleoni}}, \bibinfo {author}
  {\bibfnamefont {A.}~\bibnamefont {Ramos-Buades}}, \bibinfo {author}
  {\bibfnamefont {H.}~\bibnamefont {Estelles}},\ and\ \bibinfo {author}
  {\bibfnamefont {R.}~\bibnamefont {Jaume}},\ }\bibfield  {title} {\bibinfo
  {title} {{Setting the cornerstone for a family of models for gravitational
  waves from compact binaries: The dominant harmonic for nonprecessing
  quasicircular black holes}},\ }\href
  {https://doi.org/10.1103/PhysRevD.102.064001} {\bibfield  {journal} {\bibinfo
   {journal} {Phys. Rev. D}\ }\textbf {\bibinfo {volume} {102}},\ \bibinfo
  {pages} {064001} (\bibinfo {year} {2020})},\ \Eprint
  {https://arxiv.org/abs/2001.11412} {arXiv:2001.11412 [gr-qc]} \BibitemShut
  {NoStop}%
\bibitem [{\citenamefont {Garc\'\i{}a-Quir\'os}\ \emph
  {et~al.}(2020)\citenamefont {Garc\'\i{}a-Quir\'os}, \citenamefont {Colleoni},
  \citenamefont {Husa}, \citenamefont {Estell\'es}, \citenamefont {Pratten},
  \citenamefont {Ramos-Buades}, \citenamefont {Mateu-Lucena},\ and\
  \citenamefont {Jaume}}]{Garcia-Quiros:2020qpx}%
  \BibitemOpen
  \bibfield  {author} {\bibinfo {author} {\bibfnamefont {C.}~\bibnamefont
  {Garc\'\i{}a-Quir\'os}}, \bibinfo {author} {\bibfnamefont {M.}~\bibnamefont
  {Colleoni}}, \bibinfo {author} {\bibfnamefont {S.}~\bibnamefont {Husa}},
  \bibinfo {author} {\bibfnamefont {H.}~\bibnamefont {Estell\'es}}, \bibinfo
  {author} {\bibfnamefont {G.}~\bibnamefont {Pratten}}, \bibinfo {author}
  {\bibfnamefont {A.}~\bibnamefont {Ramos-Buades}}, \bibinfo {author}
  {\bibfnamefont {M.}~\bibnamefont {Mateu-Lucena}},\ and\ \bibinfo {author}
  {\bibfnamefont {R.}~\bibnamefont {Jaume}},\ }\bibfield  {title} {\bibinfo
  {title} {{Multimode frequency-domain model for the gravitational wave signal
  from nonprecessing black-hole binaries}},\ }\href
  {https://doi.org/10.1103/PhysRevD.102.064002} {\bibfield  {journal} {\bibinfo
   {journal} {Phys. Rev. D}\ }\textbf {\bibinfo {volume} {102}},\ \bibinfo
  {pages} {064002} (\bibinfo {year} {2020})},\ \Eprint
  {https://arxiv.org/abs/2001.10914} {arXiv:2001.10914 [gr-qc]} \BibitemShut
  {NoStop}%
\bibitem [{\citenamefont {Pratten}\ \emph {et~al.}(2021)\citenamefont {Pratten}
  \emph {et~al.}}]{pratten2021_ComputationallyEfficientModels}%
  \BibitemOpen
  \bibfield  {author} {\bibinfo {author} {\bibfnamefont {G.}~\bibnamefont
  {Pratten}} \emph {et~al.},\ }\bibfield  {title} {\bibinfo {title}
  {Computationally efficient models for the dominant and subdominant harmonic
  modes of precessing binary black holes},\ }\href
  {https://doi.org/10.1103/PhysRevD.103.104056} {\bibfield  {journal} {\bibinfo
   {journal} {Physical Review D}\ }\textbf {\bibinfo {volume} {103}},\ \bibinfo
  {pages} {104056} (\bibinfo {year} {2021})}\BibitemShut {NoStop}%
\bibitem [{\citenamefont {Buonanno}\ and\ \citenamefont
  {Damour}(1999)}]{buonanno1999_EffectiveOnebodyApproach}%
  \BibitemOpen
  \bibfield  {author} {\bibinfo {author} {\bibfnamefont {A.}~\bibnamefont
  {Buonanno}}\ and\ \bibinfo {author} {\bibfnamefont {T.}~\bibnamefont
  {Damour}},\ }\bibfield  {title} {\bibinfo {title} {Effective one-body
  approach to general relativistic two-body dynamics},\ }\href
  {https://doi.org/10.1103/PhysRevD.59.084006} {\bibfield  {journal} {\bibinfo
  {journal} {Physical Review D}\ }\textbf {\bibinfo {volume} {59}},\ \bibinfo
  {pages} {084006} (\bibinfo {year} {1999})}\BibitemShut {NoStop}%
\bibitem [{\citenamefont {Damour}(2001)}]{Damour:2001tu}%
  \BibitemOpen
  \bibfield  {author} {\bibinfo {author} {\bibfnamefont {T.}~\bibnamefont
  {Damour}},\ }\bibfield  {title} {\bibinfo {title} {{Coalescence of two
  spinning black holes: an effective one-body approach}},\ }\href
  {https://doi.org/10.1103/PhysRevD.64.124013} {\bibfield  {journal} {\bibinfo
  {journal} {Phys. Rev. D}\ }\textbf {\bibinfo {volume} {64}},\ \bibinfo
  {pages} {124013} (\bibinfo {year} {2001})},\ \Eprint
  {https://arxiv.org/abs/gr-qc/0103018} {arXiv:gr-qc/0103018} \BibitemShut
  {NoStop}%
\bibitem [{\citenamefont {Boh\'e}\ \emph {et~al.}(2017)\citenamefont {Boh\'e}
  \emph {et~al.}}]{Bohe:2016gbl}%
  \BibitemOpen
  \bibfield  {author} {\bibinfo {author} {\bibfnamefont {A.}~\bibnamefont
  {Boh\'e}} \emph {et~al.},\ }\bibfield  {title} {\bibinfo {title} {{Improved
  effective-one-body model of spinning, nonprecessing binary black holes for
  the era of gravitational-wave astrophysics with advanced detectors}},\ }\href
  {https://doi.org/10.1103/PhysRevD.95.044028} {\bibfield  {journal} {\bibinfo
  {journal} {Phys. Rev. D}\ }\textbf {\bibinfo {volume} {95}},\ \bibinfo
  {pages} {044028} (\bibinfo {year} {2017})},\ \Eprint
  {https://arxiv.org/abs/1611.03703} {arXiv:1611.03703 [gr-qc]} \BibitemShut
  {NoStop}%
\bibitem [{\citenamefont {Cotesta}\ \emph {et~al.}(2018)\citenamefont
  {Cotesta}, \citenamefont {Buonanno}, \citenamefont {Boh\'e}, \citenamefont
  {Taracchini}, \citenamefont {Hinder},\ and\ \citenamefont
  {Ossokine}}]{Cotesta:2018fcv}%
  \BibitemOpen
  \bibfield  {author} {\bibinfo {author} {\bibfnamefont {R.}~\bibnamefont
  {Cotesta}}, \bibinfo {author} {\bibfnamefont {A.}~\bibnamefont {Buonanno}},
  \bibinfo {author} {\bibfnamefont {A.}~\bibnamefont {Boh\'e}}, \bibinfo
  {author} {\bibfnamefont {A.}~\bibnamefont {Taracchini}}, \bibinfo {author}
  {\bibfnamefont {I.}~\bibnamefont {Hinder}},\ and\ \bibinfo {author}
  {\bibfnamefont {S.}~\bibnamefont {Ossokine}},\ }\bibfield  {title} {\bibinfo
  {title} {{Enriching the Symphony of Gravitational Waves from Binary Black
  Holes by Tuning Higher Harmonics}},\ }\href
  {https://doi.org/10.1103/PhysRevD.98.084028} {\bibfield  {journal} {\bibinfo
  {journal} {Phys. Rev. D}\ }\textbf {\bibinfo {volume} {98}},\ \bibinfo
  {pages} {084028} (\bibinfo {year} {2018})},\ \Eprint
  {https://arxiv.org/abs/1803.10701} {arXiv:1803.10701 [gr-qc]} \BibitemShut
  {NoStop}%
\bibitem [{\citenamefont {Pan}\ \emph {et~al.}(2014)\citenamefont {Pan},
  \citenamefont {Buonanno}, \citenamefont {Taracchini}, \citenamefont {Kidder},
  \citenamefont {Mrou\'e}, \citenamefont {Pfeiffer}, \citenamefont {Scheel},\
  and\ \citenamefont {Szil\'agyi}}]{Pan:2013rra}%
  \BibitemOpen
  \bibfield  {author} {\bibinfo {author} {\bibfnamefont {Y.}~\bibnamefont
  {Pan}}, \bibinfo {author} {\bibfnamefont {A.}~\bibnamefont {Buonanno}},
  \bibinfo {author} {\bibfnamefont {A.}~\bibnamefont {Taracchini}}, \bibinfo
  {author} {\bibfnamefont {L.~E.}\ \bibnamefont {Kidder}}, \bibinfo {author}
  {\bibfnamefont {A.~H.}\ \bibnamefont {Mrou\'e}}, \bibinfo {author}
  {\bibfnamefont {H.~P.}\ \bibnamefont {Pfeiffer}}, \bibinfo {author}
  {\bibfnamefont {M.~A.}\ \bibnamefont {Scheel}},\ and\ \bibinfo {author}
  {\bibfnamefont {B.}~\bibnamefont {Szil\'agyi}},\ }\bibfield  {title}
  {\bibinfo {title} {{Inspiral-merger-ringdown waveforms of spinning,
  precessing black-hole binaries in the effective-one-body formalism}},\ }\href
  {https://doi.org/10.1103/PhysRevD.89.084006} {\bibfield  {journal} {\bibinfo
  {journal} {Phys. Rev. D}\ }\textbf {\bibinfo {volume} {89}},\ \bibinfo
  {pages} {084006} (\bibinfo {year} {2014})},\ \Eprint
  {https://arxiv.org/abs/1307.6232} {arXiv:1307.6232 [gr-qc]} \BibitemShut
  {NoStop}%
\bibitem [{\citenamefont {Ossokine}\ \emph {et~al.}(2020)\citenamefont
  {Ossokine} \emph
  {et~al.}}]{ossokine2020_MultipolarEffectiveonebodyWaveforms}%
  \BibitemOpen
  \bibfield  {author} {\bibinfo {author} {\bibfnamefont {S.}~\bibnamefont
  {Ossokine}} \emph {et~al.},\ }\bibfield  {title} {\bibinfo {title}
  {Multipolar effective-one-body waveforms for precessing binary black holes:
  {{Construction}} and validation},\ }\href
  {https://doi.org/10.1103/PhysRevD.102.044055} {\bibfield  {journal} {\bibinfo
   {journal} {Physical Review D}\ }\textbf {\bibinfo {volume} {102}},\ \bibinfo
  {pages} {044055} (\bibinfo {year} {2020})}\BibitemShut {NoStop}%
\bibitem [{\citenamefont {Nagar}\ \emph {et~al.}(2018)\citenamefont {Nagar}
  \emph {et~al.}}]{Nagar:2018zoe}%
  \BibitemOpen
  \bibfield  {author} {\bibinfo {author} {\bibfnamefont {A.}~\bibnamefont
  {Nagar}} \emph {et~al.},\ }\bibfield  {title} {\bibinfo {title} {{Time-domain
  effective-one-body gravitational waveforms for coalescing compact binaries
  with nonprecessing spins, tides and self-spin effects}},\ }\href
  {https://doi.org/10.1103/PhysRevD.98.104052} {\bibfield  {journal} {\bibinfo
  {journal} {Phys. Rev. D}\ }\textbf {\bibinfo {volume} {98}},\ \bibinfo
  {pages} {104052} (\bibinfo {year} {2018})},\ \Eprint
  {https://arxiv.org/abs/1806.01772} {arXiv:1806.01772 [gr-qc]} \BibitemShut
  {NoStop}%
\bibitem [{\citenamefont {Nagar}\ \emph
  {et~al.}(2020{\natexlab{a}})\citenamefont {Nagar}, \citenamefont
  {Riemenschneider}, \citenamefont {Pratten}, \citenamefont {Rettegno},\ and\
  \citenamefont {Messina}}]{Nagar:2020pcj}%
  \BibitemOpen
  \bibfield  {author} {\bibinfo {author} {\bibfnamefont {A.}~\bibnamefont
  {Nagar}}, \bibinfo {author} {\bibfnamefont {G.}~\bibnamefont
  {Riemenschneider}}, \bibinfo {author} {\bibfnamefont {G.}~\bibnamefont
  {Pratten}}, \bibinfo {author} {\bibfnamefont {P.}~\bibnamefont {Rettegno}},\
  and\ \bibinfo {author} {\bibfnamefont {F.}~\bibnamefont {Messina}},\
  }\bibfield  {title} {\bibinfo {title} {{Multipolar effective one body
  waveform model for spin-aligned black hole binaries}},\ }\href
  {https://doi.org/10.1103/PhysRevD.102.024077} {\bibfield  {journal} {\bibinfo
   {journal} {Phys. Rev. D}\ }\textbf {\bibinfo {volume} {102}},\ \bibinfo
  {pages} {024077} (\bibinfo {year} {2020}{\natexlab{a}})},\ \Eprint
  {https://arxiv.org/abs/2001.09082} {arXiv:2001.09082 [gr-qc]} \BibitemShut
  {NoStop}%
\bibitem [{\citenamefont {Nagar}\ \emph
  {et~al.}(2020{\natexlab{b}})\citenamefont {Nagar}, \citenamefont {Pratten},
  \citenamefont {Riemenschneider},\ and\ \citenamefont
  {Gamba}}]{Nagar:2019wds}%
  \BibitemOpen
  \bibfield  {author} {\bibinfo {author} {\bibfnamefont {A.}~\bibnamefont
  {Nagar}}, \bibinfo {author} {\bibfnamefont {G.}~\bibnamefont {Pratten}},
  \bibinfo {author} {\bibfnamefont {G.}~\bibnamefont {Riemenschneider}},\ and\
  \bibinfo {author} {\bibfnamefont {R.}~\bibnamefont {Gamba}},\ }\bibfield
  {title} {\bibinfo {title} {{Multipolar effective one body model for
  nonspinning black hole binaries}},\ }\href
  {https://doi.org/10.1103/PhysRevD.101.024041} {\bibfield  {journal} {\bibinfo
   {journal} {Phys. Rev. D}\ }\textbf {\bibinfo {volume} {101}},\ \bibinfo
  {pages} {024041} (\bibinfo {year} {2020}{\natexlab{b}})},\ \Eprint
  {https://arxiv.org/abs/1904.09550} {arXiv:1904.09550 [gr-qc]} \BibitemShut
  {NoStop}%
\bibitem [{\citenamefont {Blackman}\ \emph {et~al.}(2015)\citenamefont
  {Blackman}, \citenamefont {Field}, \citenamefont {Galley}, \citenamefont
  {Szil{\'a}gyi}, \citenamefont {Scheel}, \citenamefont {Tiglio},\ and\
  \citenamefont {Hemberger}}]{blackman2015_FastAccuratePrediction}%
  \BibitemOpen
  \bibfield  {author} {\bibinfo {author} {\bibfnamefont {J.}~\bibnamefont
  {Blackman}}, \bibinfo {author} {\bibfnamefont {S.~E.}\ \bibnamefont {Field}},
  \bibinfo {author} {\bibfnamefont {C.~R.}\ \bibnamefont {Galley}}, \bibinfo
  {author} {\bibfnamefont {B.}~\bibnamefont {Szil{\'a}gyi}}, \bibinfo {author}
  {\bibfnamefont {M.~A.}\ \bibnamefont {Scheel}}, \bibinfo {author}
  {\bibfnamefont {M.}~\bibnamefont {Tiglio}},\ and\ \bibinfo {author}
  {\bibfnamefont {D.~A.}\ \bibnamefont {Hemberger}},\ }\bibfield  {title}
  {\bibinfo {title} {Fast and {{Accurate Prediction}} of {{Numerical Relativity
  Waveforms}} from {{Binary Black Hole Coalescences Using Surrogate Models}}},\
  }\href {https://doi.org/10.1103/PhysRevLett.115.121102} {\bibfield  {journal}
  {\bibinfo  {journal} {Physical Review Letters}\ }\textbf {\bibinfo {volume}
  {115}},\ \bibinfo {pages} {121102} (\bibinfo {year} {2015})}\BibitemShut
  {NoStop}%
\bibitem [{\citenamefont {Varma}\ \emph {et~al.}(2019)\citenamefont {Varma},
  \citenamefont {Field}, \citenamefont {Scheel}, \citenamefont {Blackman},
  \citenamefont {Gerosa}, \citenamefont {Stein}, \citenamefont {Kidder},\ and\
  \citenamefont {Pfeiffer}}]{Varma:2019csw}%
  \BibitemOpen
  \bibfield  {author} {\bibinfo {author} {\bibfnamefont {V.}~\bibnamefont
  {Varma}}, \bibinfo {author} {\bibfnamefont {S.~E.}\ \bibnamefont {Field}},
  \bibinfo {author} {\bibfnamefont {M.~A.}\ \bibnamefont {Scheel}}, \bibinfo
  {author} {\bibfnamefont {J.}~\bibnamefont {Blackman}}, \bibinfo {author}
  {\bibfnamefont {D.}~\bibnamefont {Gerosa}}, \bibinfo {author} {\bibfnamefont
  {L.~C.}\ \bibnamefont {Stein}}, \bibinfo {author} {\bibfnamefont {L.~E.}\
  \bibnamefont {Kidder}},\ and\ \bibinfo {author} {\bibfnamefont {H.~P.}\
  \bibnamefont {Pfeiffer}},\ }\bibfield  {title} {\bibinfo {title} {{Surrogate
  models for precessing binary black hole simulations with unequal masses}},\
  }\href {https://doi.org/10.1103/PhysRevResearch.1.033015} {\bibfield
  {journal} {\bibinfo  {journal} {Phys. Rev. Research.}\ }\textbf {\bibinfo
  {volume} {1}},\ \bibinfo {pages} {033015} (\bibinfo {year} {2019})},\ \Eprint
  {https://arxiv.org/abs/1905.09300} {arXiv:1905.09300 [gr-qc]} \BibitemShut
  {NoStop}%
\bibitem [{\citenamefont {Williams}\ \emph {et~al.}(2020)\citenamefont
  {Williams}, \citenamefont {Heng}, \citenamefont {Gair}, \citenamefont
  {Clark},\ and\ \citenamefont
  {Khamesra}}]{williams2020_PrecessingNumericalRelativity}%
  \BibitemOpen
  \bibfield  {author} {\bibinfo {author} {\bibfnamefont {D.}~\bibnamefont
  {Williams}}, \bibinfo {author} {\bibfnamefont {I.~S.}\ \bibnamefont {Heng}},
  \bibinfo {author} {\bibfnamefont {J.}~\bibnamefont {Gair}}, \bibinfo {author}
  {\bibfnamefont {J.~A.}\ \bibnamefont {Clark}},\ and\ \bibinfo {author}
  {\bibfnamefont {B.}~\bibnamefont {Khamesra}},\ }\bibfield  {title} {\bibinfo
  {title} {Precessing numerical relativity waveform surrogate model for binary
  black holes: {{A Gaussian}} process regression approach},\ }\bibfield
  {journal} {\bibinfo  {journal} {Physical Review D}\ }\textbf {\bibinfo
  {volume} {101}},\ \href {https://doi.org/10.1103/PhysRevD.101.063011}
  {10.1103/PhysRevD.101.063011} (\bibinfo {year} {2020})\BibitemShut {NoStop}%
\bibitem [{\citenamefont {Field}\ \emph {et~al.}(2014)\citenamefont {Field},
  \citenamefont {Galley}, \citenamefont {Hesthaven}, \citenamefont {Kaye},\
  and\ \citenamefont {Tiglio}}]{Field_2014}%
  \BibitemOpen
  \bibfield  {author} {\bibinfo {author} {\bibfnamefont {S.~E.}\ \bibnamefont
  {Field}}, \bibinfo {author} {\bibfnamefont {C.~R.}\ \bibnamefont {Galley}},
  \bibinfo {author} {\bibfnamefont {J.~S.}\ \bibnamefont {Hesthaven}}, \bibinfo
  {author} {\bibfnamefont {J.}~\bibnamefont {Kaye}},\ and\ \bibinfo {author}
  {\bibfnamefont {M.}~\bibnamefont {Tiglio}},\ }\bibfield  {title} {\bibinfo
  {title} {{Fast prediction and evaluation of gravitational waveforms using
  surrogate models}},\ }\href {https://doi.org/10.1103/PhysRevX.4.031006}
  {\bibfield  {journal} {\bibinfo  {journal} {Phys. Rev. X}\ }\textbf {\bibinfo
  {volume} {4}},\ \bibinfo {pages} {031006} (\bibinfo {year} {2014})},\ \Eprint
  {https://arxiv.org/abs/1308.3565} {arXiv:1308.3565 [gr-qc]} \BibitemShut
  {NoStop}%
\bibitem [{\citenamefont {P\"urrer}(2014)}]{Purrer:2014fza}%
  \BibitemOpen
  \bibfield  {author} {\bibinfo {author} {\bibfnamefont {M.}~\bibnamefont
  {P\"urrer}},\ }\bibfield  {title} {\bibinfo {title} {{Frequency domain
  reduced order models for gravitational waves from aligned-spin compact
  binaries}},\ }\href {https://doi.org/10.1088/0264-9381/31/19/195010}
  {\bibfield  {journal} {\bibinfo  {journal} {Class. Quant. Grav.}\ }\textbf
  {\bibinfo {volume} {31}},\ \bibinfo {pages} {195010} (\bibinfo {year}
  {2014})},\ \Eprint {https://arxiv.org/abs/1402.4146} {arXiv:1402.4146
  [gr-qc]} \BibitemShut {NoStop}%
\bibitem [{\citenamefont {Abbott}\ \emph
  {et~al.}(2020{\natexlab{a}})\citenamefont {Abbott} \emph
  {et~al.}}]{abbott2020:GW190521BinaryBlack}%
  \BibitemOpen
  \bibfield  {author} {\bibinfo {author} {\bibfnamefont {R.}~\bibnamefont
  {Abbott}} \emph {et~al.},\ }\bibfield  {title} {\bibinfo {title}
  {{{GW190521}}: {{A Binary Black Hole Merger}} with a {{Total Mass}} of 150
  {{M}}},\ }\href {https://doi.org/10.1103/PhysRevLett.125.101102} {\bibfield
  {journal} {\bibinfo  {journal} {Physical Review Letters}\ }\textbf {\bibinfo
  {volume} {125}},\ \bibinfo {pages} {1–17} (\bibinfo {year}
  {2020}{\natexlab{a}})}\BibitemShut {NoStop}%
\bibitem [{\citenamefont {Abbott}\ \emph
  {et~al.}(2020{\natexlab{b}})\citenamefont {Abbott} \emph
  {et~al.}}]{abbott2020:GW190412ObservationBinaryblackhole}%
  \BibitemOpen
  \bibfield  {author} {\bibinfo {author} {\bibfnamefont {R.}~\bibnamefont
  {Abbott}} \emph {et~al.},\ }\bibfield  {title} {\bibinfo {title}
  {{{GW190412}}: {{Observation}} of a binary-black-hole coalescence with
  asymmetric masses},\ }\href {https://doi.org/10.1103/PhysRevD.102.043015}
  {\bibfield  {journal} {\bibinfo  {journal} {Physical Review D}\ }\textbf
  {\bibinfo {volume} {102}},\ \bibinfo {pages} {043015} (\bibinfo {year}
  {2020}{\natexlab{b}})}\BibitemShut {NoStop}%
\bibitem [{\citenamefont {P{\"u}rrer}\ and\ \citenamefont
  {Haster}(2020)}]{purrer2020_GravitationalWaveformAccuracy}%
  \BibitemOpen
  \bibfield  {author} {\bibinfo {author} {\bibfnamefont {M.}~\bibnamefont
  {P{\"u}rrer}}\ and\ \bibinfo {author} {\bibfnamefont {C.~J.}\ \bibnamefont
  {Haster}},\ }\bibfield  {title} {\bibinfo {title} {Gravitational waveform
  accuracy requirements for future ground-based detectors},\ }\href
  {https://doi.org/10.1103/PhysRevResearch.2.023151} {\bibfield  {journal}
  {\bibinfo  {journal} {Physical Review Research}\ }\textbf {\bibinfo {volume}
  {2}},\ \bibinfo {pages} {1–30} (\bibinfo {year} {2020})}\BibitemShut
  {NoStop}%
\bibitem [{\citenamefont {Kumar}\ \emph {et~al.}(2016)\citenamefont {Kumar},
  \citenamefont {Chu}, \citenamefont {Fong}, \citenamefont {Pfeiffer},
  \citenamefont {Boyle}, \citenamefont {Hemberger}, \citenamefont {Kidder},
  \citenamefont {Scheel},\ and\ \citenamefont
  {Szilagyi}}]{kumar2016_AccuracyBinaryBlack}%
  \BibitemOpen
  \bibfield  {author} {\bibinfo {author} {\bibfnamefont {P.}~\bibnamefont
  {Kumar}}, \bibinfo {author} {\bibfnamefont {T.}~\bibnamefont {Chu}}, \bibinfo
  {author} {\bibfnamefont {H.}~\bibnamefont {Fong}}, \bibinfo {author}
  {\bibfnamefont {H.~P.}\ \bibnamefont {Pfeiffer}}, \bibinfo {author}
  {\bibfnamefont {M.}~\bibnamefont {Boyle}}, \bibinfo {author} {\bibfnamefont
  {D.~A.}\ \bibnamefont {Hemberger}}, \bibinfo {author} {\bibfnamefont {L.~E.}\
  \bibnamefont {Kidder}}, \bibinfo {author} {\bibfnamefont {M.~A.}\
  \bibnamefont {Scheel}},\ and\ \bibinfo {author} {\bibfnamefont
  {B.}~\bibnamefont {Szilagyi}},\ }\bibfield  {title} {\bibinfo {title}
  {Accuracy of binary black hole waveform models for aligned-spin binaries},\
  }\href {https://doi.org/10.1103/PhysRevD.93.104050} {\bibfield  {journal}
  {\bibinfo  {journal} {Physical Review D}\ }\textbf {\bibinfo {volume} {93}},\
  \bibinfo {pages} {1–25} (\bibinfo {year} {2016})}\BibitemShut {NoStop}%
\bibitem [{\citenamefont {Abbott}\ \emph
  {et~al.}(2021{\natexlab{c}})\citenamefont {Abbott} \emph
  {et~al.}}]{theligoscientificcollaboration2021_PopulationMergingCompact}%
  \BibitemOpen
  \bibfield  {author} {\bibinfo {author} {\bibfnamefont {R.}~\bibnamefont
  {Abbott}} \emph {et~al.} (\bibinfo {collaboration} {LIGO Scientific
  Collaboration, VIRGO Collaboration, KAGRA Collaboration}),\ }\href@noop {}
  {\bibinfo {title} {{The population of merging compact binaries inferred using
  gravitational waves through GWTC-3}}} (\bibinfo {year}
  {2021}{\natexlab{c}}),\ \Eprint {https://arxiv.org/abs/2111.03634}
  {arXiv:2111.03634 [astro-ph.HE]} \BibitemShut {NoStop}%
\bibitem [{\citenamefont {Hannam}\ \emph {et~al.}(2021)\citenamefont {Hannam},
  \citenamefont {Hoy}, \citenamefont {Thompson}, \citenamefont {Fairhurst},\
  and\ \citenamefont {Raymond}}]{Hannam:2021pit}%
  \BibitemOpen
  \bibfield  {author} {\bibinfo {author} {\bibfnamefont {M.}~\bibnamefont
  {Hannam}}, \bibinfo {author} {\bibfnamefont {C.}~\bibnamefont {Hoy}},
  \bibinfo {author} {\bibfnamefont {J.~E.}\ \bibnamefont {Thompson}}, \bibinfo
  {author} {\bibfnamefont {S.}~\bibnamefont {Fairhurst}},\ and\ \bibinfo
  {author} {\bibfnamefont {V.}~\bibnamefont {Raymond}} (\bibinfo
  {collaboration} {VIRGO}),\ }\href@noop {} {\bibinfo {title} {{Measurement of
  general-relativistic precession in a black-hole binary}}} (\bibinfo {year}
  {2021}),\ \Eprint {https://arxiv.org/abs/2112.11300} {arXiv:2112.11300
  [gr-qc]} \BibitemShut {NoStop}%
\bibitem [{\citenamefont {{Jan}}\ \emph {et~al.}(2020)\citenamefont {{Jan}},
  \citenamefont {{Yelikar}}, \citenamefont {{Lange}},\ and\ \citenamefont
  {{O'Shaughnessy}}}]{2020PhRvD.102l4069J}%
  \BibitemOpen
  \bibfield  {author} {\bibinfo {author} {\bibfnamefont {A.~Z.}\ \bibnamefont
  {{Jan}}}, \bibinfo {author} {\bibfnamefont {A.~B.}\ \bibnamefont
  {{Yelikar}}}, \bibinfo {author} {\bibfnamefont {J.}~\bibnamefont {{Lange}}},\
  and\ \bibinfo {author} {\bibfnamefont {R.}~\bibnamefont {{O'Shaughnessy}}},\
  }\bibfield  {title} {\bibinfo {title} {{Assessing and marginalizing over
  compact binary coalescence waveform systematics with RIFT}},\ }\href
  {https://doi.org/10.1103/PhysRevD.102.124069} {\bibfield  {journal} {\bibinfo
   {journal} {\prd}\ }\textbf {\bibinfo {volume} {102}},\ \bibinfo {eid}
  {124069} (\bibinfo {year} {2020})},\ \Eprint
  {https://arxiv.org/abs/2011.03571} {arXiv:2011.03571 [gr-qc]} \BibitemShut
  {NoStop}%
\bibitem [{\citenamefont {{Williamson}}\ \emph {et~al.}(2017)\citenamefont
  {{Williamson}}, \citenamefont {{Lange}}, \citenamefont {{O'Shaughnessy}},
  \citenamefont {{Clark}}, \citenamefont {{Kumar}}, \citenamefont
  {{Calder{\'o}n Bustillo}},\ and\ \citenamefont
  {{Veitch}}}]{2017PhRvD..96l4041W}%
  \BibitemOpen
  \bibfield  {author} {\bibinfo {author} {\bibfnamefont {A.~R.}\ \bibnamefont
  {{Williamson}}}, \bibinfo {author} {\bibfnamefont {J.}~\bibnamefont
  {{Lange}}}, \bibinfo {author} {\bibfnamefont {R.}~\bibnamefont
  {{O'Shaughnessy}}}, \bibinfo {author} {\bibfnamefont {J.~A.}\ \bibnamefont
  {{Clark}}}, \bibinfo {author} {\bibfnamefont {P.}~\bibnamefont {{Kumar}}},
  \bibinfo {author} {\bibfnamefont {J.}~\bibnamefont {{Calder{\'o}n
  Bustillo}}},\ and\ \bibinfo {author} {\bibfnamefont {J.}~\bibnamefont
  {{Veitch}}},\ }\bibfield  {title} {\bibinfo {title} {{Systematic challenges
  for future gravitational wave measurements of precessing binary black
  holes}},\ }\href {https://doi.org/10.1103/PhysRevD.96.124041} {\bibfield
  {journal} {\bibinfo  {journal} {\prd}\ }\textbf {\bibinfo {volume} {96}},\
  \bibinfo {eid} {124041} (\bibinfo {year} {2017})},\ \Eprint
  {https://arxiv.org/abs/1709.03095} {arXiv:1709.03095 [gr-qc]} \BibitemShut
  {NoStop}%
\bibitem [{\citenamefont {Ferguson}\ \emph {et~al.}(2021)\citenamefont
  {Ferguson}, \citenamefont {Jani}, \citenamefont {Laguna},\ and\ \citenamefont
  {Shoemaker}}]{ferguson2021_AssessingReadinessNumerical}%
  \BibitemOpen
  \bibfield  {author} {\bibinfo {author} {\bibfnamefont {D.}~\bibnamefont
  {Ferguson}}, \bibinfo {author} {\bibfnamefont {K.}~\bibnamefont {Jani}},
  \bibinfo {author} {\bibfnamefont {P.}~\bibnamefont {Laguna}},\ and\ \bibinfo
  {author} {\bibfnamefont {D.}~\bibnamefont {Shoemaker}},\ }\bibfield  {title}
  {\bibinfo {title} {Assessing the readiness of numerical relativity for
  {{LISA}} and {{3G}} detectors},\ }\href
  {https://doi.org/10.1103/PhysRevD.104.044037} {\bibfield  {journal} {\bibinfo
   {journal} {Physical Review D}\ }\textbf {\bibinfo {volume} {104}},\ \bibinfo
  {pages} {044037} (\bibinfo {year} {2021})}\BibitemShut {NoStop}%
\bibitem [{\citenamefont {Kunert}\ \emph {et~al.}(2022)\citenamefont {Kunert},
  \citenamefont {Pang}, \citenamefont {Tews}, \citenamefont {Coughlin},\ and\
  \citenamefont {Dietrich}}]{Kunert:2021hgm}%
  \BibitemOpen
  \bibfield  {author} {\bibinfo {author} {\bibfnamefont {N.}~\bibnamefont
  {Kunert}}, \bibinfo {author} {\bibfnamefont {P.~T.~H.}\ \bibnamefont {Pang}},
  \bibinfo {author} {\bibfnamefont {I.}~\bibnamefont {Tews}}, \bibinfo {author}
  {\bibfnamefont {M.~W.}\ \bibnamefont {Coughlin}},\ and\ \bibinfo {author}
  {\bibfnamefont {T.}~\bibnamefont {Dietrich}},\ }\bibfield  {title} {\bibinfo
  {title} {{Quantifying modeling uncertainties when combining multiple
  gravitational-wave detections from binary neutron star sources}},\ }\href
  {https://doi.org/10.1103/PhysRevD.105.L061301} {\bibfield  {journal}
  {\bibinfo  {journal} {Phys. Rev. D}\ }\textbf {\bibinfo {volume} {105}},\
  \bibinfo {pages} {L061301} (\bibinfo {year} {2022})},\ \Eprint
  {https://arxiv.org/abs/2110.11835} {arXiv:2110.11835 [astro-ph.HE]}
  \BibitemShut {NoStop}%
\bibitem [{\citenamefont {Gamba}\ \emph
  {et~al.}(2021{\natexlab{a}})\citenamefont {Gamba}, \citenamefont {Breschi},
  \citenamefont {Bernuzzi}, \citenamefont {Agathos},\ and\ \citenamefont
  {Nagar}}]{PhysRevD.103.124015}%
  \BibitemOpen
  \bibfield  {author} {\bibinfo {author} {\bibfnamefont {R.}~\bibnamefont
  {Gamba}}, \bibinfo {author} {\bibfnamefont {M.}~\bibnamefont {Breschi}},
  \bibinfo {author} {\bibfnamefont {S.}~\bibnamefont {Bernuzzi}}, \bibinfo
  {author} {\bibfnamefont {M.}~\bibnamefont {Agathos}},\ and\ \bibinfo {author}
  {\bibfnamefont {A.}~\bibnamefont {Nagar}},\ }\bibfield  {title} {\bibinfo
  {title} {Waveform systematics in the gravitational-wave inference of tidal
  parameters and equation of state from binary neutron-star signals},\ }\href
  {https://doi.org/10.1103/PhysRevD.103.124015} {\bibfield  {journal} {\bibinfo
   {journal} {Phys. Rev. D}\ }\textbf {\bibinfo {volume} {103}},\ \bibinfo
  {pages} {124015} (\bibinfo {year} {2021}{\natexlab{a}})}\BibitemShut
  {NoStop}%
\bibitem [{\citenamefont
  {Finn}(1992)}]{finn1992_DetectionMeasurementGravitational}%
  \BibitemOpen
  \bibfield  {author} {\bibinfo {author} {\bibfnamefont {L.~S.}\ \bibnamefont
  {Finn}},\ }\bibfield  {title} {\bibinfo {title} {Detection, measurement, and
  gravitational radiation},\ }\href {https://doi.org/10.1103/PhysRevD.46.5236}
  {\bibfield  {journal} {\bibinfo  {journal} {Physical Review D}\ }\textbf
  {\bibinfo {volume} {46}},\ \bibinfo {pages} {5236} (\bibinfo {year}
  {1992})}\BibitemShut {NoStop}%
\bibitem [{\citenamefont {Cutler}\ and\ \citenamefont
  {Flanagan}(1994)}]{cutler1994_GravitationalWavesMerging}%
  \BibitemOpen
  \bibfield  {author} {\bibinfo {author} {\bibfnamefont {C.}~\bibnamefont
  {Cutler}}\ and\ \bibinfo {author} {\bibfnamefont {I.~E.}\ \bibnamefont
  {Flanagan}},\ }\bibfield  {title} {\bibinfo {title} {Gravitational waves from
  merging compact binaries: {{How}} accurately can one extract the binary's
  parameters from the inspiral waveform?},\ }\href
  {https://doi.org/10.1103/PhysRevD.49.2658} {\bibfield  {journal} {\bibinfo
  {journal} {Physical Review D}\ }\textbf {\bibinfo {volume} {49}},\ \bibinfo
  {pages} {2658} (\bibinfo {year} {1994})}\BibitemShut {NoStop}%
\bibitem [{\citenamefont {Damour}\ \emph {et~al.}(2011)\citenamefont {Damour},
  \citenamefont {Nagar},\ and\ \citenamefont {Trias}}]{Damour:2010zb}%
  \BibitemOpen
  \bibfield  {author} {\bibinfo {author} {\bibfnamefont {T.}~\bibnamefont
  {Damour}}, \bibinfo {author} {\bibfnamefont {A.}~\bibnamefont {Nagar}},\ and\
  \bibinfo {author} {\bibfnamefont {M.}~\bibnamefont {Trias}},\ }\bibfield
  {title} {\bibinfo {title} {{Accuracy and effectualness of closed-form,
  frequency-domain waveforms for non-spinning black hole binaries}},\ }\href
  {https://doi.org/10.1103/PhysRevD.83.024006} {\bibfield  {journal} {\bibinfo
  {journal} {Phys. Rev. D}\ }\textbf {\bibinfo {volume} {83}},\ \bibinfo
  {pages} {024006} (\bibinfo {year} {2011})},\ \Eprint
  {https://arxiv.org/abs/1009.5998} {arXiv:1009.5998 [gr-qc]} \BibitemShut
  {NoStop}%
\bibitem [{\citenamefont {Jaranowski}\ \emph {et~al.}(1998)\citenamefont
  {Jaranowski}, \citenamefont {Kr{\'o}lak},\ and\ \citenamefont
  {Schutz}}]{jaranowski1998_DataAnalysisGravitationalwave}%
  \BibitemOpen
  \bibfield  {author} {\bibinfo {author} {\bibfnamefont {P.}~\bibnamefont
  {Jaranowski}}, \bibinfo {author} {\bibfnamefont {A.}~\bibnamefont
  {Kr{\'o}lak}},\ and\ \bibinfo {author} {\bibfnamefont {B.~F.}\ \bibnamefont
  {Schutz}},\ }\bibfield  {title} {\bibinfo {title} {Data analysis of
  gravitational-wave signals from spinning neutron stars: {{The}} signal and
  its detection},\ }\href {https://doi.org/10.1103/PhysRevD.58.063001}
  {\bibfield  {journal} {\bibinfo  {journal} {Physical Review D}\ }\textbf
  {\bibinfo {volume} {58}},\ \bibinfo {pages} {063001} (\bibinfo {year}
  {1998})}\BibitemShut {NoStop}%
\bibitem [{\citenamefont {Santamaria}\ \emph {et~al.}(2010)\citenamefont
  {Santamaria} \emph {et~al.}}]{Santamaria:2010yb}%
  \BibitemOpen
  \bibfield  {author} {\bibinfo {author} {\bibfnamefont {L.}~\bibnamefont
  {Santamaria}} \emph {et~al.},\ }\bibfield  {title} {\bibinfo {title}
  {{Matching post-Newtonian and numerical relativity waveforms: systematic
  errors and a new phenomenological model for non-precessing black hole
  binaries}},\ }\href {https://doi.org/10.1103/PhysRevD.82.064016} {\bibfield
  {journal} {\bibinfo  {journal} {Phys. Rev. D}\ }\textbf {\bibinfo {volume}
  {82}},\ \bibinfo {pages} {064016} (\bibinfo {year} {2010})},\ \Eprint
  {https://arxiv.org/abs/1005.3306} {arXiv:1005.3306 [gr-qc]} \BibitemShut
  {NoStop}%
\bibitem [{\citenamefont {McWilliams}\ \emph {et~al.}(2010)\citenamefont
  {McWilliams}, \citenamefont {Kelly},\ and\ \citenamefont
  {Baker}}]{McWilliams:2010eq}%
  \BibitemOpen
  \bibfield  {author} {\bibinfo {author} {\bibfnamefont {S.~T.}\ \bibnamefont
  {McWilliams}}, \bibinfo {author} {\bibfnamefont {B.~J.}\ \bibnamefont
  {Kelly}},\ and\ \bibinfo {author} {\bibfnamefont {J.~G.}\ \bibnamefont
  {Baker}},\ }\bibfield  {title} {\bibinfo {title} {{Observing mergers of
  non-spinning black-hole binaries}},\ }\href
  {https://doi.org/10.1103/PhysRevD.82.024014} {\bibfield  {journal} {\bibinfo
  {journal} {Phys. Rev. D}\ }\textbf {\bibinfo {volume} {82}},\ \bibinfo
  {pages} {024014} (\bibinfo {year} {2010})},\ \Eprint
  {https://arxiv.org/abs/1004.0961} {arXiv:1004.0961 [gr-qc]} \BibitemShut
  {NoStop}%
\bibitem [{\citenamefont {Sun}\ \emph {et~al.}(2020)\citenamefont {Sun} \emph
  {et~al.}}]{Sun:2020wke}%
  \BibitemOpen
  \bibfield  {author} {\bibinfo {author} {\bibfnamefont {L.}~\bibnamefont
  {Sun}} \emph {et~al.},\ }\bibfield  {title} {\bibinfo {title}
  {{Characterization of systematic error in Advanced LIGO calibration}},\
  }\href {https://doi.org/10.1088/1361-6382/abb14e} {\bibfield  {journal}
  {\bibinfo  {journal} {Class. Quant. Grav.}\ }\textbf {\bibinfo {volume}
  {37}},\ \bibinfo {pages} {225008} (\bibinfo {year} {2020})},\ \Eprint
  {https://arxiv.org/abs/2005.02531} {arXiv:2005.02531 [astro-ph.IM]}
  \BibitemShut {NoStop}%
\bibitem [{\citenamefont {Abbott}\ \emph
  {et~al.}(2021{\natexlab{d}})\citenamefont {Abbott} \emph
  {et~al.}}]{theligoscientificcollaboration2021:ObservationGravitationalWaves}%
  \BibitemOpen
  \bibfield  {author} {\bibinfo {author} {\bibfnamefont {R.}~\bibnamefont
  {Abbott}} \emph {et~al.} (\bibinfo {collaboration} {LIGO Scientific
  Collaboration, KAGRA Collaboration, Virgo Collaboration}),\ }\bibfield
  {title} {\bibinfo {title} {{Observation of Gravitational Waves from Two
  Neutron Star\textendash{}Black Hole Coalescences}},\ }\href
  {https://doi.org/10.3847/2041-8213/ac082e} {\bibfield  {journal} {\bibinfo
  {journal} {Astrophys. J. Lett.}\ }\textbf {\bibinfo {volume} {915}},\
  \bibinfo {pages} {L5} (\bibinfo {year} {2021}{\natexlab{d}})},\ \Eprint
  {https://arxiv.org/abs/2106.15163} {arXiv:2106.15163 [astro-ph.HE]}
  \BibitemShut {NoStop}%
\bibitem [{\citenamefont {Cutler}\ and\ \citenamefont
  {Vallisneri}(2007)}]{cutler2007_LISADetectionsMassive}%
  \BibitemOpen
  \bibfield  {author} {\bibinfo {author} {\bibfnamefont {C.}~\bibnamefont
  {Cutler}}\ and\ \bibinfo {author} {\bibfnamefont {M.}~\bibnamefont
  {Vallisneri}},\ }\bibfield  {title} {\bibinfo {title} {{{LISA}} detections of
  massive black hole inspirals: Parameter extraction errors due to inaccurate
  template waveforms},\ }\href {https://doi.org/10.1103/PhysRevD.76.104018}
  {\bibfield  {journal} {\bibinfo  {journal} {Physical Review D}\ }\textbf
  {\bibinfo {volume} {76}},\ \bibinfo {pages} {104018} (\bibinfo {year}
  {2007})},\ \Eprint {https://arxiv.org/abs/0707.2982} {arXiv:0707.2982}
  \BibitemShut {NoStop}%
\bibitem [{\citenamefont {Abbott}\ \emph {et~al.}(2019)\citenamefont {Abbott}
  \emph {et~al.}}]{abbott2019:PropertiesBinaryNeutron}%
  \BibitemOpen
  \bibfield  {author} {\bibinfo {author} {\bibfnamefont {B.~P.}\ \bibnamefont
  {Abbott}} \emph {et~al.},\ }\bibfield  {title} {\bibinfo {title} {Properties
  of the {{Binary Neutron Star Merger GW170817}}},\ }\href
  {https://doi.org/10.1103/PhysRevX.9.011001} {\bibfield  {journal} {\bibinfo
  {journal} {Physical Review X}\ }\textbf {\bibinfo {volume} {9}},\ \bibinfo
  {pages} {011001} (\bibinfo {year} {2019})}\BibitemShut {NoStop}%
\bibitem [{\citenamefont {Biscoveanu}\ \emph {et~al.}(2021)\citenamefont
  {Biscoveanu}, \citenamefont {Isi}, \citenamefont {Varma},\ and\ \citenamefont
  {Vitale}}]{Biscoveanu:2021nvg}%
  \BibitemOpen
  \bibfield  {author} {\bibinfo {author} {\bibfnamefont {S.}~\bibnamefont
  {Biscoveanu}}, \bibinfo {author} {\bibfnamefont {M.}~\bibnamefont {Isi}},
  \bibinfo {author} {\bibfnamefont {V.}~\bibnamefont {Varma}},\ and\ \bibinfo
  {author} {\bibfnamefont {S.}~\bibnamefont {Vitale}},\ }\bibfield  {title}
  {\bibinfo {title} {{Measuring the spins of heavy binary black holes}},\
  }\href {https://doi.org/10.1103/PhysRevD.104.103018} {\bibfield  {journal}
  {\bibinfo  {journal} {Phys. Rev. D}\ }\textbf {\bibinfo {volume} {104}},\
  \bibinfo {pages} {103018} (\bibinfo {year} {2021})},\ \Eprint
  {https://arxiv.org/abs/2106.06492} {arXiv:2106.06492 [gr-qc]} \BibitemShut
  {NoStop}%
\bibitem [{\citenamefont {Varma}\ and\ \citenamefont
  {Ajith}(2017)}]{Varma:2016dnf}%
  \BibitemOpen
  \bibfield  {author} {\bibinfo {author} {\bibfnamefont {V.}~\bibnamefont
  {Varma}}\ and\ \bibinfo {author} {\bibfnamefont {P.}~\bibnamefont {Ajith}},\
  }\bibfield  {title} {\bibinfo {title} {{Effects of nonquadrupole modes in the
  detection and parameter estimation of black hole binaries with nonprecessing
  spins}},\ }\href {https://doi.org/10.1103/PhysRevD.96.124024} {\bibfield
  {journal} {\bibinfo  {journal} {Phys. Rev. D}\ }\textbf {\bibinfo {volume}
  {96}},\ \bibinfo {pages} {124024} (\bibinfo {year} {2017})},\ \Eprint
  {https://arxiv.org/abs/1612.05608} {arXiv:1612.05608 [gr-qc]} \BibitemShut
  {NoStop}%
\bibitem [{\citenamefont {Colleoni}\ \emph {et~al.}(2021)\citenamefont
  {Colleoni}, \citenamefont {Mateu-Lucena}, \citenamefont {Estell\'es},
  \citenamefont {Garc\'\i{}a-Quir\'os}, \citenamefont {Keitel}, \citenamefont
  {Pratten}, \citenamefont {Ramos-Buades},\ and\ \citenamefont
  {Husa}}]{Colleoni:2020tgc}%
  \BibitemOpen
  \bibfield  {author} {\bibinfo {author} {\bibfnamefont {M.}~\bibnamefont
  {Colleoni}}, \bibinfo {author} {\bibfnamefont {M.}~\bibnamefont
  {Mateu-Lucena}}, \bibinfo {author} {\bibfnamefont {H.}~\bibnamefont
  {Estell\'es}}, \bibinfo {author} {\bibfnamefont {C.}~\bibnamefont
  {Garc\'\i{}a-Quir\'os}}, \bibinfo {author} {\bibfnamefont {D.}~\bibnamefont
  {Keitel}}, \bibinfo {author} {\bibfnamefont {G.}~\bibnamefont {Pratten}},
  \bibinfo {author} {\bibfnamefont {A.}~\bibnamefont {Ramos-Buades}},\ and\
  \bibinfo {author} {\bibfnamefont {S.}~\bibnamefont {Husa}},\ }\bibfield
  {title} {\bibinfo {title} {{Towards the routine use of subdominant harmonics
  in gravitational-wave inference: Reanalysis of GW190412 with generation X
  waveform models}},\ }\href {https://doi.org/10.1103/PhysRevD.103.024029}
  {\bibfield  {journal} {\bibinfo  {journal} {Phys. Rev. D}\ }\textbf {\bibinfo
  {volume} {103}},\ \bibinfo {pages} {024029} (\bibinfo {year} {2021})},\
  \Eprint {https://arxiv.org/abs/2010.05830} {arXiv:2010.05830 [gr-qc]}
  \BibitemShut {NoStop}%
\bibitem [{\citenamefont {Vitale}\ \emph {et~al.}(2014)\citenamefont {Vitale},
  \citenamefont {Lynch}, \citenamefont {Veitch}, \citenamefont {Raymond},\ and\
  \citenamefont {Sturani}}]{Vitale:2014mka}%
  \BibitemOpen
  \bibfield  {author} {\bibinfo {author} {\bibfnamefont {S.}~\bibnamefont
  {Vitale}}, \bibinfo {author} {\bibfnamefont {R.}~\bibnamefont {Lynch}},
  \bibinfo {author} {\bibfnamefont {J.}~\bibnamefont {Veitch}}, \bibinfo
  {author} {\bibfnamefont {V.}~\bibnamefont {Raymond}},\ and\ \bibinfo {author}
  {\bibfnamefont {R.}~\bibnamefont {Sturani}},\ }\bibfield  {title} {\bibinfo
  {title} {{Measuring the spin of black holes in binary systems using
  gravitational waves}},\ }\href
  {https://doi.org/10.1103/PhysRevLett.112.251101} {\bibfield  {journal}
  {\bibinfo  {journal} {Phys. Rev. Lett.}\ }\textbf {\bibinfo {volume} {112}},\
  \bibinfo {pages} {251101} (\bibinfo {year} {2014})},\ \Eprint
  {https://arxiv.org/abs/1403.0129} {arXiv:1403.0129 [gr-qc]} \BibitemShut
  {NoStop}%
\bibitem [{\citenamefont {Fairhurst}\ \emph {et~al.}(2020)\citenamefont
  {Fairhurst}, \citenamefont {Green}, \citenamefont {Hoy}, \citenamefont
  {Hannam},\ and\ \citenamefont {Muir}}]{Fairhurst:2019vut}%
  \BibitemOpen
  \bibfield  {author} {\bibinfo {author} {\bibfnamefont {S.}~\bibnamefont
  {Fairhurst}}, \bibinfo {author} {\bibfnamefont {R.}~\bibnamefont {Green}},
  \bibinfo {author} {\bibfnamefont {C.}~\bibnamefont {Hoy}}, \bibinfo {author}
  {\bibfnamefont {M.}~\bibnamefont {Hannam}},\ and\ \bibinfo {author}
  {\bibfnamefont {A.}~\bibnamefont {Muir}},\ }\bibfield  {title} {\bibinfo
  {title} {{Two-harmonic approximation for gravitational waveforms from
  precessing binaries}},\ }\href {https://doi.org/10.1103/PhysRevD.102.024055}
  {\bibfield  {journal} {\bibinfo  {journal} {Phys. Rev. D}\ }\textbf {\bibinfo
  {volume} {102}},\ \bibinfo {pages} {024055} (\bibinfo {year} {2020})},\
  \Eprint {https://arxiv.org/abs/1908.05707} {arXiv:1908.05707 [gr-qc]}
  \BibitemShut {NoStop}%
\bibitem [{\citenamefont {Krishnendu}\ and\ \citenamefont
  {Ohme}(2022)}]{Krishnendu:2021cyi}%
  \BibitemOpen
  \bibfield  {author} {\bibinfo {author} {\bibfnamefont {N.~V.}\ \bibnamefont
  {Krishnendu}}\ and\ \bibinfo {author} {\bibfnamefont {F.}~\bibnamefont
  {Ohme}},\ }\bibfield  {title} {\bibinfo {title} {{Interplay of
  spin-precession and higher harmonics in the parameter estimation of binary
  black holes}},\ }\href {https://doi.org/10.1103/PhysRevD.105.064012}
  {\bibfield  {journal} {\bibinfo  {journal} {Phys. Rev. D}\ }\textbf {\bibinfo
  {volume} {105}},\ \bibinfo {pages} {064012} (\bibinfo {year} {2022})},\
  \Eprint {https://arxiv.org/abs/2110.00766} {arXiv:2110.00766 [gr-qc]}
  \BibitemShut {NoStop}%
\bibitem [{\citenamefont {Abbott}\ and\ \citenamefont {{et
  al}}(2020)}]{abbott2020_ProspectsObservingLocalizing}%
  \BibitemOpen
  \bibfield  {author} {\bibinfo {author} {\bibfnamefont {B.~P.}\ \bibnamefont
  {Abbott}}\ and\ \bibinfo {author} {\bibnamefont {{et al}}},\ }\bibfield
  {title} {\bibinfo {title} {Prospects for observing and localizing
  gravitational-wave transients with {{Advanced LIGO}}, {{Advanced Virgo}} and
  {{KAGRA}}},\ }\href {https://doi.org/10.1007/s41114-020-00026-9} {\bibfield
  {journal} {\bibinfo  {journal} {Living Reviews in Relativity}\ }\textbf
  {\bibinfo {volume} {23}},\ \bibinfo {pages} {3} (\bibinfo {year}
  {2020})}\BibitemShut {NoStop}%
\bibitem [{\citenamefont {Pannarale}\ \emph {et~al.}(2015)\citenamefont
  {Pannarale}, \citenamefont {Berti}, \citenamefont {Kyutoku}, \citenamefont
  {Lackey},\ and\ \citenamefont {Shibata}}]{Pannarale:2015jka}%
  \BibitemOpen
  \bibfield  {author} {\bibinfo {author} {\bibfnamefont {F.}~\bibnamefont
  {Pannarale}}, \bibinfo {author} {\bibfnamefont {E.}~\bibnamefont {Berti}},
  \bibinfo {author} {\bibfnamefont {K.}~\bibnamefont {Kyutoku}}, \bibinfo
  {author} {\bibfnamefont {B.~D.}\ \bibnamefont {Lackey}},\ and\ \bibinfo
  {author} {\bibfnamefont {M.}~\bibnamefont {Shibata}},\ }\bibfield  {title}
  {\bibinfo {title} {{Aligned spin neutron star-black hole mergers: a
  gravitational waveform amplitude model}},\ }\href
  {https://doi.org/10.1103/PhysRevD.92.084050} {\bibfield  {journal} {\bibinfo
  {journal} {Phys. Rev. D}\ }\textbf {\bibinfo {volume} {92}},\ \bibinfo
  {pages} {084050} (\bibinfo {year} {2015})},\ \Eprint
  {https://arxiv.org/abs/1509.00512} {arXiv:1509.00512 [gr-qc]} \BibitemShut
  {NoStop}%
\bibitem [{\citenamefont {Gamba}\ \emph
  {et~al.}(2021{\natexlab{b}})\citenamefont {Gamba}, \citenamefont {Ak\c{c}ay},
  \citenamefont {Bernuzzi},\ and\ \citenamefont {Williams}}]{Gamba:2021ydi}%
  \BibitemOpen
  \bibfield  {author} {\bibinfo {author} {\bibfnamefont {R.}~\bibnamefont
  {Gamba}}, \bibinfo {author} {\bibfnamefont {S.}~\bibnamefont {Ak\c{c}ay}},
  \bibinfo {author} {\bibfnamefont {S.}~\bibnamefont {Bernuzzi}},\ and\
  \bibinfo {author} {\bibfnamefont {J.}~\bibnamefont {Williams}},\ }\href@noop
  {} {\bibinfo {title} {{Effective-one-body waveforms for precessing coalescing
  compact binaries with post-Newtonian Twist}}} (\bibinfo {year}
  {2021}{\natexlab{b}}),\ \Eprint {https://arxiv.org/abs/2111.03675}
  {arXiv:2111.03675 [gr-qc]} \BibitemShut {NoStop}%
\bibitem [{\citenamefont {Chiaramello}\ and\ \citenamefont
  {Nagar}(2020)}]{Chiaramello:2020ehz}%
  \BibitemOpen
  \bibfield  {author} {\bibinfo {author} {\bibfnamefont {D.}~\bibnamefont
  {Chiaramello}}\ and\ \bibinfo {author} {\bibfnamefont {A.}~\bibnamefont
  {Nagar}},\ }\bibfield  {title} {\bibinfo {title} {{Faithful analytical
  effective-one-body waveform model for spin-aligned, moderately eccentric,
  coalescing black hole binaries}},\ }\href
  {https://doi.org/10.1103/PhysRevD.101.101501} {\bibfield  {journal} {\bibinfo
   {journal} {Phys. Rev. D}\ }\textbf {\bibinfo {volume} {101}},\ \bibinfo
  {pages} {101501} (\bibinfo {year} {2020})},\ \Eprint
  {https://arxiv.org/abs/2001.11736} {arXiv:2001.11736 [gr-qc]} \BibitemShut
  {NoStop}%
\bibitem [{\citenamefont {Dietrich}\ \emph {et~al.}(2019)\citenamefont
  {Dietrich}, \citenamefont {Samajdar}, \citenamefont {Khan}, \citenamefont
  {Johnson-McDaniel}, \citenamefont {Dudi},\ and\ \citenamefont
  {Tichy}}]{Dietrich:2019kaq}%
  \BibitemOpen
  \bibfield  {author} {\bibinfo {author} {\bibfnamefont {T.}~\bibnamefont
  {Dietrich}}, \bibinfo {author} {\bibfnamefont {A.}~\bibnamefont {Samajdar}},
  \bibinfo {author} {\bibfnamefont {S.}~\bibnamefont {Khan}}, \bibinfo {author}
  {\bibfnamefont {N.~K.}\ \bibnamefont {Johnson-McDaniel}}, \bibinfo {author}
  {\bibfnamefont {R.}~\bibnamefont {Dudi}},\ and\ \bibinfo {author}
  {\bibfnamefont {W.}~\bibnamefont {Tichy}},\ }\bibfield  {title} {\bibinfo
  {title} {{Improving the NRTidal model for binary neutron star systems}},\
  }\href {https://doi.org/10.1103/PhysRevD.100.044003} {\bibfield  {journal}
  {\bibinfo  {journal} {Phys. Rev. D}\ }\textbf {\bibinfo {volume} {100}},\
  \bibinfo {pages} {044003} (\bibinfo {year} {2019})},\ \Eprint
  {https://arxiv.org/abs/1905.06011} {arXiv:1905.06011 [gr-qc]} \BibitemShut
  {NoStop}%
\end{thebibliography}%

\end{document}